\documentclass{aa}  
\usepackage{graphicx}
\usepackage{epstopdf}
\usepackage{caption3}
\usepackage{subfig}

\usepackage{txfonts}

\usepackage{color}

\begin{document}

   \title{Barlenses and X-shape features compared: different manifestations of Boxy/Peanut bulges}
   

   \author{E. Laurikainen
          \inst{1}
          \and
          H. Salo\inst{1}
          }

   \institute{Astronomy Research Unit, University of Oulu, FI-90014 Finland  \\
              \email{eija.laurikainen@oulu.fi}
             }

   \date{Received: 14.5.2016, accepted: 31.8.2016}

   \abstract
{Morphological characteristics of Boxy/Peanut bulges are studied,
  in particular whether most of the flux associated to bulges in
  galaxies with masses similar to those of the Milky Way at redshift z$\sim$0, could
  belong to vertically thick inner part of the bar, in a
  similar manner as in the Milky Way itself.  At high galaxy
  inclinations such structures manifest as Boxy/Peanut/X-shape
  features, and near to face-on view as barlenses. We also study the
  possibility that bulges in some fraction of unbarred galaxies could
  form in a similar manner as the bulges in barred galaxies.}
{We use the Spitzer Survey of Stellar Structure in Galaxies (S$^4$G) and the
  Near-IR S0 galaxy Survey (NIRS0S), to compile complete samples of
  galaxies with barlenses (N = 85), and X-shape features (N = 88). A
  sample of unbarred galaxies (N = 41) is also selected, based on
  similarity in their surface brightness profiles with those of
  barlens galaxies. Sizes and minor-to-major axis ratios (b/a) of these presumably
  vertically thick inner bar components are compared, and interpreted
  by means of synthetic images using N-body simulation models.
  Barlenses and their parent galaxies are divided into different
  sub-groups. Their possible parent galaxy counterparts in 
  galaxies where the barlenses are manifested as X-shape features, are also identified.}
 {Unsharp  mask images are created for all 214 sample galaxies. These
   images are used to recognize the X-shape features, and to measure
   their linear sizes, both along and perpendicular to the bar. For
   detecting possible boxy isophotes (using B$_4$ -parameter), isophotal
   analysis is also performed for the barlens galaxies. In the
   interpretation N-body simulations from \citet{salo2016} 
are used:
   the models, exhibiting Boxy/Peanut/X/barlens morphologies, are viewed from
   isotropically chosen directions, covering the full range of galaxy
   inclinations in the sky. The created synthetic images are
   analyzed in a similar manner as the observations.}
%
{This is the first time that the observed properties of 
  barlenses and X-shape features are directly compared, over a large range
  of galaxy inclinations. A comparison with the
simulation models shows that the differences in their apparent sizes ,
a/r$_{\rm bar} \gtrsim$
  0.5 for barlenses and a/r$_{\rm bar}$ $\lesssim$ 0.5 for X-shapes, can be
  explained by projection effects. Observations at various
  inclinations are consistent with intrinsic 
a$_{\rm bl} \approx$ a$_{\rm X} \approx$ 0.5 r$_{\rm bar}$: 
here intrinsic size means the face-on
  semimajor axis length for bars and barlenses, and the semilength of
  X-shape when the bar is viewed exactly edge on. While X-shapes are
  quite common at intermediate galaxy inclinations (for $i$ = 40$^\circ$ -
  60$^\circ$ their frequency is $\sim$ half of barlenses), they are
  seldom observed at smaller inclinations. { This is consistent
    with our simulation models which have a small compact classical bulge
    producing a steep inner rotation slope, whereas bulgeless shallow rotation
    curve models predict that X-shapes should be visible even in face-on
    geometry. The steep rotation curve models are also consistent with
    the observed trend with B$_4$ being positive at low inclination, and 
getting negative values
    for $i$ $\gtrsim$ 40$^\circ$-60$^\circ$, thus implying boxy isophotes}. In total,
    only about one quarter of barlenses (with $i$ $\le$ 60$^\circ$) show boxy
    isophotes.}
%
 {Our
analysis are consistent with the idea that barlenses and X-shape
features are physically the same phenomenon. However, which of the two
features is observed in a galaxy depends, not only on galaxy inclination,
but also on its central flux concentration.  The
  observed nearly round face-on barlens morphology is expected when at
  least a few percents of the disk mass is in a central component,
  within a region much smaller than the size of the barlens itself.
Barlenses participate to secular evolution of galaxies, and might even
act as a transition phase between barred and unbarred galaxies.  We
also discuss that the large range of stellar population ages obtained
for the photometric bulges in the literature, are consistent with our
interpretation.  }


   \maketitle
%

\section{Introduction}

What is the amount of baryonic mass confined into the bulges of
galaxies and how was that mass accumulated, is a critical question to
answer while constructing models of galaxy formation and evolution.
The answer to this question depends on how well the different bulge
components can be recognized, and assigned to possible physical
processes making those structures.  Most of the bulge mass associated
to photometric bulges (ie. flux above the disk) is generally assumed
to reside in classical bulges. These are relaxed, velocity dispersion
supported structures, presumably formed by galaxy mergers \citep{white1978,hopkins2009}, 
or by coalescence of massive star
forming clumps at high redshifts, drifted towards the central regions
of the galaxies  
(Bournaud et al. 2008; Elmegreen et al. 2009; see also
review by Kormendy 2016). 
This picture has been challenged by the
discovery that most of the bulge mass in the Milky Way actually
resides in a Boxy/Peanut (B/P) bar, showing also evidence of an
X-shape morphology, without any clear evidence of a classical bulge
\citep{mcwilliam2010,nataf2010,wegg2013,ness2016}.
Whether such bar-related inner structures could form most of
the bulge mass also in external Milky Way mass galaxies is a topic of
this study.

Boxy/Peanut (B/P) bulges are easy to distinguish in the edge-on view and it
has been shown that even 2/3 of all disk galaxies in S0-Sd types
have B/Ps 
(L\"utticke, Dettmar $\&$ Pohlen 2000; Bureau et al. 2006;
but see also Yoshino $\&$ Yamauchi 2014). Many B/P bulges also show
cylindrical rotation \citep{kormendy1982,bureau1999,falco2006,falco2016,ianuzzi2015}, 
which generally confirms their bar
origin.  Verification of a galaxy as barred is difficult in the
edge-on view, but it has been shown that, at an optimal range of
viewing angles, B/Ps are visible even in less inclined galaxies, as
revealed by their boxy isophotes 
(Beaton et al. 2007; Erwin $\&$ Debattista 2013, hereafter ED2013). A new morphological feature, a
barlens (bl), was recognized by \citet{lauri2011}, 
and it has been
suggested (Laurikainen et al. 2014, hereafter L+2014; Athanassoula et
al. 2015, hereafter A+2015; see also Laurikainen et al. 2007) that
they might be the face-on counterparts of B/P bulges. Association of a
barlens to the Milky Way bulge has been recently made by  \citet{gerhard2016}.

Because of their fairly round appearance barlenses are often
erroneously associated with classical bulges (see the review by
Laurikainen $\&$ Salo 2016), but there is cumulative evidence
showing that barlenses might indeed form part of the bar. Their
optical colors are very similar to the colors of bars (Herrera-Endoqui
et al. 2016, hereafter HE+2016), and in particular, their surface brightness profiles are
  very similar to those predicted for the B/P-bulges in
hydrodynamical simulation models when viewed face-on (A+2015).
The first indirect observational evidence connecting barlenses with B/P
bulges (which often have X-shape features in unsharp mask images), was
based on the axial ratio distribution of the combined sample of their parent galaxies,
which appeared to be flat
(L+2014).  However, it remained unclear why barlenses concentrate
  on earlier Hubble types than the B/P/X-shape bulges (peak values are T = -1
and T = +1, respectively). 
Is this simply an observational bias when classifying galaxies
   at low and high inclinations, or could it indicate some
   intrinsic difference between the parent galaxies hosting barlenses
   and X-shape features? The latter possibility is suggested by the
   recent N-body simulations by Salo $\&$ Laurikainen (2016; submitted to
   ApJ), who demonstrated that a steep inner rotation curve
   leads to realistic-looking round barlens morphology, with no
   trace of an X-shape in the face-on geometry. However, reducing the
   central mass concentration, and thus shifting the galaxy
   to a later Hubble type, produced more elongated barlenses, which
   exhibited X-features at a much larger range of galaxy inclinations.

As barred and unbarred galaxies presumably appear in similar galaxy
environments (see Aquerri, M\'endez-Abreu $\&$ Corsini 2009), 
it is not plausible that bulges in barred
galaxies form smoothly by secular evolution, and bulges in unbarred
galaxies by some violent processes, like major galaxy
mergers. Therefore, our hypothesis that many classical bulges are
misclassified B/P/X features can be valid only if an explanation is
found also for the bulges of unbarred galaxies, in
the same line with the explanation for the barred galaxies.  In fact,
there is observational evidence which hints to that direction.
Namely, the inner lenses (normalized to galaxy size) in unbarred
galaxies are shown to have similar sizes as barlenses in barred
galaxies (Laurikainen et al. 2013; Herrera-Endoqui et al. 2015,
hereafter HE+2015). Inner lenses in unbarred galaxies might therefore represent evolved bars where the thin
bar component has been completely dissolved, or the classical elongated
bar never formed.  However, whether those lenses
are also vertically thick needs to be shown. 

In this study the properties of 85 barlenses and their parent
galaxies are studied, and compared with the properties of 88 
galaxies hosting bars with X-shape inner feature.
An additional sample of 41 unbarred galaxies is also
selected.  As a database we use the Spitzer Survey of Stellar
Structure in Galaxies \citep{sheth2010}, and the Near-IR S0
galaxy Survey \citep{lauri2011}.
The properties of the analyzed features are compared with those obtained for synthetic
images, created from 
simulation models taken from \citet{salo2016}. 
To obtain a fair comparison, the analysis for the
synthetic images is done in a similar manner as for the observations.


 \section{Data and the sample selection}

{ The Spitzer Survey of Stellar Structure in Galaxies (S$^4$G, Sheth
  et al. 2010) is a sample of 2352 nearby galaxies observed at mid-IR
  wavelengths, covering all Hubble types and disk inclinations. The
  galaxies have HI radial velocities V$_{\rm radio}$ $<$ 3000
  kms$^{-1}$ corresponding to the distance of D $<$ 40 Mpc for
  H$_{\circ}$ = 75 km s$^{-1}$, and blue photographic magnitudes
  B$_{\rm T}$ $\le$ 15.5 mag.  The Near-IR S0 galaxy Survey (NIRS0S,
  Laurikainen et al. 2011) is a survey of 185 S0-Sa galaxies, having
  magnitudes B$_{\rm T}$ $\le$ 12.5, and galaxy inclinations of $i$ $\le$ 65$^\circ$ 
(N = 215 if included are also the galaxies which slightly exceed
  the magnitude-limit). The morphological classifications of the
  S$^4$G galaxies are from \citet{buta2015}, 
and those of NIRS0S from \citet{lauri2011}
using the same classification
  criteria. The wavelengths used in the above classifications are 3.6
  $\mu$m in S$^4$G, and 2.2 $\mu$m in NIRS0S: both are fairly dust free
  regimes allowing to recognize the morphological features of the old
  stellar population.

The quality of the images is explained by \citet{lauri2011} for NIRS0S, and by \citet{salo2015},
  \citet{querejeta2015}, and \citet{munozmateos2013} for the S$^4$G
  images. The NIRS0S images typically have pixel resolution of 0.25
  arcsec, FWHM $\sim$ 1 arcsec, and FOV of 4-5 arcmin.  The
  images typically reach a surface brightnesses of 23 mag arcsec$^{-2}$
  in $K_s$, equivalent to 27 mag arcsec$^{-2}$ in the B-band. The S$^4$G
  images have a pixel resolution of 0.75 arcsec, and FWHM 2.1 arcsec,
  and they reach a surface brightness of 27 AB (1$\sigma$) mag
  arcsec$^{-2}$ at 3.6 $\mu$m (equivalent to roughly 28 mag
  arcsec$^{-2}$ at the B band). For large galaxies 
  the S$^4$G images are mosaics, covering at least 1.5 x D$_{\rm 25}$, where
  D$_{\rm 25}$ is the isophotal size of the galaxy in B-band.

From the combined S$^4$G + NIRS0S sample all barred galaxies with a
barlens (bl) in the classification were selected, which makes 84
barlens galaxies. We added to this category also NGC 1433 which
clearly has a barlens, although it is missing in the original
  classification. For the recognition of the X-shape features unsharp
masks were first done for all the S$^4$G and NIRS0S galaxies:
weak X-features can be recognized from unsharp mask images even if
they were not visible in the direct images.  This makes 88 galaxies
with identified X-shape features. In 6 of the galaxies both a barlens
and an X-shape feature were identified.  We also selected a sample of
41 largely unbarred galaxies which have similar exponential surface brightness
profiles as barlenses typically have.  Those galaxies can have inner
lenses or ringlenses (34 galaxies), but not all of them have.  A few
of them are classified as weakly barred (AB) by \citet{buta2015}. The final samples are:
\vskip 0.15cm

(1) galaxies with barlenses (N = 85)

(2) galaxies with X-shape features (N = 88) 

(3) unbarred galaxies (N = 41)
\vskip 0.15cm

Compared to the total number of galaxies in our starting
  S$^4$G+NIRS0S sample the numbers for the barlens and X-shape
  galaxies are fairly small.  This is mainly because S$^4$G, being a magnitude
  limited sample, is dominated by low luminosity late-type galaxies, whereas
  the B/P/bl features typically appear in bright galaxies with strong
  bars: of the bright galaxies $\sim$ 2/3 have bars and only a
  half of the bars are strong. It was shown by L+2014 that while
  concentrating on galaxies with -3 $\le$ T $\le$ -2,
  $i<$65$^{\circ}$, and B$_{\rm T} <$ 12.5 mag, $\sim$ 46$\%$ of the barred
  galaxies in our sample have either a barlens or an X-shape feature.

 The inclination distribution of the galaxies in our combined sample has been
 previously studied by L+2014. Although barlenses are preferentially
 concentrated to galaxies with low inclinations, and the X-shapes to galaxies 
with high inclinations,
 there is a large overlap in their parent galaxy inclinations. This
 makes our sample ideal for comparing the properties of these structures.  The 
three selected samples are shown in Tables 1, 2 and 3.  In the tables
 given are also the morphological classifications from \citet{buta2015}.
If the galaxy does not appear in S$^4$G the classification
is from Laurikainen et al. (2011). For the following galaxies differences
appear in the above classifications: for NGC 584 {SA(l)0$^-$(NIRS0S) / E(d)(S$^4$G)}, 
for NGC 5631 {SA(l)0$^-$(NIRS0S) / E0-1(S$^4$G)}, and for NGC 5646 {SA(l)0$^-$(NIRS0S) / E0-1(S$^4$G)}.
For these galaxies both classifications are given.

 The sizes of bars and barlenses are
 from HE+2015 and HE+2016, respectively.  The orientation parameters
 and the scale lengths of the disks are from \citet{salo2015}.
 For bars visual length estimates are used, because they
 are measured in a homogeneous manner for all galaxies in our sample.
 In HE+2015 and in \citet{diaz2016a} it was shown that
 the visual bar length estimates are fairly similar to those obtained from the
 maximum ellipticity in the bar region.

\section{Methods}

\subsection{Unsharp masks}
\vskip 0.5cm

In this study unsharp mask images for the complete sample of 214
galaxies were done. For the galaxies with X-shaped bars we end up to
the same sample as used by L+2014.  The images were first convolved
with a Gaussian kernel (mean $\sigma \sim 4$ arcsec), and the
original images were then divided with the convolved images. In order
to show possible low surface brightness structures special attention
was paid to find optimal parameters to illustrate the
morphologies. Widths of the Gaussian kernels that best
  illustrated the faint features were found empirically by inspecting
  a large range of values for each galaxy.  Our Gaussian convolution
method avoids possible artifacts that might appear in the simple
approach where the images are divided by rebinned images. The original
and unsharp mask images, as well as the surface brightness profiles
for the sample of 214 galaxies  are shown in electronic form 
({/www.oulu.fi/astronomy/BLX/}) The electronic file is
organized in the following manner, with an increasing NGC number in
each group: 
\vskip 0.15cm

   1. Strongly barred with a barlens (bl$_{\rm B}$)

   2. Weakly barred 

 \hskip 0.35cm     a) barlens in the classification (bl$_{\rm AB}$)

\hskip 0.35cm      b) no barlens in the classification, but has a barlens-like 

 \hskip 0.65cm surface brightness profile (AB)

   3. Unbarred

 \hskip 0.35cm     a) have an inner lens (A$_{\rm l}$)

\hskip 0.35cm      b) no inner lens appears, but similar surface brightness 

\hskip 0.65cm profile as in a barlens (A$_{\rm expo}$) 

   4. Have bl in classification and X in the unsharp mask (bl-X)

   5. X-shaped bar (X)
\vskip 0.15cm

\noindent 
The first three primary groups are selected based on the bar
  family (B, AB, A), whereas the last two groups appear to be mixtures of
  strongly and weakly barred galaxies, having similar numbers of both
  families. NGC 3384 is shown using both in K$_{\rm s}$ and 3.6 $\mu$m
  images, in which bands the galaxy shows very different morphology in the
  central regions: in K$_{\rm s}$ a central peak appears, whereas at 3.6 $\mu$m
the galaxy has a bright dispersed central region and a drop of flux in the very center.
The same web-page contains also files illustrtaing various barlens sub-groups
(defined in Section 6.1), and the sub-groups for their parent galaxies 
(defined in Section 6.2). For the galaxies both sky and deprojected 
images are shown.  

\subsection{Size measurements of the X-shape features}

 Using the obtained unsharp mask images (in the sky plane) the sizes
 of the X-shape features were measured and collected to Table 2, where the
orientation in respect of the ``thin bar'' is also given. An
 example illustration is shown in Figure \ref{fig:x_measure}: the four
 corners of the X are marked on the image, and the semi-lengths of the
 feature along the bar (a), and perpendicular to it (b) are obtained
 as mean values of the extents of the two sides.  In order to
 facilitate the measurements the images were first rotated so that the
 bar appeared horizontally.  Generally the X-shape features are clear
 (see NGC 2654), but particularly in the end-on view the 'horns'
 making the X are very weak. The measurements were repeated
 three times, and the mean values together with their errors (standard
 deviation of measurements divided by $\sqrt{3}$; typically less than
 0.5 arcseconds) are indicated.  Sizes of barlenses
 are taken from the previous measurements (HE+2016; their Table 2),
 where they were obtained by fitting ellipses to the points
 delineating the outer isophotes of the barlenses: this gives the
   semilengths along the major (a) and minor axis (b) of the barlens,
   and the orientation of the major axis.  The uncertainties were
   estimated in a similar manner as our uncertainties for the X-shape
   dimensions.

\subsection{Making synthetic images and showing them at different viewing angles}

For making the synthetic images we use N-body simulation models from
\citet{salo2016}. These simulations, performed with Gadget-2
\citep{springel2005}, addressed the influence of central mass
concentration on the formation of barlens features. In comparison to
the N-body + SPH simulation models by Athanassoula et al. (2013,
2015), the models are much simpler, consisting of only the stellar
components: a small pre-existing classical bulge, an exponential disk,
and a spherical halo.  No gas or star formation were included.  At two
scale-lengths (h$_{\rm r}$), the disk accounts for 65$\%$ of the total
radial force, the initial vertical thickness of the disc is 0.2
$\cdot$ h$_{\rm r}$, and the Toomre parameter Q $\sim$ 1.3.  We select
snapshots from two simulations, both about 3 Gyrs after the bar has
been formed and stabilized in strength. The two models differ in their
bulge-to-disk (B/D) mass ratios at the beginning of the simulation, so
that the model with B/D = 0.01 is practically bulgeless, whereas the
other model has a small bulge with B/D = 0.08.  The effective radius
of the bulge was fixed to r$_{\rm eff}$/h$_{\rm r}$ = 0.07, which is the 
typical observed value to T = 3 galaxies (Salo et al. 2015).  Both
models develop a B/P/X bulge, and in particular in the model with
larger B/D the resemblance to the typical face-on barlens morphology
is very good.  The only difference here to the simulations displayed
in \citet{salo2016} is that we have increased the number of particles
by a factor of 5, in order to improve the quality of the synthetic
images: the behavior of the models is practically unaffected by the
increased particle number.

In order to have a representative sample of galaxy orientations in the
sky-plane, the two simulation snapshots were viewed from 100
isotropically chosen directions.  In practice this was done by viewing
the galaxy first from its pole ($i$ = 0$^{\circ}$), and then
spiraling around the galaxy with suitably selected constant steps in
azimuthal angle $\phi$ and cos $i$ (we used $\Delta \phi$ =
41.4$^\circ$, and |$\Delta$ \ cos $i$| = 0.02 ). The angle $\phi$ is
counted from the direction of the bar major axis. These images for
the model with B/D = 0.08 are shown in Figure
\ref{fig:simu_montage}. Barlenses and X-shape features in the
synthetic images were measured in a similar manner as in the
observations.

In Figures \ref{fig:simu_tilt}-\ref{fig:simu_azi_inc60} we show
how the morphology of the vertically thick inner bar component varies
with the viewing angle. In all these figures the left panels show the
images, in the middle panels the isophotal contours are overlaid, and
the right panels show the unsharp masks of the same images.  The
line-of-node is always horizontal.  The simulation had $5 \cdot
10^6$ disk particles, and to increase the $S/N$ of the synthetic
images, three simulation snapshots were superposed, after rotating the
bar to the same orientation. Moreover, we made use of the reflection
symmetry with respect to the equatorial plane, and the m = 2
rotational symmetry with respect to 180$^\circ$ rotation in the
equatorial plane. The effective number of disk particles in the
synthetic image thus corresponds to 60 $\cdot$ 10$^6$.  
In Figure \ref{fig:simu_tilt} the simulation models for B/D = 0.01 and 0.08
are shown at five different inclinations keeping the
azimuthal viewing angle fixed to $\phi$ = 90$^{\circ}$. 
In Figure {\ref{fig:simu_azi_inc90}} the inclination is fixed to $i$ = 90$^{\circ}$: in the
different panels the azimuthal angle varies from the end-on
($\phi$ = 0$^\circ$) to side-on view ($\phi$ = 90$^\circ$).  
Also in Figure {\ref{fig:simu_azi_inc60}}  the azimuthal angle varies, but the inclination is
fixed to $i$ = 60$^{\circ}$.

The sizes of the X-shape and barlens features measured from the
simulated images are compared with the observations in Section 5. Here
we emphasize some morphological differences depending on the viewing
angle. Figure \ref{fig:simu_tilt} shows that in the
simulation model with B/D = 0.01 the X-shape is at some level
visible at all galaxy inclinations in the unsharp mask images,
showing also peanut-shaped isophotes in the direct images.
In the model with B/D = 0.01 the size of the X-feature also
decreases towards lower galaxy inclinations.
However, when B/D = 0.08 the X-shape disappears when the inclination
gets smaller than $i$ = 30$^\circ$- 45$^\circ$ (also, the isophotes are not
boxy anymore). 

When fixing the inclination to $i$ = 90$^{\circ}$ and allowing the
azimuthal angle to vary (Fig. \ref{fig:simu_azi_inc90})  
some morphological differences also appear: as
expected the size of the X-feature shrinks towards the end-on view
($\phi$ = 0$^\circ$).  The X-feature is always present in the B/D = 0.01
model, whereas in the model with B/D = 0.08 it disappears in the end-on view.
Another comparison with varying azimuthal angle, but this time fixing
the inclination to $i$ = 60$^{\circ}$, is also interesting: in the
bulgeless model the X-shape gradually decreases in size, and finally
disappears near the end-on view (Fig. \ref{fig:simu_azi_inc60}). 
On the other hand, in the B/D = 0.08
model the X-shape feature rapidly disappears with a decreasing azimuthal
angle and it starts to resemble a barlens: depending on the
azimuthal angle it looks like a spheroidal ($\phi$ = 0$^{\circ}$), or
the ``thin bar'' appears as two twisted spiral-like features outside
the barlens ($\phi$ = 30$^{\circ}$- 40$^{\circ}$). Also ansae can be
identified is some of those bars at $\phi$ = 30$^{\circ}$- 40$^{\circ}$.

\subsection{Isophotal analysis of barlenses}

 \noindent We made isophotal analysis for the sample galaxies using
 the IRAF ellipse routine. It provides the parameters A$_4$ and
   B$_4$ which are associated to the sin 4$\theta$ and cos 4$\theta$
   terms of the Fourier expansion of the isophotal shape,
   respectively. The fourth order coefficients (A$_4$ and B$_4$) are
 generally used as descriptors of the deviations of the isophotes from
 simple ellipses: they are boxy when B$_4$ $<$ 0 and A$_4$ $>$ 0, and
 disky when B$_4$ $>$ 0 and A$_4$ $<$ 0. The best evidence of boxiness
 can be obtained using the B$_4$ parameter. In the boxy bar region PA
 is maintained nearly constant, and $\epsilon$ gradually increases
 towards the outer edge of the bar (see Beaton et al. 2007).  In this
 study the radial profiles of B$_4$, together with the profiles of the
 position angle (PA) and ellipticity ($\epsilon$ = 1-b/a) in the bar
 regions were derived for all barlens galaxies, of which an example is
 shown in Figure \ref{fig:b4_measure}.  In the surface brightness
 profile the boxy bar forms part of the photometric bulge. If the
 barlens is boxy it is marked in Table 1, based on visual inspection
 of the isophotes and the B$_4$ profile. The table indicates also
   the mean and standard deviation of $B_4$ in the barlens region.
 The importance of higher order Fourier modes for identifying X-shape
 features, particularly at high galaxy inclinations, has been
 discussed by \citet{chiambur2015}. However, barlens galaxies in our
 sample do not have such high galaxy inclinations (the highest
   galaxy inclination is 72$^{\circ}$). We find by visual inspection
 that only roughly one-quarter (19/79) of the barlenses have boxy
 isophotes.

In Figure \ref{fig:b4_compare} the obtained $<B_4>$ values 
are shown: the red and green
colors indicate our visual detection/non-detection of the boxy
isophotes, respectively, being in good agreement with the mean
$<B_4>$. Also shown are the $B_4$ values for the synthetic images
using the simulation model with B/D = 0.08.
It appears that both in observations and in synthetic images the
detection fraction of boxy isophotes increases with galaxy
inclination, so that the isophotes start to appear boxy at $i \gtrsim$ 45$^{\circ}$.  
We have chosen the model with B/D = 0.08,
  because for this model the vertically thick bar component manifests
  as a barlens for a large range of galaxy inclinations.  This B/D
  value is also close to that obtained for the 'true' bulge components
  of barlens and X-shaped galaxies in the multi-component
  decompositions by L+2014, carried out for a small but representative
  sample of 29 barlens/X-shaped galaxies.  For the $B/D=0.01$
  model the mean $B_4$ in the inner bar region would be negative for all galaxy inclinations,
  consistent with its X-shape morphology (see Figs. 3--5, and Fig. 21
  below).

Similar figures as Figure 6 are given for the all barlens galaxies
in electronic form at /www/astronomy/BLX..

\section{Comparison of barlenses and X-shape features}

The sizes of barlenses and X-shape features are compared in Figure
\ref{fig:size_compare}. 
The parameters are shown in the sky plane
because it is not possible to deproject the highly inclined X-shaped
galaxies in a reliable manner. It appears that the sizes of both
features correlate with r$_{\rm bar}$ (upper panel) so that the size
increases with r$_{\rm bar}$. However, the X-shape features are
clearly smaller than barlenses (the uncertainty of both bl and X measurements
is comparable to the size of the plotting symbols).
 The scatter is also larger for the X-shape
features, which is not unexpected having in mind that they appear at
larger galaxy inclinations, and because the apparent size depends also on the
angle between the X-shape and the bar major axis in a specific viewing angle.
For both features the
normalized (normalization to r$_{\rm bar}$) sizes are constant as a
function of the parent galaxy mass (lower panel), which means that the
size difference is not a mass effect. The galaxy masses are from \citet{munozmateos2013}, 
derived from the 3.6 $\mu$m and 4.5 $\mu$m images,
based on mass-to-luminosity ratios from \citet{eskew2012}.

Our size measurements for the X-shape features (a/r$_{\rm bar}$
$\sim$0.2--0.5, $<$a/r$_{\rm bar}$>$\sim$0.35) agree well with
a/r$_{\rm bar}$ $\sim$ 0.4 given for the B/P structures, in the
edge-on view by \citet{dettmar2000}, and at intermediate galaxy
inclinations by ED2013.  It will be discussed in Section 9.1,
that the boxy bulges by ED2013 are actually the same entities as what we
call X-shapes, for which reason such an agreement is expected.

Histograms of the minor-to-major axis ratios of barlenses and
X-shape features are compared in Figure \ref{fig:ba_compare}.
In our combined S$^4$G + NIRS0S sample barlenses have b/a =
0.4--1.0 (upper panel), which is in agreement with that shown
previously for the NIRS0S galaxies by A+2015.  The peak value in the
sky-plane is $\sim$ 0.75, which in the disk plane is shifted to
$\sim$0.85. A majority of the X-shape features appear in the same
b/a-range with barlenses (lower panel). However, there is no reason 
why the ratios should be exactly the same. For example, in the
X-shape features there is a wing towards larger b/a. The galaxies in
this wing are IC 1711, IC 3806, NGC 4419, NGC 4565, NGC 5145, NGC
5746, NGC 5757 and NGC 5777. Five of these galaxies have high parent
galaxy inclinations ($i$ = 70$^{\circ}$- 80$^{\circ}$), in which galaxies the
bar most probably is seen nearly end-on. A well known example of nearly 
end-on galaxies is NGC
4565, discussed also by \citet{kormendy2010}. In
fact, the morphology of the X-shape in NGC 4565 is very similar to our
B/D = 0.08 synthetic image in Figure \ref{fig:simu_azi_inc90},
seen close to the end-on view ($\phi$ = 30$^{\circ}$, $i$ =
90$^{\circ}$).  The three remaining galaxies have lower inclinations
($i$ = 33$^{\circ}$, 40$^{\circ}$ and 65$^{\circ}$, respectively).  One of
them, NGC 5145, has no detection of a bar, and in the two barred
galaxies, IC 3806 and NGC 5757, the X-shape features are among the
weakest detected in our sample. 

\section{Comparison of observations and synthetic images}

We compare the observations and synthetic images in order to 
study to which extent galaxy inclination affects the major-to-minor axis ratios,
and the normalized sizes of barlenses and X-shape
features. 
The comparison is shown in Figure \ref{fig:obs_simu_compare}
where the observations are shown on the left, and the synthetic images
on the right.

The axis ratios of barlenses and X-shape features (in the sky plane)
are shown as a function of galaxy inclination in the lower panels of
Figure \ref{fig:obs_simu_compare}.  A qualitative agreement
between the observations and synthetic images is good: b/a gradually
decreases towards higher galaxy inclinations until the images are seen
nearly edge-on, where b/a increases again. Also, barlenses and X-shape
features, both in the observations and in the synthetic images, form a
continuation as a function of galaxy inclination.  This behavior is
independent of the simulation model used.
  
More informative is to look at the normalized sizes of the structures (upper
panels).  It appears that the size of a barlens is on average constant
at $i$ = 0$^{\circ}$ -- 50$^{\circ}$, increasing towards higher galaxy
inclinations ($i$ = 50$^{\circ}$ -- 65$^{\circ}$). Also, the X-shapes have
a constant size at low galaxy inclinations, whereas at high
inclinations both small and large sizes appear.  Qualitatively similar
tendencies can be found also for the synthetic images. However, it is
important to consider the two models separately: although in both
models barlenses have similar sizes, differences appear in the sizes
of their X-shape features. In the bulge model (B/D = 0.08) the
X-shapes are manifested only at $i$ $>$ 50$^{\circ}$, where the sizes are
also more similar to those of barlenses. The arrows indicate where
barlenses in the face-on view, and X-shape features in the edge-on
view in this model, have similar sizes.  On the other
hand, in the bulgeless model (B/D = 0.01) small X-shapes appear even
in almost face-on view ($i$ = 20$^{\circ}$ -- 40$^{\circ}$). 
The size gradually increases towards higher galaxy inclinations.
Inspecting the morphology of the vertically thick inner bar
components in the synthetic images helps to better understand the above
differences between the two models: the bulge model (B/D = 0.08) is
lacking small X-shapes in the face-on view because at those inclinations
the morphology is turned into a barlens morphology (see 
  Fig. \ref{fig:simu_tilt}). Also, even at higher galaxy inclinations ($i$ =
60$^{\circ}$ inspected in Fig. \ref{fig:simu_azi_inc60}) small
X-shapes are not visible if the azimuthal angle of the bar is large.

\section{Morphology of barlens galaxies} 

  \subsection{Division of barlenses to sub-groups} 

Barlenses do not form a homogeneous group of features, most probably
reflecting the fact that they consist of a combination of orbital
families of bars, with a range of different orbital energies. In order
to further investigate their morphologies barlenses are divided to
sub-groups. Our intention is not to classify all barlens galaxies, but
rather to pick up prototypical cases with clear morphological
characteristics. Examples of those groups are shown in Figures
  \ref{fig:obs_groups1} and \ref{fig:obs_groups2}: the left panels
show the original 3.6 $\mu$m or K$_{\rm s}$-band images, cut in such a
manner that they best show the bar region. With the same image cuts
also shown are the unsharp mask images, in the sky-plane and when deprojected to the disk
plane. The surface brightness profiles as a function of the isophotal
semimajor axis are also shown, together with the profiles along the bar
major and minor axis. However, if i $>$ 65$^{\circ}$ only the
bar major axis profiles are shown. The following sub-groups were
recognized:
\vskip 0.15cm

Group a: {\it a regular ``thin bar'' is a characteristic feature; on
  top of that a round barlens appears.}  Outside the central peak the
barlens has an exponential surface brightness profile, both along the
bar major and minor axis. The ``thin bar'' is prominent and penetrates
deep into the central regions of the galaxy.  The surface brightness
profile along the bar major axis continues without cutoffs until the
end of the bar.  Good examples of this group are: NGC 1015, NGC 1452,
NGC 4608 and NGC 4643. As an example shown is NGC 4643 in Figure
 \ref{fig:obs_groups1}, discussed previously also by L+2014.
                                   
\vskip 0.15cm

Group b: {\it a large barlens dominates the bar; it has a small-scale
  structure at low surface brightness levels, which structure is
  typically elongated along the bar major-axis}. Prototypical cases
are NGC 5101 shown in Figure \ref{fig:obs_groups1}, and NGC 4314
  discussed by L+2014 (see their Fig. 1).  Other galaxies belonging to
  this group are: NGC 1512, NGC 4245, NGC 4394, NGC 4596, NGC 5375,
  and with some reservation also NGC 1640.  In many galaxies in this
  group the ``thin bar'' is manifested only as tips at the two ends of
  the bar. In NGC 4314 the ``thin bar'' is clumpy at low surface brightness levels.
\vskip 0.15cm

Group c: {\it barlens has two components, a bright ``inner disk'' and
  a low surface brightness structure outside that feature.}  The
``inner disk'' is still larger than typical nuclear bars or rings, and
it is generally oriented along the underlying large-scale disk.
Examples are NGC 1398, NGC 2787, NGC 3945, NGC 4262, NGC 4371, NGC
4754 and NGC 3384, and possibly also NGC 3489. As an example shown is
NGC 1398 (Fig.  \ref{fig:obs_groups1}). This is a group of
  barlens galaxies with the lowest fraction of inner rings or
  ringlenses (only 33$\%$ have r/rl). 
\vskip 0.15cm

Group d: {\it barlens and the ``thin bar'' have low surface brightnesses
  compared to that of the underlying disk}. The central regions of all
these galaxies are dominated by strong nuclear bars, nuclear rings
or lenses, or by a few star forming clumps as in NGC 7552.
The strong star forming regions are well visible at 3.6 $\mu$m wavelength. 
Good examples are NGC 613, NGC 1097, NGC 1300,
NGC 5728, and NGC 7552, of which NGC 1300 is shown in Figure  \ref{fig:obs_groups1}.
\vskip 0.15cm

Group e: {\it barlens dominates the bar; it is round and covers most of the
  bar size. The ``thin bar'' often ends up to ansae (appear in classification by
  Buta et al. 2015) at the two ends of the bar}. Examples of strong
bars (B) are: NGC 936, NGC 1440, NGC 1533, NGC 2983,
NGC 3941, NGC 3992, NGC 4548, NGC 4340, NGC4579, NGC 5770 and
NGC 6654. Examples of weak bars (AB) are: NGC1291,
NGC 1326, NGC 2859, and NGC 6782. All these weak bars have also nuclear bars,
often surrounded by nuclear rings. With uncertainty, included to
this group are also the galaxies: NGC 3892, NGC 3953 and
NGC 4143. Our example of strong bars is NGC 4548, and of weak bars 
NGC 1291 (Fig.  \ref{fig:obs_groups2}).
\vskip 0.15cm

Group f: {\it barlens dominates the bar which ends up to two tightly wound
  spiral arms}.  Good examples are NGC 1079, NGC 1350, NGC
2273, NGC 2543, NGC 3368 and NGC 5026. In  Figure  \ref{fig:obs_groups2} NGC 2273 is shown. 
\vskip 0.15cm

Group g: {\it barlens dominates the bar to such a level that barely no
  ``thin bar'' appears.} All these galaxies are weakly barred (AB),
and except for NGC 5750, have no inner rings. Examples are: NGC 1302,
NGC 2293, NGC 4503, NGC 4659, NGC 4984, NGC 5750, and NGC 6684, and
with some reservation also NGC 1022. NGC 4503 has no barlens in the
classification by \citet{buta2015}, but the galaxy has similar
morphology as the other galaxies in this group. NGC 4659 has also an
X-shape feature in the unsharp mask image. The surface brightness profiles of
these galaxies resemble those of unbarred early-type galaxies (see Laurikainen
et al. 2009, 2010). As an example NGC 2293 is shown (Fig.  \ref{fig:obs_groups2}). 
\vskip 0.15cm

Galaxies in the groups a -- d are largely strongly barred (B), in
the group e both strong and weak bars appear, whereas galaxies in
the groups f-g are weakly barred (AB). The classified galaxies in the different barlens
groups are shown at /www/astronomy/BLX.

 \subsection{Division of barlens parent galaxies to sub-groups} 

As barlenses, also their parent galaxies can be divided to sub-groups,
based on their characteristic morphological features. The following
groups were recognized  (see Fig.  \ref{fig:obs_parents}), ordered according to an
increasing dominance of later Hubble types:
\vskip 0.15cm

Group 1: {\it S0-S0/a, mainly S0$^{\circ}$;} the ``thin bar'' is very weak
(1a), or it is dominated by ansae at the two ends of the bar 
(1b).  Most of these galaxies have outer lenses (L) or ringlenses (RL)
(in 88$\%$; in 100$\%$ if uncertain galaxies are
excluded). Example galaxies having a shallow thin bar are: NGC 1440,
NGC 1533, NGC 3266, NGC 3489, NGC 4659, NGC 5750, NGC 5838, NGC
6684. Bars with ansae are: NGC 2787, NGC 2983, NGC 3941, NGC 4143, NGC
4262, NGC 4754, NGC 7079. Uncertain cases are NGC 1201 and NGC 2293,
which galaxies have no L or RL.  In Figure  \ref{fig:obs_parents} shown are NGC 1440
(1a) and NGC 2787 (1b).
\vskip 0.15cm

Group 2: {\it S0$^+$-S0/a, mainly S0$^+$}; an inner lens (or
ringlens) fills the space inside the bar radius. The galaxies in this
group almost always have also outer rings (R) or outer lenses (L).
Good examples are: NGC 1079, NGC 1291, NGC 1302, NGC 1326, NGC 2859,
NGC 2968, NGC 3380, NGC 3637, NGC 3945, NGC 4984, NGC 5134, NGC 5701,
NGC 5728, and NGC 6782.  In the galaxies NGC 1512 and NGC 1350 the
region inside the bar radius is somewhat less crowded.  As an example
NGC 2859 is shown.
\vskip 0.15cm

Group 3: {\it S0-S0/a, mainly S0/a}; these are strongly barred galaxies in which the
bar ends up to sharp features, which can be arcs or rs-type inner rings. 
As in group 2, also all these galaxies have outer
rings (R) or lenses (L).  Good examples are: NGC 936, NGC 1015, NGC 4596, NGC
4643, NGC 5101, and NGC 5375, of which NGC 4643 is shown. 
\vskip 0.15cm

Group 4: {\it S0-Sab, mainly Sa}; a strong bar ends up to a prominent,
fully developed inner ring (r), which is the dominant feature of the
galaxy.  In some of the galaxies ansae also appear in the bar, but due to a
superposition with the prominent inner rings they are not always
clear. Galaxies in this group have a large range of Hubble types. Good
examples are: NGC 1452, NGC 4245, NGC 4340, NGC 4371, NGC 4454, NGC
4608, and NGC 5770, of which NGC 1452 is shown. 
\vskip 0.15cm

Group 5: {\it S0-Sb, mainly Sa, Sab;} galaxies in this group have two
extremely open spiral arms, and a lens-like structure (not an inner lens 
in galaxy classification) at the bar
radius. In some cases the inner feature is a lens 
(NGC 4314, shown in Fig.  \ref{fig:obs_parents}), and in
some cases spiral arm segments around the bar (NGC 613 and NGC 1097). The
lens in NGC 4314 is not as inner lenses in galaxy classification, being more
elongated along the bar major axis.  Other examples of this group are
NGC 3368, NGC 4593 and NGC 7552. 
\vskip 0.15cm

Group 6: {\it Sa-Sb}; the dominant features are two
prominent, tightly wound spiral arms, starting from the two ends of the
bar. In distinction to the previous group the inner lens-lihe structure is missing,
and the spiral arms are more tightly wound.  Good examples are NGC
1300, NGC 2273, NGC 2543, NGC 4795 and NGC 5026. As an example
NGC 2273 is shown.
\vskip 0.15cm

Group 7: {\it Sab-Sb}; these galaxies have fully or partly developed inner
ring (r) surrounding the bar, and the outer disk is dominated by
multiple spiral arms. Good examples are NGC 1433, NGC 3351, NGC 3953, NGC 3992,
NGC 4639:, NGC 4394, NGC 4902, NGC 5339, NGC 5850, NGC 5957, NGC 7421,
IC 1067, and IC 2051.  As an example NGC 4902 is shown: the barlens in
this galaxy shows also a weak X-shape feature in the unsharp mask
image.

The above galaxies identified in the different barlens parent galaxy groups
are shown at //www/astronomy/BLX.

  \subsection{\bf Cross-correlating barlens groups with their parent galaxies} 

Statistics of the morphological features in the galaxies with
different barlens groups are collected to Table 4, and those in
the parent galaxy groups to Table 5. A general tendency is that
barlens galaxies very often have inner rings or ringlenses. For the
barlens sample as a whole the fraction is 78$\%$, and for none of the sub-groups 
the fractions are below 50$\%$.  This fraction is higher in strongly barred and
early-type galaxies, being even 100$\%$ in the barlens groups a and
b. In these particular groups the fraction of ansae is smaller than
in the other groups (in groups a and b 12-17$\%$ have ansae, in comparison to 33-57$\%$
in the other groups). On the other hand, the fraction of nuclear
features is typically high (25-42$\%$ have nuclear bars or
rings). There is a peak in the fraction of nuclear features in the
barlens group d where all galaxies have nuclear bars or rings.

Looking at the parent galaxy groups there is a tendency of increasing
galaxy mass from the group 1a towards the group 4, the mean mass
increasing from  log(M*/M$_{\odot}$) = 10.30$\pm$0.19 to log(M*/M$_{\odot}$) =
10.83$\pm$0.13. A minimum in the parent galaxy mass appears in group 4, where
$<$log(M*/$M_{\odot})>$ = 10.22$\pm$0.10. The uncertainties are stdev/$\surd$N.
Also the groups 2 and 6 have
something in common: the bars in these galaxies are typically weak
(only 33$\%$ belong to B family, compared to 67-100$\%$ of strong bars
in the other groups).  This is in spite of the fact that the galaxies
are fairly massive, ie.  $<$log(M*/M$_{\odot}$) $>$ = 10.58$\pm$0.07 and 10.70$\pm$0.10
for the groups 2 and 6, respectively.  Common to these two groups is
also that the region inside the bar radius is crowded (even more in
group 2), and the bars are often surrounded by inner (78$\%$, 50$\%$)
and outer (89$\%$, 50$\%$) rings or ringlenses. Inner rings and
ringlenses are even more common in the groups 3--5 and 7, where actually
all galaxies have such features: in the galaxies of these groups the
``thin bars'' are prominent which might explain the large number of
rings.  Bars in the groups 3--5 and 7 do not have frequently ansae
(0-33$\%$ have ansae, in comparison to 17-100$\%$ in other
groups). Nuclear features have a peak in the parent galaxy group 5
(86$\%$ have nuclear features) .

Barlens and parent galaxy groups are cross-checked in Table
  6. The numbers of galaxies in which both groups were identified is
fairly small, and there is also a large scatter, but some tendencies can be
seen:

Group 5 -- Group d: There is a connection between the barlens group d and the
parent galaxy group 5: in both groups the bar has a low surface brightness,
and the parent galaxy shows two open spiral arms
and a lens-like structure at the bar radius. Most probably, due to a
fairly shallow potential well these galaxies are efficient in
transferring material towards the central regions of the galaxies,
triggering nuclear bars, rings or lenses, which are typical in these
galaxies.  The bars never have ansae.  These galaxies have the
largest masses among the groups studied by us, ie.
$<$log(M*/M$_{\odot})>$ = 10.86$\pm$0.15 and $<$log(M*/M$_{\odot}) >$ = 10.71$\pm$0.14 for the
groups d and 5, respectively.

Group 1b -- Groups c, e: There is also a connection between barlenses that have
bright ``inner disks'' (barlens group c), and the parent galaxy
morphology. Namely, in 3/6 of the groups c/e the ``thin bar'' is
manifested mainly as ansae at the edges of the bar potential (parent
galaxy group 1b). The dominant outer features are lenses (L), which
appear even in 86$\%$ of these galaxies (in comparison to 5-25$\%$ in
the other parent galaxy groups).  On the other hand, inner rings (only
14$\%$ have r or rl) and nuclear features (only 14$\%$ have nuclear
features) are less common than in any of the other parent galaxy
groups.

Groups 3,4,7 -- Groups a, b: The parent galaxy groups 3, 4, and 7 are
associated to barlens groups a and b (4/6, 4/7 and 4/5 of the
parent galaxy groups, respectively).  These are largely strongly
barred galaxies, where the bar is a combination of a prominent
classical bar and a prominent barlens. All these galaxies have partly
or fully developed inner rings or ringlenses (100$\%$), whereas outer
rings appear only in 33-50$\%$ of the galaxies.

\section{Morphology of the galaxies with X-shaped bars} 
 
An interesting question is do the parent galaxies of the X-shaped bars
have similar morphologies as the galaxies with barlenses?  Because the
X-shapes appear in galaxies with higher galaxy inclinations, the observations
are more susceptible to dust and therefore any statistics of their
structure components is less reliable. However, a general trend is
that, in a similar manner as barlenses, also the galaxies with
X-shapes typically have inner rings and small inner disks (see also
Bureau et al. 2006 for X-shapes in the edge-on view). Such an inner
disk is particularly prominent in an X-shaped galaxy NGC 4216, with $i$=79$^{\circ}$}, shown in Figure \ref{fig:obs_weakests}. 

In Figure \ref{fig:obs_similars} we show three galaxies having
bars that manifest X-shape features: below each of these galaxies their barlens galaxy
counterparts are shown. 
The first galaxy pair is NGC 7179 (X) / NGC 5101 (bl): both are strongly
barred, have a Hubble stage S0/a, and either rl or rs surrounds the
``thin bar''.  In the unsharp mask images the ``thin bars'' 
appear mainly as tips in flux at the two ends of the bar potential. 
Another example pair is IC 1067 (X) / NGC 4643 (bl): in these galaxies
the inner rings are complete, and the ``thin bars'' appear as
classical elongated features penetrating deep into the central regions
of the galaxies.  The third pair is NGC 3673 (X) / NGC 2273 (bl): both
are early-type spirals, dominated by a barlens or an X-shape feature,
which end up to two tightly wound spiral arms.  While looking at the
surface brightness profiles of the first two pairs, it is obvious that
barlenses have central peaks, which are missing in their X-shaped
counterparts.  The galaxy with an X-shaped bar in NGC 3673 has stronger central
flux concentration than the other X-shaped bars
discussed above, but it is still less prominent than in its barlens
galaxy counterpart.

In the simulation models discussed in the literature, the X-shape
features are generally associated to strong bars in massive galaxies
\citep{atha2005,valpuesta2006}, but our 
examples show that X-features can appear also in weak bars in low mass
galaxies. The weakest X-shapes in our sample appear in NGC 5145, and
in the two low mass galaxies IC 3806 and IC 0335 (see 
  Fig. \ref{fig:obs_weakests}), having masses of
log(M*/M$_{\odot}$) = 9.49 and log(M*/M$_{\odot}$) = 9.94,
respectively. In IC 3806 and NGC 4145 the X-shape appears mainly as
four blops in the four corners of the X-feature. All three galaxies
can be considered as bulgeless (i.e have no photometric bulge).  NGC 5145 has a shallow flux
concentration in the surface brightness profile, but it is actually 
an inner disk in the disk plane. In
the unsharp mask images we identify X-shape features also in a few
strongly interacting galaxies (NGC 3227 and NGC 4302), in a warped
galaxy (NGC 660), and in some otherwise peculiar galaxies (NGC 3190,
NGC 3628).

We have six galaxies in our sample, which have a barlens in the
classification by \citet{buta2015}, and in which galaxies an X-shape
feature is identified in our unsharp mask image. These galaxies are
NGC 3185, NGC 3380, NGC 4902, NGC 5957, NGC 7421, and IC 1067 (IC 1067
is shown in Fig. \ref{fig:obs_similars}). These are naturally
also galaxies which appear in the overlapping inclination region
of $i$ $\sim$ 45$^{\circ}$ -- 60$^{\circ}$, where galaxies exhibit both barlenses
and X-shape features.  The most face-on of these galaxies are NGC 3185 and
IC 1067 with $i$ = 38$^{\circ}$ and 49$^{\circ}$, respectively.

\section{Morphology of unbarred galaxies}    

Unbarred galaxies in our sample were selected based on their inner
surface brightness profiles, which resemble those of barlens and
X-shaped galaxies as much as possible. As an example we show a pair
NGC 3599 (unbarred) / NGC 4643 (bl) in Figure
  \ref{fig:obs_nonbars_selection}.  Both galaxies have a prominent
central peak within 7 -- 10'', and an exponential sub-section outside
that region.  In NGC 4643 this sub-section at r $\sim$ 30'' corresponds
to the barlens, and in NGC 3599 a similar sub-section extends to
r $\sim$ 20''. In NGC 4643 the longer and more elongated part of the
bar, ie. the ``thin bar'', is manifested as a bump in the surface
brightness profile at r $\sim$ 50'', which bump is naturally lacking in
the unbarred galaxy. More examples of the unbarred galaxies are shown
in Figure \ref{fig:obs_nonbars}. All these galaxies have
sub-structure in the unsharp mask images, and related to their
surface brightness profiles, also counterparts among
the barlens galaxies can be found. For example, a fairly good
correspondence in the surface brightness profiles exists between the
unbarred galaxy NGC 3065 (Fig. \ref{fig:obs_nonbars}, uppermost
panel) and the barlens galaxy NGC 1398 (see 
  Fig. \ref{fig:obs_groups1}, barlens group $c$). Both galaxies have
prominent central mass concentrations, and a shallower surface
brightness profiles immediately outside that peak.  Additional
wiggles in the surface brightness profile of NGC 1398 are due to an
inner ring and spiral arms.

We have used the surface brightness profiles of the unbarred galaxies
to estimate the sizes of the regions corresponding the structures
associated to barlenses in barred galaxies. In some galaxies this is
the radius defining the photometric bulge (ie. flux above the
exponential disk; NGC 4489), whereas in some galaxies it is the
exponential sub-section outside the central peak (NGC 3599). These
radial distances are marked with dotted vertical lines in Figures
  \ref{fig:obs_nonbars_selection} and \ref{fig:obs_nonbars}.

While deciding where to put that radius we inspected the original
and unsharp mask images, in order to recognize the morphological
structures behind these profiles. The sizes of these ``bulges'',
together with the sizes of barlenses are plotted
as a function of galaxy mass in Figure \ref{fig:obs_sizes_nonbars}. It appears that the
sizes of ``bulges'' in the unbarred galaxies follow
a similar relation as the sizes of barlenses.  The number of the unbarred galaxies in
the figure is smaller than the total number of unbarred galaxies in
our sample, because the radius of the ``bulge'' was measured only for
those galaxies in which it was possible to define it in a reliable
manner.

In order to further study the nature of ``bulges'' of the unbarred
galaxies we looked at possible fine-structures in their unsharp mask
images. Indeed, many kind of faint features can be recognized.  For
example, IC 2764 (Fig. \ref{fig:obs_nonbars}) shows three blops
at r $\sim$ 12'', which is also the radius where the nearly exponential
sub-section of the surface brightness profile ends. Characteristic for
NGC 4489 and NGC 3599 (Fig. \ref{fig:obs_nonbars}) is that both
galaxies have weak two-armed spiral-like features inside the
exponential sub-sections at r $\sim$ 12'' and 22'', respectively.  In
NGC 5311 (not shown) spiral features appear inside r $\sim$ 22'', which
also marks the size of the nuclear lens in the classification by \citet{buta2015}.  
Also, in IC 5267 (Fig. \ref{fig:obs_nonbars}) an
elongated feature appears at r = 10'' -- 15''. All the
above galaxies have low inclinations of $i$ = 23$^{\circ}$ -
35$^{\circ}$. It is unlikely that the faint features discussed above
could form part of a dynamically hot spheroidal component, ie. a
classical bulge.

\section{Discussion} 

It is widely accepted that the vertically thick B/P structures are
common in the edge-on galaxies (L\"utticke, Dettmar $\&$ Pohlen 2000; see
also the review by Laurikainen $\&$ Salo 2016),
appearing in $\sim$2/3 of the S0-Sd galaxies in the nearby universe. Many of
them show also X-shape features in unsharp mask images \citep{bureau2006}
confirming the bar-like origin of these structures. However,
the photometric bulges of barred galaxies in less inclined galaxies
are repeatedly interpreted as classical bulges.  Echoing \citet{kormendy2010}:
``as long as face-on and edge-on
galaxies appear to show physical differences we cannot be sure that we
understand them''.  As a possible solution to this ambiguity it has
been suggested by us that the bulges in barred Milky Way mass galaxies
are actually the face-on counterparts of B/P bulges (L+2014;
A+2015). Taking this view would considerably change the paradigm of
bulge formation in the Milky Way mass galaxies. However, before such a
view can be adopted, a more in-depth understanding of the properties of
these features is needed.

\subsection{What are barlenses?} 

Barlenses have been recognized as lens-like structures embedded in bars in
low and moderately inclined galaxies, covering nearly half of the bar
size \citep{lauri2011}. The given name was somewhat
unfortunate because barlenses were actually assumed to be vertically
thick, in a similar manner as the B/P bulges in the edge-on view. This
has lead to some confusion in the literature where barlenses are
sometimes considered as structures in the disk plane (see for example
Gadotti et al. 2015). Looking at their surface brightness profiles in
detail shows that barlenses appear as exponential sub-sections, both
along the bar major and minor axis (L+2014, their Fig. 1; A+2015, their Fig. 2;
Figs. 11, 12 and 15 in this work). These exponential sub-sections can
penetrate into the central regions of the galaxies, but more often
additional central flux concentrations also appear. We have shown
examples indicating that such central concentrations are
characteristic to barlenses, but are generally lacking in the X-shaped
bars, which is fundamental to understand the nature of these
structures.

Related to this matter is also our finding
that only $\sim$ 24$\%$ of the barlenses have boxy isophotes.  
We have 38 galaxies in common with the sample by ED2013 who discussed
boxy bar isophotes at intermediate galaxy inclinations ($i$ $>$ 45$^{\circ}$).  For 22 of
these galaxies they find evidence of boxy isophotes, based on similar
isophotal analysis as carried out by us. ED2013 interpret this as
evidence of B/P bulges.  We identify an X-shape feature in 14 of these
galaxies (ie. 64$\%$ of the B/Ps by ED2013). For the remaining 8 galaxies 
we confirm the boxy isophotes.  In 16 of the galaxies common with our sample
ED2013 did not find any evidence of boxiness, in
agreement with our analysis (except for NGC 3489 for which galaxy we
find boxy isophotes).  

It appears that boxy isophotes at $i$ $>$ 45$^{\circ}$ is an
  efficient tool to find the vertically thick inner bar components,
  which at these galaxy inclinations are manifested as X-shape
  features.  However, most barlenses,
  which typically appear at lower galaxy inclinations, do not exhibit boxy isophotes.

\subsection{Barlenses form only in centrally concentrated galaxies} 

Barlenses have been studied already before they were called as such.
In \citet{lauri2007} they were called as lenses, which in the
structural decompositions were fitted with a separate function
(usually in addition to the main bar component). In Fourier analysis the
same structures were manifested as flat or double peaked in the m =
2 density amplitude profiles. The resemblance of such profiles 
with the simulation models by \citet{atha2002} made the authors
to suggest that those ``lenses'' might actually be vertically thick inner bar
components.  That barlenses indeed can be vertically thick was later
shown by L+2014 and A+2015. In
the first paper the observed axial ratio distribution of the galactic disks in the 
combined sample of the parent galaxies of barlenses and X-shape features was shown to be
flat, as expected if they are the same features seen at different
viewing angles.  \citet{atha2015} showed the connection
between barlenses and B/Ps 
using hydrodynamical simulations. They looked at the vertically thick
inner bar components at face-on view, and compared the
surface brightness profiles of the model snap-shots with those seen in the observations. In the
simulation models the same size was measured for the barlens in the
face-on view, and for the X-shape feature in the edge-on view.  Consistent
with this picture is also the fact that even 88$\%$ of the B/P bulges
in in edge-on view show X-shape features in the unsharp
mask images \citep{bureau2006}.  However, it still remained a puzzle
why barlenses appear in earlier Hubble types than the X-shape
features.  This is shown even more clearly in our Figure
  \ref{fig:obs_histo} (middle panel) using the same inclination bin
for both type of objects. In this study we further showed that
barlenses appear systematically larger than the X-shape features, in
the same galaxy inclination bin (see Figure
  \ref{fig:obs_simu_compare}).

These apparent ambiguities can be understood due to the orientations
effects, and by the fact that barlenses form mainly in galaxies with
peaked central mass concentrations. \citet{salo2016}
showed that a steep rotation curve is needed
to make a barlens morphology in face-on view, while a more shallow rotation
curve may lead to boxy or even X-shaped face-on morphology.  In principle such
central mass concentrations can be associated to classical bulges or
other central mass concentrations, which are more pronounced in the
early-type galaxies where barlenses generally appear.  With our
simulation models the effect of the central mass concentration on the
bar morphology is illustrated in Figure 20: shown are five models from
\citet{salo2016} with increasing relative mass, which
varies between B/D = 0.01 and 0.16. The models are shown at
different galaxy inclinations (keeping the azimuthal angle fixed to
$\phi$ = 90$^\circ$).  It appears that in the face-on view the barlens
morphology becomes increasingly evident when the bulge dominance
increases. We can also see that the galaxy inclination where the
barlens becomes evident depends on B/D: with large B/D the
barlens is visible even at fairly high galaxy inclinations, whereas with low
B/D it can be seen only in nearly face-on view. Note that the effective radius of
the bulge is fixed to the same value in all these models, being less
than $10\%$ of the barlens radius: thus the direct contribution of the
bulge flux to the apparent barlens morphology is insignificant.

Further observational evidence for our interpretation can be found
from the bulge-disk-bar decompositions made for the S$^4$G sample by
\citet{salo2015}. In Figure \ref{fig:obs_histo} shown separately are
the galaxies with barlenses and X-shape features: it appears that
barlens galaxies indeed are more centrally concentrated (right panel),
in spite of the fact that they are not more massive than the galaxies
with X-shaped bars (left panel).  The comparison is made within
an inclination bin $i$ = 45$^{\circ}$ -- 60$^{\circ}$ where both
features appear. Note that although in these decompositions even 4
components were used, the inner bar components were not fitted
separately. More sophisticated decompositions were made by L+2014, who
used a sample of 29 galaxies, fitting besides bars, bulges and disks,
also the inner bar component (bl or X), with a separate function. They
found that most of the photometric bulge actually consists of barlenses or
X-shape features having $<$B(X-feature)/T)$>$ = 0.08$\pm$0.02 and 
$<$B(barlens)/T$>$ = 0.18$\pm$0.11. For the central peaks they found
B/T = 0.08$\pm$0.01 and 0.12$\pm$0.02 for the X-shapes and barlenses,
respectively (the original paper has
less decimals). These values are not far from those used in our
simulation models with B/D = 0.01 and 0.08 (ie. B/T = 0.01 and 0.09,
respectively). Again, this comparison qualitatively shows that
barlenses have at least slightly higher central flux concentrations
than the X-shaped bars.

Morphology of the Milky Way (Hubble type T = 3) bar/bulge, showing
 an X-shape in nearly end-on view ($\phi$ $\sim$ 30$^{\circ}$), has been reconstructed by
  \citet{wegg2015}, based on the best-fitting star count model in the
  near-IR. More recently the X-shape has been detected also directly
  by \citet{ness2016}.  Morphology of the Milky Way bulge has been
  compared with one of the barlens galaxies in our sample, NGC 4314
  (T=1), by Bland-Hawthorn $\&$ Gerhard (2016; see their Figs. 9 and
  10): they suggest that in face-on view the projected bar/bulge of the
  Milky Way would resemble the barlens morphology of NGC 4314 ($i$ = 20$^{\circ}$). In the
Milky Way the normalized size of the X-feature is 0.3 (a$_{\rm X}$/r$_{\rm bar}$ = 1.5/5.0), 
which is the same as the sizes of the
X-shape features in our sample. However, as we are looking at the
Milky Way boxy bulge nearly end-on view, it is possible that its
relative size is underestimated.  Both galaxies have also small inner
disks (ie. disky pseudobulges) embedded in the vertically thick inner
bar component.

\subsection{Optical colors of barlenses}

Optical colors of barlenses have been recently studied by HE+2016, and
it is interesting to look at if the barlens groups recognized in this study
have any association with those colors. HE+2016 used Sloan Digital Sky
Survey images (u, g, r, i, z) to study the colors of 43 barlens
galaxies. Major and minor axis profiles along the bar were derived
using the (g-r) and (i-z) color index maps. They found that barlenses have
on average similar colors as the surrounding vertically ``thin bars''. Barlenses
were divided to sub-groups, based on the major axis color
profiles.  The largest group were those with completely flat color
profiles (10/43 galaxies). Interestingly, all these galaxies are 
early-type systems (7 SO$^{\circ}$- SO$^+$, and 3 SO/a). It appears that barlenses of
these galaxies have strong ``thin bars'',
ie. they belong to our barlens groups a, b or c. The
favorite bar type is a classical rectangular bar (ie. barlens group
a). The galaxies with dusty barlenses (8/43) are typically early-type
spirals. Prominent nuclear rings appear in 4 of the galaxies: these systems
belong either to our barlens group a (N = 1) or b (N = 3), and to our
parent galaxy groups 4 or 5. The fractions of inner rings and
ringlenses (91-100$\%$, respectively) in these galaxies are
exceptionally high. Of the four galaxies that have both a barlens and an
X-shape feature, the barlens structure is either dusty (2 galaxies),
or has a blue nuclear region (2 galaxies).

One of the messages of this comparison is that the early-type
galaxies in our sample, with prominent barlenses, do not have redder
central regions (compared to the color of the ``thin bar''): 
if such red central regions (with sizes of barlenses) were seen, that 
 could be interpreted as prominent classical
bulges. It is also interesting that barlenses in the early-type
spirals can be dusty, in spite of the fact that the mean colors of the
barlenses correspond to the colors of typical elliptical galaxies. It
means that barlenses are capable of capturing gas and convert that
into stars, ie. not all gas is transferred to the nuclear regions of
the galaxies.  However, colors give us only hints of the mean stellar
populations, and spectroscopy is needed to distinguish possible range
of stellar ages and metallicities in these structures.

\subsection{Individual galaxies with detailed spectroscopy available in the literature}

Having in mind that the Milky Way might have a barlens in face-on view, it is worth
looking at in which way the stellar populations and kinematics of its
bulge have been interpreted in the literature. Recent reviews of the Milky Way bulge are
given by \citet{gerhard2016}, \citet{dimatteo2015}, \citet{shen2016}, and \citet{gadotti2016}.
We look at also similar properties of two external galaxies forming part of our
barlens galaxy sample, studied in detail in the literature.

\vskip 0.3cm
Milky Way (MW): 
\vskip 0.1cm

\noindent For the Milky Way bulge the early stellar population
analysis pointed to a fairly massive classical bulge.  The stars of
the bulge were found to be metal poor and older than 10 Gyr 
\citep{terndrup1988,ortolani1995,zoccali2003,brown2010}. 
Those stars were also more $\alpha$-enhanced than the thick
disk stars of the same metallicity 
\citep{mcwilliam1994,rich2005,zoccali2006,lacureur2007,hill2011,johnson2011}. 
These observations lead to the idea
that the Milky Way bulge is a fairly massive classical bulge that
formed in a rapid event at high redshift, out of the gas that was not
yet enriched by the cycle of star formation and feedback. The idea
that the bulge could have been formed via bar buckling episodes was
therefore abandoned.

However, this picture has changed more recently. The Milky Way bulge
has turned out to have an X-shape morphology \citep{mcwilliam2010,nataf2010,wegg2013,gonzalez2015}.
The bulge also rotates cylindrically \citep{howard2008,kunder2012},
as expected for a vertically thick bar component.
Nearly 50$\%$ of the stellar mass at r $<$ 10 kpc was indeed in place
already at high redshift, but the most metal-rich stars ([Fe/H] $>$ -0.4
dex) show a range of stellar ages of 3 -- 12 Gyr. The age decreases
with increasing metallicity \citep{bensby2013,ness2014}.
Such observations are not expected in classical
bulges. Although even 60$\%$ of the stars in the Milky Way bulge are
metal poor, the dynamical models for the bulge do not predict
existence of a massive classical bulge (Shen et al. 2010; Di Matteo et
al. 2015; see review by Shen $\&$ Lin 2016): nowadays the Milky Way
bulge is interpreted to largely form part of the bar so that at most
10$\%$ of the total galaxy mass belongs to a classical bulge, if at
all.

A problem in this interpretation is how to explain the observed age,
metallicity, and $\sigma$-gradients in the vertical direction 
\citep{zoccali2008,Gonzalez2011,johnson2011,johnson2013}: the
most metal-rich and youngest stars appear at low galactic latitudes,
whereas the fraction of dynamically hotter, metal-poor stars
(-1$<$[Fe/H] $<$ -0.5 dex) increases towards higher galactic
latitudes. It has been speculated that bar buckling would dilute such
stellar population gradients, but they can be explained by assuming
that the oldest stars at high galactic latitudes correspond to those
originally formed in the thick disk \citep{ness2014,dimatteo2016}.
For external galaxies such metallicity gradients in the vertical
direction have not much studied yet.

We can compare the observations of the Milky Way bulge with the
stellar populations and kinematics of two barlens galaxies, NGC 5701
and NGC 7552, studied by \citet{sedel2015}. The IR-images of these galaxies, their
unsharp mask images, and the surface brightness profiles are shown 
in Figure \ref{fig:obs_popu}.

\vskip 0.3cm
NGC 5701  [(R'1)SA$\underline{\rm B}$(rl,bl)0/a]:
\vskip 0.1cm 

\noindent This galaxy belongs to our barlens group b  (Fig. \ref{fig:obs_groups1}) and
parent galaxy group 2  (Fig. \ref{fig:obs_parents}).  Integral-field spectroscopy was made by Seidel
et al. (2015) using a field-of-view of 36'', which covers most of the barlens
size having a radius r = 25'' in our definition. We use this radius to evaluate the
stellar population measurements by Seidel et al.  The optical colors are studied by
HE+2016: this galaxy shows flat (g-r) and (i-z) color profiles
throughout the bar major axis, which colors are also typical for
the elliptical galaxies. Using full spectral fitting Seidel et al. find
flat (old) age and metallicity profiles in the barlens region (their
Fig. 6). Deviations from that appear only in the innermost 4'', where
nuclear spiral arms reside \citep{erwin2002}, manifested also
as a sigma-drop in the same region. Seidel et al. divide the stars
also into three sub-populations with different stellar ages (their
Fig. 8). It appears that the mass of the barlens is dominated by the
old stellar population (70 -- 85$\%$ of the mass have ages $>$10
Gyr). However, outside r = 5 -- 10'' the intermediate age (1.5 -- 10 Gyr)
stellar population becomes increasingly important towards the edge of the
barlens.  In the barlens region there is a metallicity gradient: the
metallicity goes from solar or slightly sub-solar ([Fe/H]$\sim$ 0 -- -0.2) 
from the center to the outer parts. This corresponds to the
intermediate metallicities observed in the Milky Way bulge.

\vskip 0.25cm
NGC 7552 [(R')SB(r$\underline{\rm s}$,bl,nr)a)]: 
\vskip 0.15cm

\noindent This galaxy belongs to our barlens group d (
  Fig. \ref{fig:obs_groups1}), and parent galaxy group 5 (
  Fig. \ref{fig:obs_parents}). There is a lot of structure in the
unsharp mask image in the barlens region at r $<$ 30'': a nuclear star
forming ring appears at r $<$ 5'', and a weak ring-like feature at
r$\sim$20''. These features are manifested also in the stellar
populations, metallicities and kinematics, as analyzed by Seidel et
al. (2015).  The nuclear starburst is dominated by stars younger than
1.5 Gyr, with solar-to-subsolar metallicities.  The outer ring-like
feature contains a non-negligible amount of old stars ($\sim$ 14 Gyr),
with lower metallicities than the stars in the inner parts of the
galaxy. An important thing is that, although old stars ($>$ 10 Gyr)
appear throughout the barlens, the fraction of intermediate age stars
(1.5 -- 10 Gyr) exceeds that fraction in many regions. Also very young
stars ($<$ 1.5 Gyr) appear throughout the barlens. Also in this galaxy
a metallicity gradient appears: the metallicities in the barlens
region are similar or slightly lower than those in NGC 5701. The
rotation curve in the barlens region also shows a double hump,
generally associated to B/P bulges.
\vskip 0.25cm

The interpretation that massive classical bulges appear in external
Milky Way mass early-type galaxies has largely based on photometry
(bulges have large S\'ersic indexes and B/T-values), and on the
observation that their stars are on average as old as those in 
elliptical galaxies.  Looking at the mass weighted mean stellar ages
of the above two barlens galaxies (excluding the nuclear starburst in
NGC 7552), in principle we could make the same conclusion. In NGC 5701
the bulge stars are old also using the light weighted stellar
ages. The typical stellar ages of $\sim$10 Gyr or older, are similar
as observed in the Milky Way bulge (Sanchez-Bl\'azquez et
al. 2011, hereafter SB+2011, and references there). And also, the metallicites of these galaxies
are not much different from those of the Milky Way bulge (in NGC 5701
it is slightly higher than in the Milky Way bulge). However, looking at the
stellar populations in more detail also shows that both galaxies have
a range of stellar ages and  metallicity gradients, again similar to
those found in the Milky Way bulge.  NGC 7552 has also kinematic
evidence associating the barlens to a vertically thick bar
component. 

As in the Milky Way, also in NGC 5701 and NGC 7552 the stellar
populations of the barlenses are difficult to interpret by means of
classical bulges, inspite of the fact that even 80$\%$ (in NGC 5701)
and 50$\%$ (in NGC 5772) of their stars were formed at high
redshifts. In the Milky Way bulge that fraction is $\sim$60$\%$.  In
photometric decompositions the values of the S\'ersic index and 
B/T largely depend on how the decompositions are made (see L+2014):
low values are obtained when both the ``thin bar'' and the
barlens are fitted with separate functions. That is what actually
needs to be done if we are interested in isolating bulges which do not form
part of the bar itself.

Kinematic evidence of a B/P bulge has been found also for the barlens
galaxy NGC 1640 in our sample by \citet{mendez2014}: they
used the h$_4$ coefficient of the Gauss-Hermite parametrization of the
line-of-sight velocity distribution, and showed that NGC 1640 has a
double minimum before the end of the bar, interpreted as evidence of a
vertically thick bar component. Similarly, kinematic evidence of B/P
is found by \citet{mendezabreu2008} for NGC 98, which galaxy we
would classify as having a barlens (but does not form part of our
sample due to its large distance).  It appears that detailed stellar
population, kinematic, and morphological analysis is needed for more galaxies with
barlens and X-shape bars.

\subsection{Stellar populations and kinematics in the samples of barred and unbarred galaxies} 

We can also look at what is known about stellar populations of bulges
in major galaxy samples. B/P bulges in the edge-on S0-Sa galaxies has
been studied by \citet{williams2011} and \citet{williams2012}. They found that the main
body of the B/P bulges lack a correlation between metallicity gradient
and velocity dispersion $\sigma$, although such a correlation exists
in elliptical galaxies, and is indeed expected in highly relaxed
systems. In many studies photometric bulges in the S0-Sbc galaxies are
found to be on average old ($>$ 10 Gyr), similar to those in
elliptical galaxies \citep{proctor2002,falco2006,mcarthur2009}.
Also bars and photometric bulges
seem to have similar stellar populations, dominated by old metal-rich
stars (Perez et al. 2011, SB+2011).
The mass weighted age-gradients
are flat at all radii, and the metallicity decreases from the center
outwards (S\'anchez-Bl\'azquez et al. 2014, hereafter SB+2014). Bulges also have older stars and higher
metallicities than the disks (SB+2014), although relatively old stars
(age $<$ 4 Gyr) dominate even the disks in spiral galaxies \citep{morelli2015}.
The above stellar population ages of bulges and
bars are consistent with the colors obtained for barlenses by
HE+2016.

However, as discussed in the previous section, photometric bulges can
have also young stellar populations, which is obvious either by
comparing the mean mass and light weighted ages, or by
dividing the stellar ages into different bins, as was done for example
by \citet{sedel2015}. Young stellar ages dominate particularly
  the bulges of low surface brightness galaxies \citep{morelli2012}. 
The stellar populations of bulges in the Calar Alto Legacy
Integral Field Spectroscopy Area (CALIFA) survey of $\sim$ 300
galaxies, covering the redshifts of z = 0.005 -- 0.03, and the galaxy mass range
of log(M*/M$_{\odot}$) = 9.1 -- 11.8, has been recently analyzed by
Gonz\'alez-Delgado et al. (2015). They found that the photometric
bulges in Sa-Sb galaxies and in the cores of E/S0 galaxies have
similar old metal-rich stars. But they also found that the
light-weighted mean stellar ages of the bulges in Sa -- Sb galaxies are
only $\sim$ 6 Gyr old, compared to $\sim$ 10 Gyr obtained from the mass
weighted stellar ages. It is only in Sc -- Scd Hubble types where
both the light and mass weighted stellar population ages of bulges are
younger than those of the bulges in earlier Hubble types.

Stellar population studies of barred and unbarred galaxies have shown
apparent controversial results, but as discussed by \citet{lauri2016}, 
a critical point is what do we mean by bulge. More
metal-rich and $\alpha$-enhanced bulges in barred galaxies are found
by \citet{perez2011}, who considered as bulges the central regions with
similar sizes as nuclear rings. It is well known that the central
regions of barlenses have frequently nuclear features (Laurikainen et
al. 2011; discussed also in this study): in case of nuclear rings they
are star-forming regions, whereas nuclear bars typically have
old stellar populations. Therefore, relying on these regions would not
tell us anything about the stellar populations of possible spheroidals
or B/P bulges of bars.  On the other hand, similar stellar populations in
barred and unbarred galaxies has been found by SB+2014 for 62 face-on
galaxies, and also by \citet{jablonka2007} for 32 edge-on galaxies,
without restricting to the innermost regions of bars. These
two studies used synthetic stellar population methods in their analysis. 
It is worth noticing that the above results do not rule out the possibility that
bulges in unbarred galaxies were formed in similar processes as the
vertically thick inner bar components.

Kinematics of bulges have generally been studied only in small galaxy
regions, not yet covering the sizes of barlenses (see a reviews by
Falc\'on-Barroso 2016, and M\'endez-Abreu 2016). In Atlas3D \citep{emsellem2011,cappellari2007} 
most of the bulges were found to
be fast rotating, which is consistent with the idea that they are
features of the disk. Only 15$\%$ of the bulges in Atlas3D have
signatures of B/Ps, in terms of having double humped rotation curves
or twisting isophotes \citep{krajnovic2011}.  However, while
restricting to barlenses and X-shaped structures, as identified in
our study, even 36$\%$ of the X-shapes and 56$\%$ of barlenses in Atlas3D
have boxy or twisted isophotes.  Most
probably the kinematic analysis performed for the Atlas3D has
recognized only a small fraction of possible B/P/X-shape bulges in that
sample.

Concerning the kinematics of unbarred galaxies, the recent study by
\citet{holmes2015}, based on the CALIFA survey, is interesting. They
used H$\alpha$ velocity fields to search for bar-like non-circular
flows in barred and unbarred systems. Weakly barred (AB) systems are
typically under their detection limit, but in spite of that clear
non-circular flows were detected in a few unbarred galaxies, which
have no photometric evidence of a bar. These galaxies are not
interacting, and have no isophotal twists.  Having in mind that only
strong amplitudes were detected, most probably only the top of the
iceberg was recognized.  These photometrically unbarred galaxies could
be similar to the unbarred galaxies studied by us. They are not
classified as barred, but their photometric bulges might be similar to
the inner parts of strong bars, manifested as B/P/X in the edge-on view,
and as barlenses in face-on view.

\subsection{How relevant is the idea that ``bulges'' in the Milky Way mass galaxies
are largely inner parts of bars?} 

Above we have discussed that the stars of bulges in the Milky Way mass
S0s and early-type spirals in the CALIFA survey \citep{gontzalezdelgado2015}
are older and more metal rich than the stars of their
disks. A large majority of the stars in bulges are as old as in the cores of elliptical galaxies.
Excluding the nuclear regions bars and bulges also have similar mean
stellar population ages and metallicities (SB+2014). Detailed analysis
of some individual barlens galaxies have also shown that the
photometric bulges actually have a range of stellar ages
between 1.5 -- 14 Gyr, which means that the masses of bulges must have
been accumulated in a large period of time, or at least more than
  one starburts event has occurred. In barlens galaxies the photometric
bulge consists mostly the barlens itself.
Consistent with the idea that bulges were not formed in a single event is also the
observation that their mass correlates with the
galaxy mass \citep{gontzalezdelgado2015}. When evolved in
  isolation the central mass concentrations are smaller than in galaxies of clusters, but even in
  isolated galaxies bulges have old stellar
  populations (ie. are red in g-i) \citep{fernandez2014}. As bars
and photometric bulges are largely made of the same stellar
populations, it is possible that bars have played an important role in
accumulating the central mass concentrations in galaxies, in some
process which rises the stars to higher vertical distances, ie. makes
B/P/bl bulges.

What are the stellar populations and metallicities of the formed
bulges depends, besides on the formation and evolution of the bar
itself, also on possible interaction between the ``thin bar'' and the
thick disk, and on how efficiently the bar collects gas from the
surrounding disk.  It has been shown by the simulation models that
when a barlens forms, a range of stellar population ages appears in
the barlens. In the models by \citet{atha2013}, analyzed by A+2015,
barlenses form first in the oldest stellar population, to which mass
added later are stars formed from the gas which is gradually
accumulated to the bar and the barlens. A different approach was taken
by \citet{atha2016b} who studied mergers of two disk galaxies with hot
gaseous halos, ending up to Milky Way mass galaxies with B/T = 0.1 -- 0.2
for the classical bulge. The classical bulge formed during a violent
relaxation has the oldest stars, while the stars in the barlens are
younger and cover a range of stellar ages (7.8 -- 8.6 Gyr), which stars
were largely formed during the assembling of the disk.  Also these
simulations are qualitatively consistent with the barlens
observations, although the barlenses discussed in our study can have
also younger stars, probably related to later gas accretion to these
galaxies.  However, it is good to have in mind that even based on stellar
populations there is no unambiguous way of distinguishing barlenses and
classical bulges: namely, also classical bulges formed by wet minor mergers
can have young stellar populations, or, old stars in the central
  regions of bars might appear, originated from the thick
  disks. Also, although promising, even the above major merger simulations,
with relatively small B/T, still have a problem of making too much
bulge during the violent relaxation phase (see also the review by
Brooks $\&$ Christensen 2016).

It is predicted by the cosmological simulation models \citep{debuhr2012}
that bars which form inside the dark matter halos at
z = 1 -- 1.3 buckle at z = 0.5 -- 0.7, thus forming B/P bulges. These bars are
long-lasting and are maintained until z = 0. In principle, earlier bar
formation is also possible, but their formation is restricted by large
gas fractions observed in high redshift galaxies (gas cannot cool to
form stars), and also by a threshold in the relative disk-to-halo mass
needed to trigger the disk instability. The epoch predicted for the
formation of boxy bulges by \citet{debuhr2012} is not far away from
z$\sim$1, when most of the central mass concentration in galaxies
seems to be assembled \citep{vandokkum2013}.  In fact, although the
Hubble sequence might be in place at some level already at z = 2.5 \citep{wuyts2011},
many galaxies at z = 1 -- 3 still have irregular clumpy
appearance \citep{abraham1996,berg1996,elm2007}.
Based on Fourier analysis of bars it was shown by \citet{salo2016},
using stacked S$^4$G density profiles, that the
bars with barlenses or X-shape features are mode centrally
concentrated than bars in general, indicating that bars and bulges
in these galaxies are coupled (see also D\'iaz-Garc\'ia et al. 2016b for barred and
unbarred galaxies).

Using a volume-limited sample it has been shown by
  \citet{aguerri2009} that the local galaxy densities of barred and
  unbarred galaxies are similar, which in our view calls for an
  explanation for the formation of bulges in unbarred galaxies. Living
  in similar environments with barred galaxies their formative
  processes cannot be completely different. In this study we have
  discussed potential candidates of unbarred galaxies which might have
  bar-like potential wells.  That kind of bar potentials in unbarred
  galaxies have been discussed from the theoretical point of view by
  \citet{patsis2002}: those bulges follow similar orbital families as
  the vertically thick bar components, but are lacking the more
  extended vertically thin bar components.  Indeed, it seems that
there is room for the interpretation that most of the bulge mass in
the Milky Way mass galaxies actually resides in bars.

\section{Summary and conclusions} 

We use the Spitzer
Survey of Stellar Structure in Galaxies (S$^4$G, Sheth et al. 2010),
and the Near-IR S0 galaxy survey (NIRS0S, Laurikainen et al. 2011),
to compare the properties of barlenses and X-shape features in the
infrared.  The sample consists of 88 X-shape features identified in
the unsharp mask images, and 85 barlenses that appear in the
classifications by \citet{buta2015} and \citet{lauri2011}.
Additionally, 41 unbarred galaxies are selected having similar
surface brightness profiles with the other galaxies studied by us.  The observations
are also compared with synthetic images using N-body simulation
models.  
  
Unsharp mask images were created for all 214 galaxies, which are used
to measure the sizes and ellipticities of the X-shape features, and to
inspect the low surface brightness features of bars.  For barlenses
isophotal analysis is also carried out obtaining the radial profiles
of the position angles (PA), ellipticities ($\epsilon$) and B$_4$
cosine terms. Synthetic images are created using two simulation
models taken from Salo $\&$ Laurikainen (2016), one representing a
bulgeless galaxy (B/D = 0.01), and another where the galaxy had a
small bulge (B/D = 0.08) at the beginning of the simulation. The model images
are rotated so that a full range of galaxy inclinations in the sky
is obtained, which images are measured in a similar manner as the
observations. The following results are obtained:
\vskip 0.3cm

1. Barlenses in the combined S$^4$G+NIRS0S sample have sizes of
a/r$_{\rm bar} \sim$ 0.5, confirming the previous result by A+2015 for
NIRS0S. We find that the X-shape features appear almost a
factor two smaller than barlenses, which is the case even 
at $i$ = 45$^{\circ}$ -- 65$^{\circ}$, where both features appear.  
We show that this size difference is mainly a projection effect,
and due to the fact that barlenses form in more centrally
concentrated galaxies: observations and simulations show very similar
trends, even if in the models the intrinsic sizes of barlenses and X-shape features
are the same. Our simulation models with different
bulge masses suggest that in order to see an X-shape at
$i$ $\lesssim$ 40$^\circ$, the central mass concentration 
needs to be small.  This is consistent with the observation
that the X-shape features appear predominantly in galaxies with 
smaller B/T than the barlenses do. 
\vskip 0.15cm
 
2. Minor-to-major axis ratios of barlenses appear at b/a = 0.5 -- 1.0, in
agreement with those given by A+2015 for NIRS0S.  Our analysis further
shows that the b/a-distribution for the X-shape features is similar,
although not necessarily the same as for barlenses. 
A comparison with the synthetic images show very similar b/a variations
as a function of galaxy inclination.
\vskip 0.15cm

3. We show that only 24$\%$ of barlenses have boxy isophotes, which
fraction steadily increases towards higher galaxy inclination.  This
is shown using the B$_4$ parameter, which becomes on average negative for $i$
$\gtrsim$ 50$^\circ$.  A remarkably similar trend is obtained for the
vertically thick bar component in our simulation model with B/D =
0.08.  We also verified that the observations of B/P bulges of bars by
ED2013 are consistent with this picture.  Also, at intermediate galaxy
inclinations we find 6 galaxies, which have a barlens in the
classification by \citet{buta2015}, and an X-shape feature in our
unsharp mask image.
\vskip 0.15cm 

4. Barlenses are divided to morphological sub-groups,
based on their fine structures visible in the unsharp mask
images. Barlenses can be round featureless structures, or they can have
additional low surface brightness features along the bar major axis.
Most of the barlenses appear in strong bars of bright galaxies, but
they are recognized also in weakly barred galaxies.  In fact, our
group g, representing the weakest bars among the barlenses in our
sample, are morphologically close to unbarred galaxies. Examples of galaxies with 
X detected in
weak bars in low mass galaxies are IC 3806, IC 335 and NGC 5145.
\vskip 0.15cm
  
5. The sizes of ``bulges'' in unbarred galaxies are measured,
mimicking the barlens regions of typical barlens galaxies. We find
that the sizes of such photometric bulges correlate with the galaxy mass, in a
similar manner as the sizes of barlenses.  We speculate that such
bulges in unbarred galaxies might form in nearly bar-like potentials, as
predicted by \citet{patsis2002}.
\vskip 0.15cm

6. The parent galaxies of barlenses are also divided to sub-groups.
Characteristic features are inner rings and ringlenses, which appear
in 50-100$\%$ of the barlens galaxies. The fractions of inner disks
(disky pseudobulges) and ansae (at the two ends of the bar) vary among
the different parent galaxy groups. Galaxy mass steadily increases
from the group 1 to 4, which is also associated to a decreasing
fraction of early-type galaxies among these groups.  Morphological
counterparts of some barlens parent galaxies are identified among the galaxies
with X-shaped bars.
\vskip 0.3cm

Conclusion: we have shown evidence that barlenses at low galaxy
  inclinations are physically the same inner bar components as
  B/P/X-shape features in more inclined galaxies. Whether these
  structures are barlenses or show boxy/peanut/X-shape
  features depends, besides galaxy orientation, also on the central
  mass concentration of the parent galaxy.  This is shown by comparing
  directly the properties of barlenses and X-shaped features, and is also
  verified by our simulation models.

For two barlens galaxies detailed stellar populations and kinematics,
given in the literature, are discussed in the context of the
identified barlenses. The properties of these galaxies are also compared with those of
the Milky Way bulge.
We conclude that the stellar populations of barlenses in these galaxies
are similar to those of the Milky Way bulge.
   
\begin{acknowledgements}
We acknowledge Panos Patsis of valuable discussions while prepearing
this manuscript, and the anonymous referee who's comments have
considerable increased the quality of this paper. We also acknowledge
the Spitzer Space Telescope of the allocated observing for the
S$^4$G-project, and also of the observing time allocated to NIRS0S
project during 2003-2009, based on observations made with several
telescopes.  They include the New Technology Telescope (NTT), operated
at the Southern European Observatory (ESO), William Herschel Telescope
(WHT), the Italian Telescopio Nazionale Galileo (TNG), and the Nordic
Optical Telescope (NOT), operated on the island of La Palma.  This
work was also supported by the DAGAL network: Programmme (Marie Curie
Actions) of the European Unions Seventh Framework Programme
FP7/2007-2013 under REA grant agreement number PITN-GA-2011-289313. We
also acknowledge Riku Rautio of participating in making some of the X-shape
measurements.

All the electgronic figures of this paper are available in a web-page: /www.oulu.fi/astronomy/BLX.

\end{acknowledgements}

\clearpage
\newpage


   \begin{figure}
   \centering
   \includegraphics[width=\hsize]{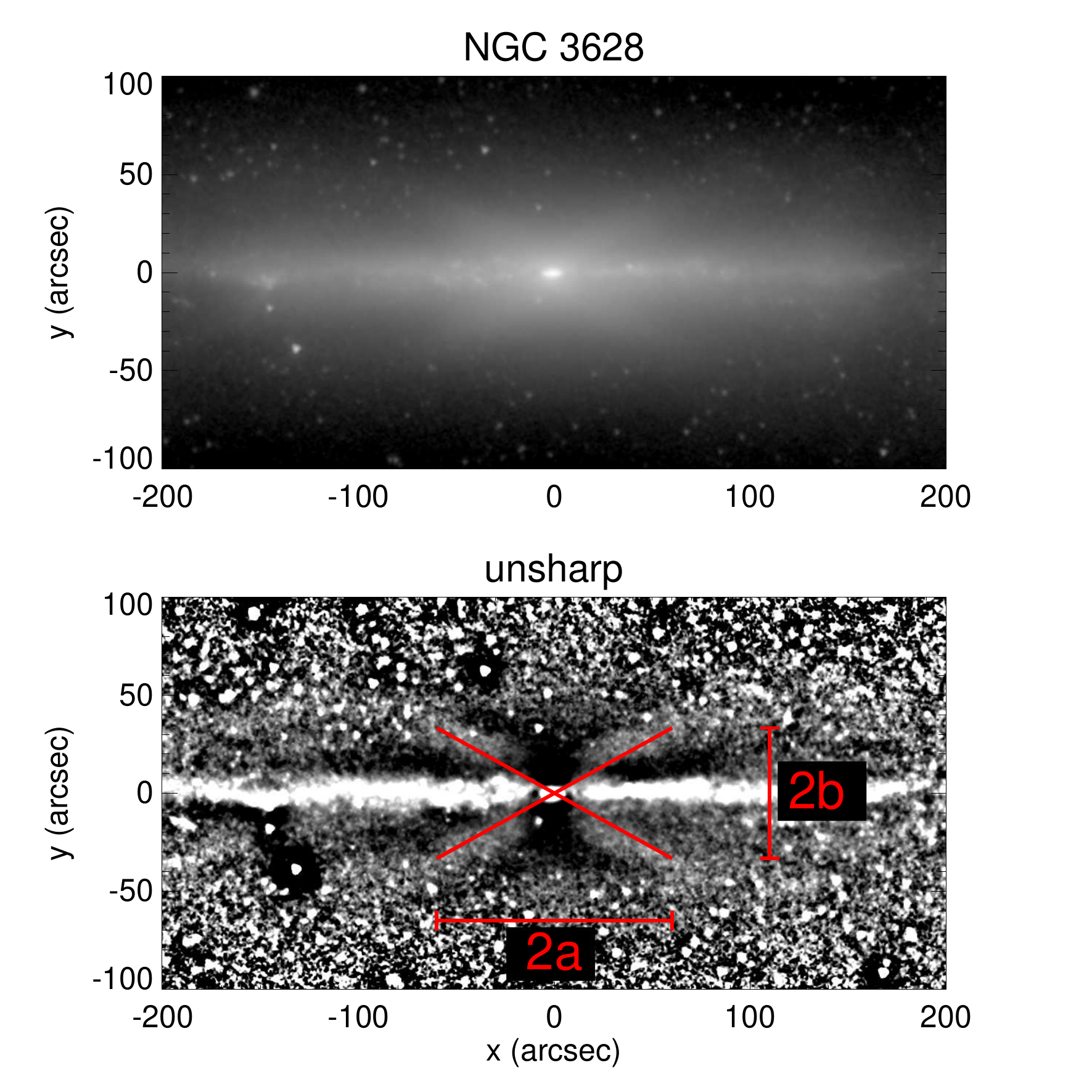} 
      \caption{NGC 3628 ({\it upper panel}) is used as an example to
        demonstrate how the sizes of the X-shapes are measured. The
        unsharp mask image ({\it lower panel}) is used, which is rotated so
        that the bar major axis appears horizontally. The extent
          of the feature is measured both along the bar direction and
          perpendicular to it, and the semilenghts are denoted by a
          and b, respectively.  }
         \label{fig:x_measure}
   \end{figure}

\clearpage
\newpage
\begin{figure*}
\includegraphics[angle=0,width=17.0cm]{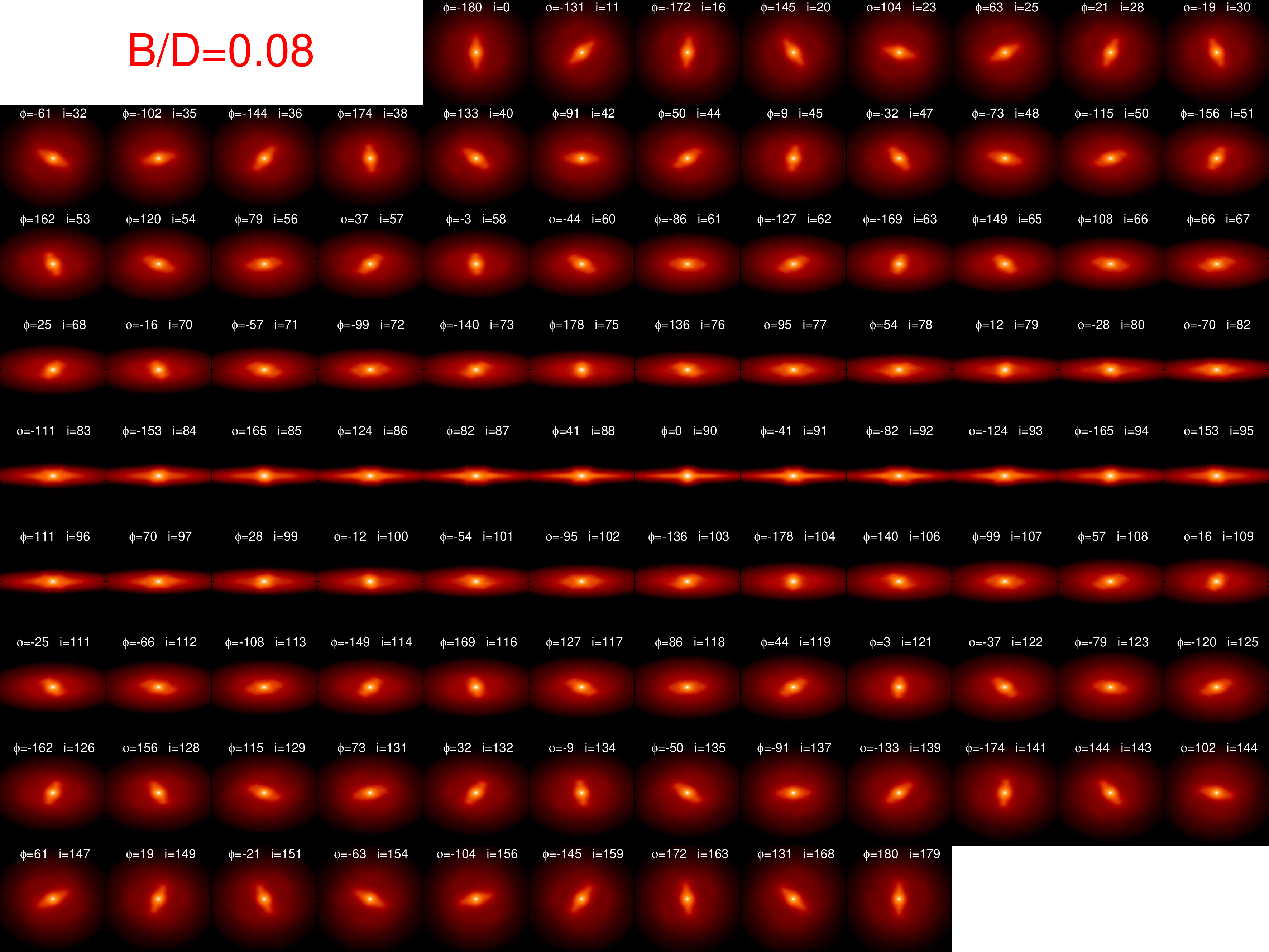} 
\caption{Synthetic images used in comparison with observations, from
  the simulation with B/D = 0.08. The same simulation snapshot is
  viewed from 100 isotropically chosen directions.  The labels in the
  frames indicate the viewing azimuth $\phi$ with respect to bar major
  axis, and the viewing inclination $i$. The line-of-nodes are
  horizontal.  The simulation model is explained in Section 3.3.}
\label{fig:simu_montage}
\end{figure*}

\clearpage
\newpage
\begin{figure*}
\subfloat[]{\includegraphics[angle=0,width=9.5cm]{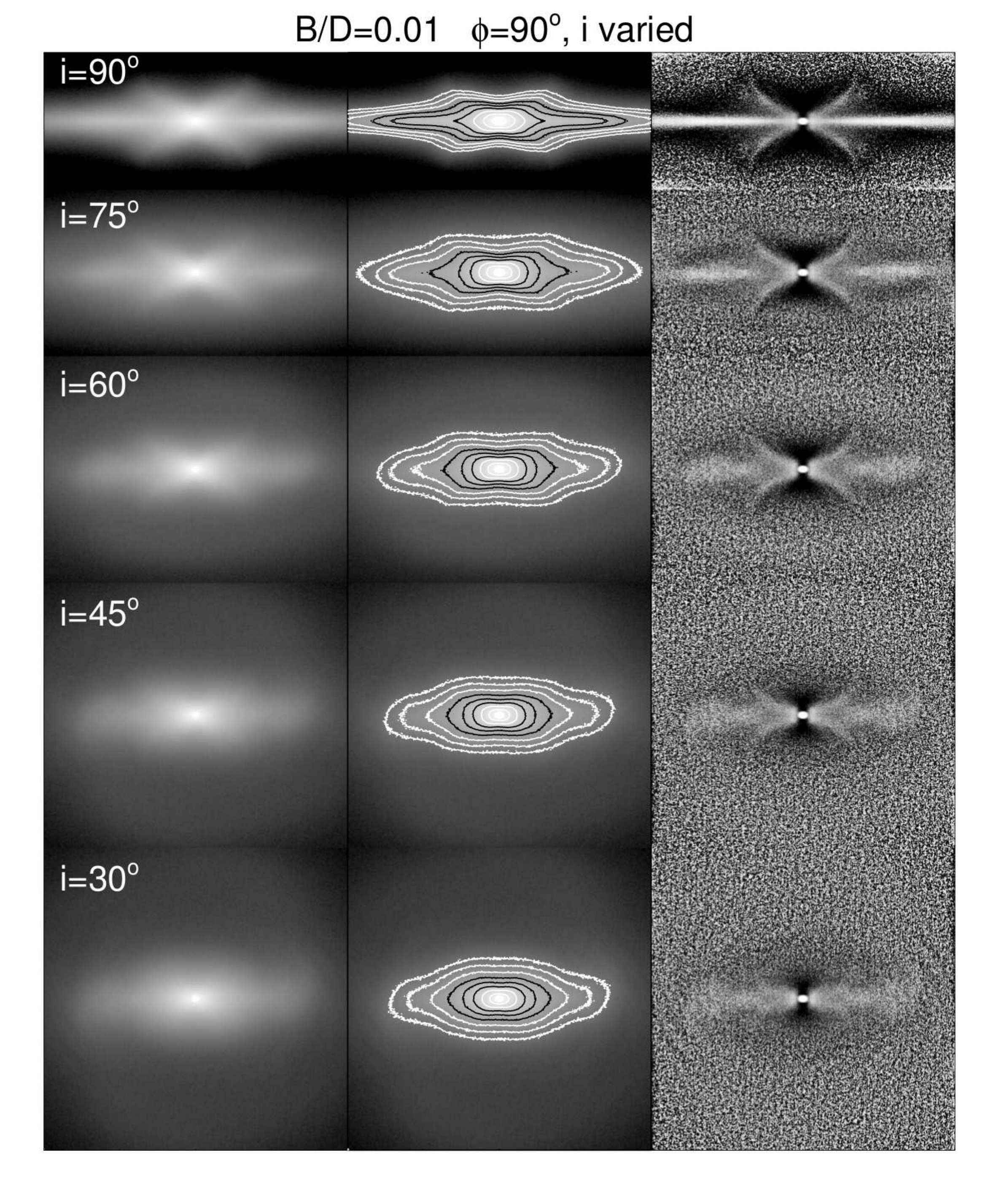}} 
\subfloat[]{\includegraphics[angle=0,width=9.5cm]{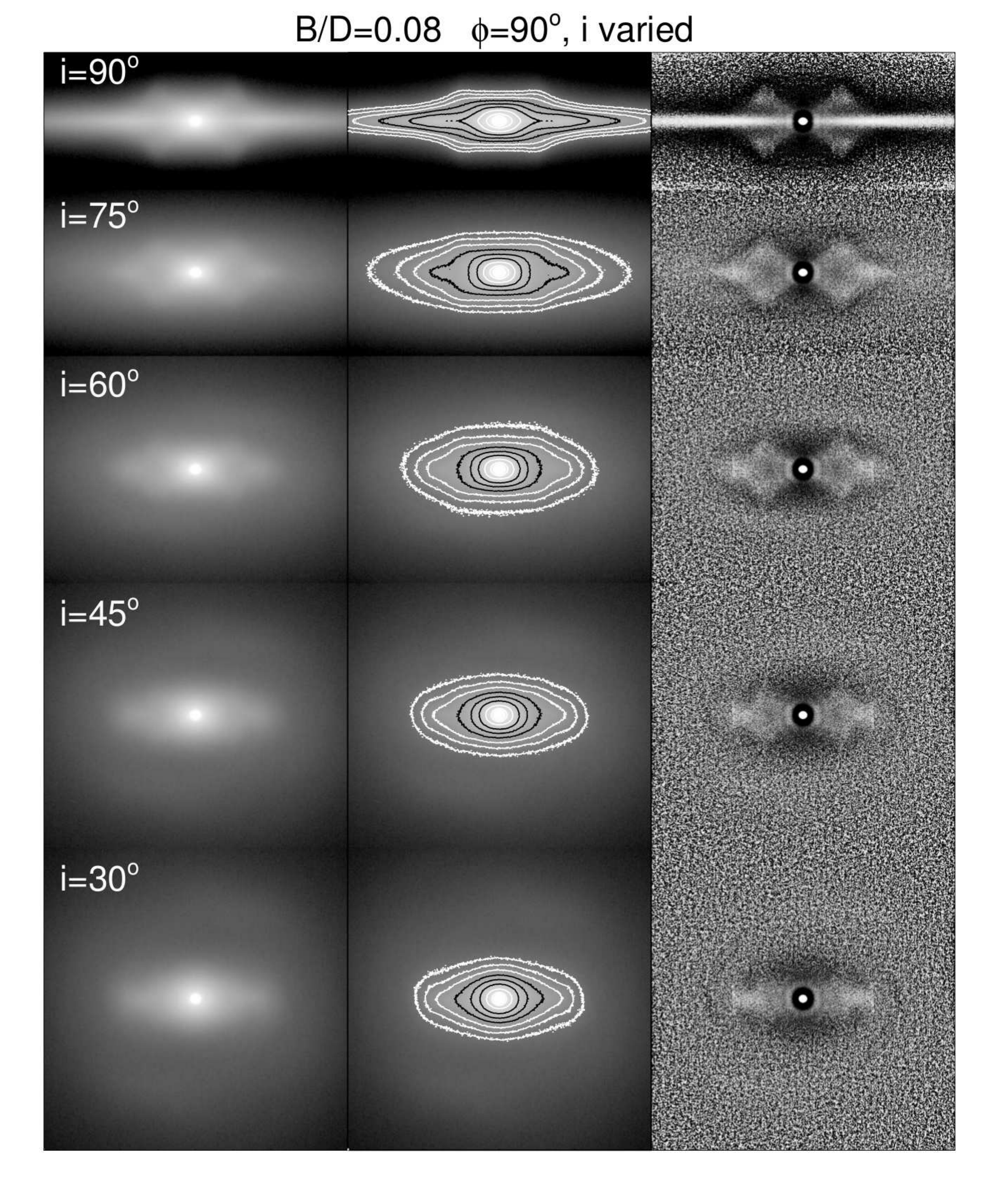}} 
\caption{Snapshot of the simulation with {\it (a)} B/D = 0.01 and {\it (b)}
  $B/D$ = 0.08 are viewed at azimuthal angle $\phi$ = 90$^\circ$, from five
  galaxy inclinations ($i$ = 90$^\circ$ corresponds the side-on-view
  of the bar).  The line-of-node is horizontal. The left panels show
  the synthetic images, in the middle panel the isophotal contours,
  separated by 0.5 mags, are overlayed on the images, while the right
  panels show the unsharp mask images.  }
 \label{fig:simu_tilt}
\end{figure*}

\clearpage
\newpage
\begin{figure*}
\subfloat[]{\includegraphics[angle=0,width=9.5cm]{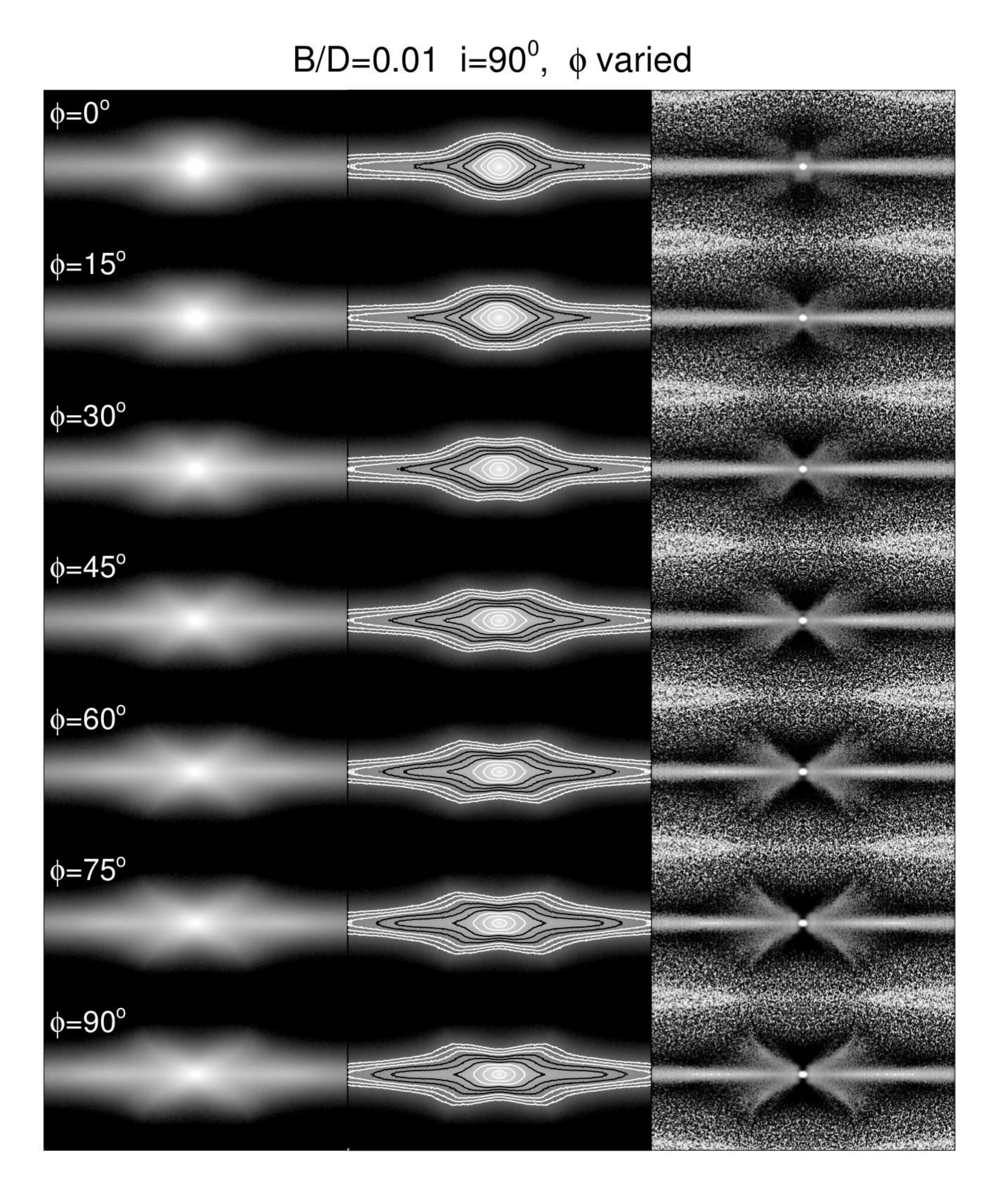}}
\subfloat[]{\includegraphics[angle=0,width=9.5cm]{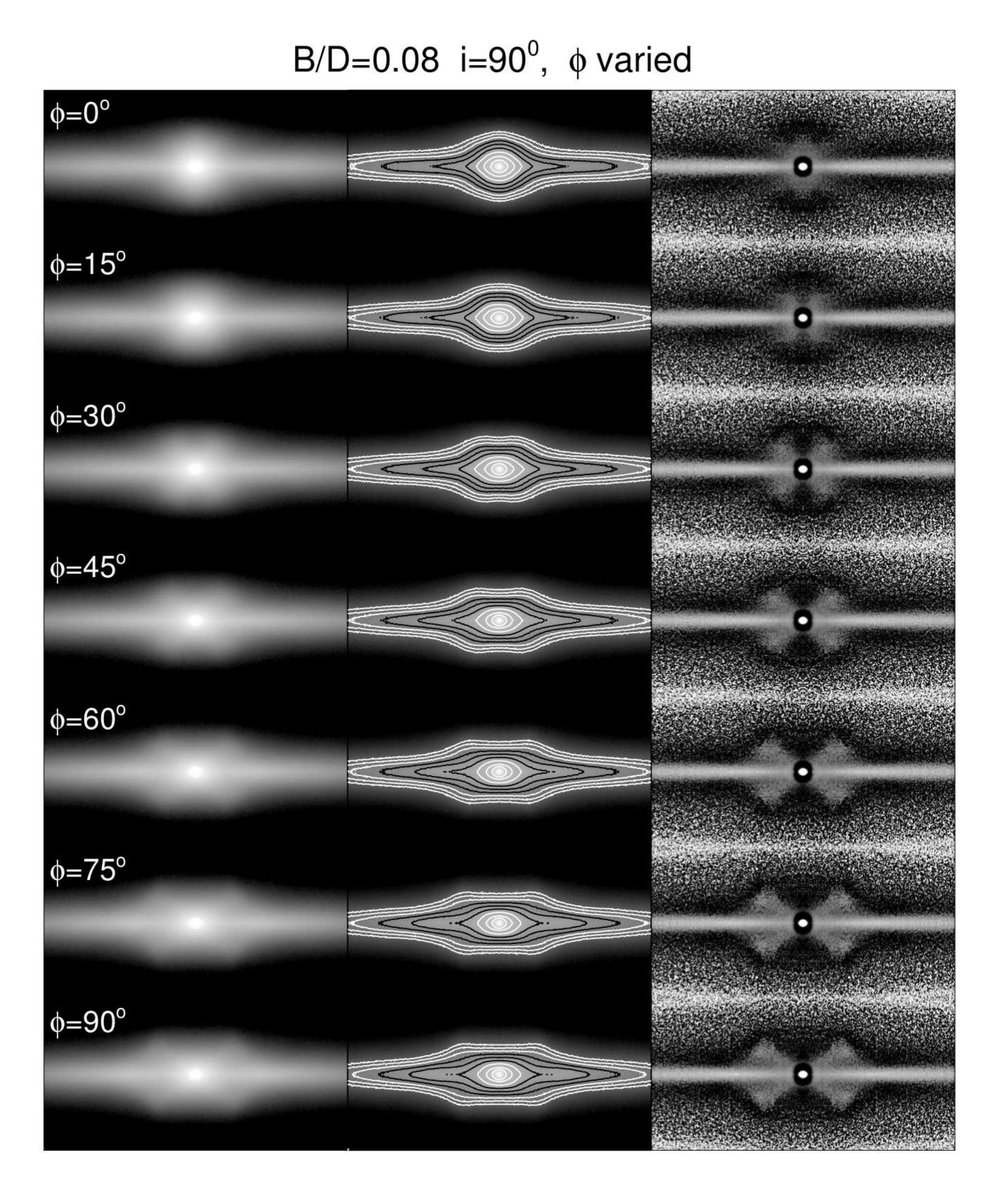}}

\caption{Same as Figure 3, except that the inclination is fixed to
  $i$ = 90$^\circ$, and the viewing azimuth is varied. In the upper panels the
  bar is seen end-on ($\phi$ = 0$^\circ$), and in the lowest panels side-on ($\phi$ = 90$^\circ$).}
\label{fig:simu_azi_inc90}
\end{figure*}

\clearpage
\newpage
\begin{figure*}
\subfloat[]{\includegraphics[angle=0,width=9.5cm]{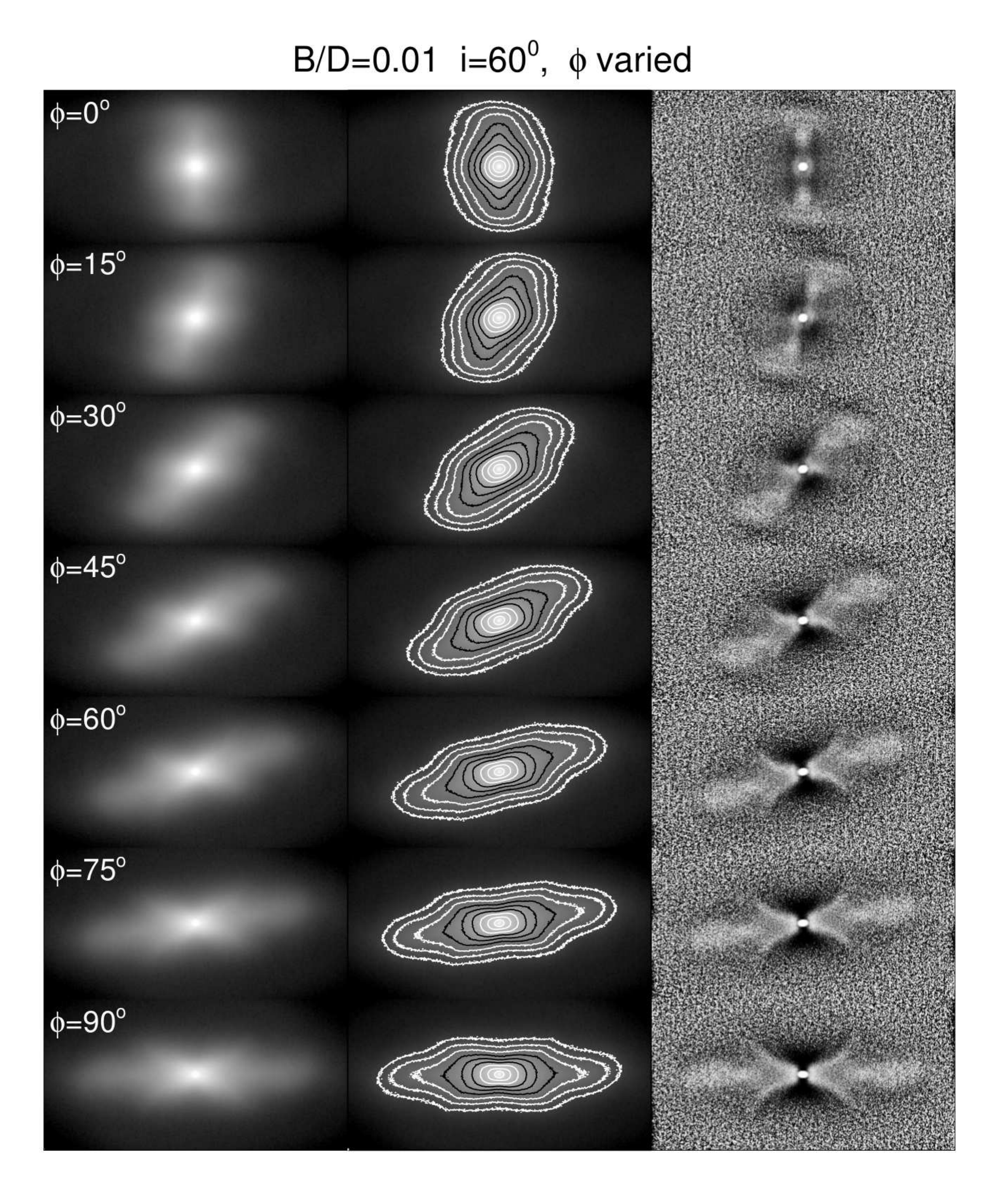}} 
\subfloat[]{\includegraphics[angle=0,width=9.5cm]{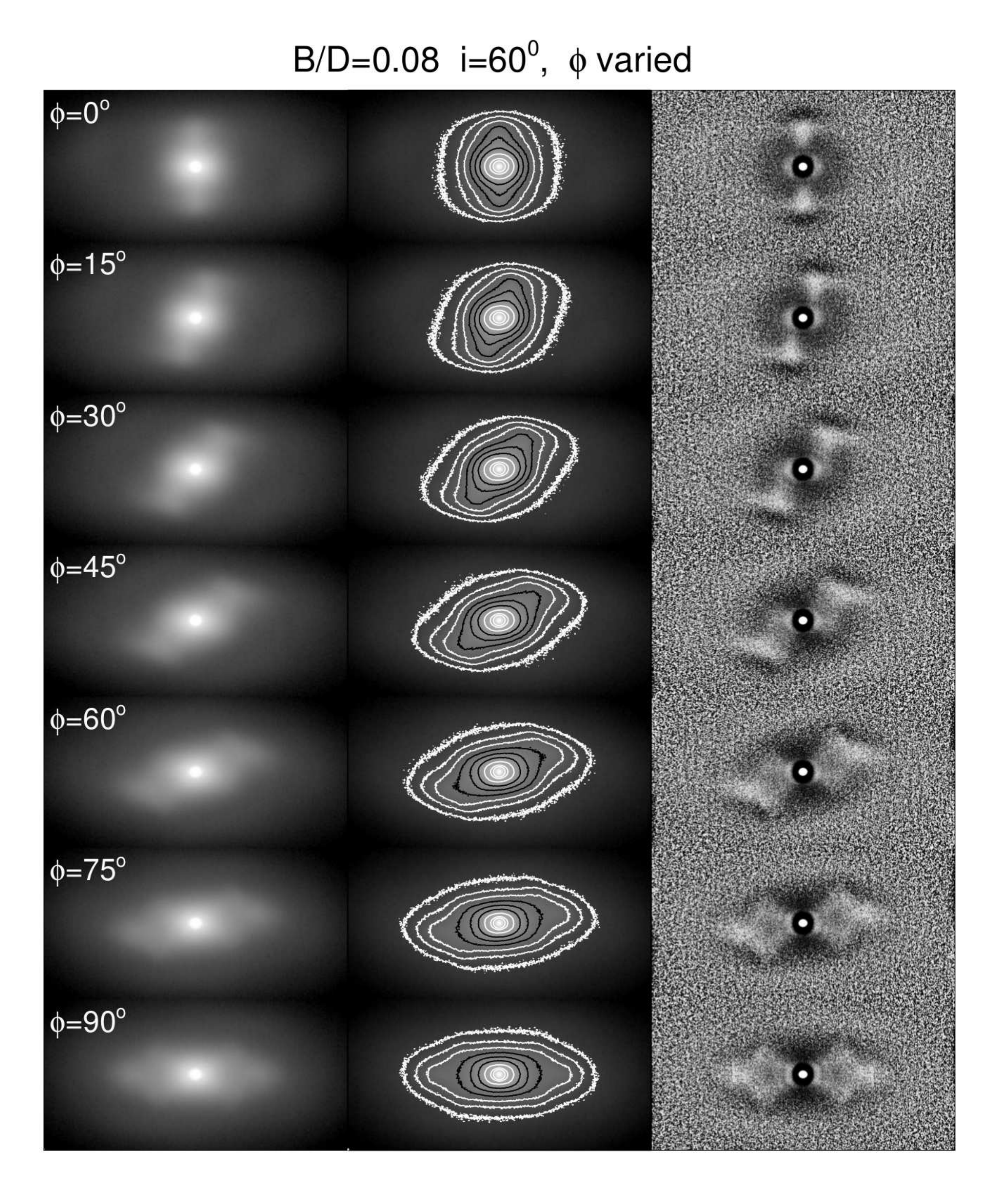}} 
\caption{Same as Figure 4, except that the
  inclination is fixed to $i$ = 60$^{\circ}$. In the different panels the
  azimuthal angle $\phi$ varies.
}
\label{fig:simu_azi_inc60}
\end{figure*}

\clearpage
\newpage

  \begin{figure}
   \centering
   \includegraphics[width=\hsize]{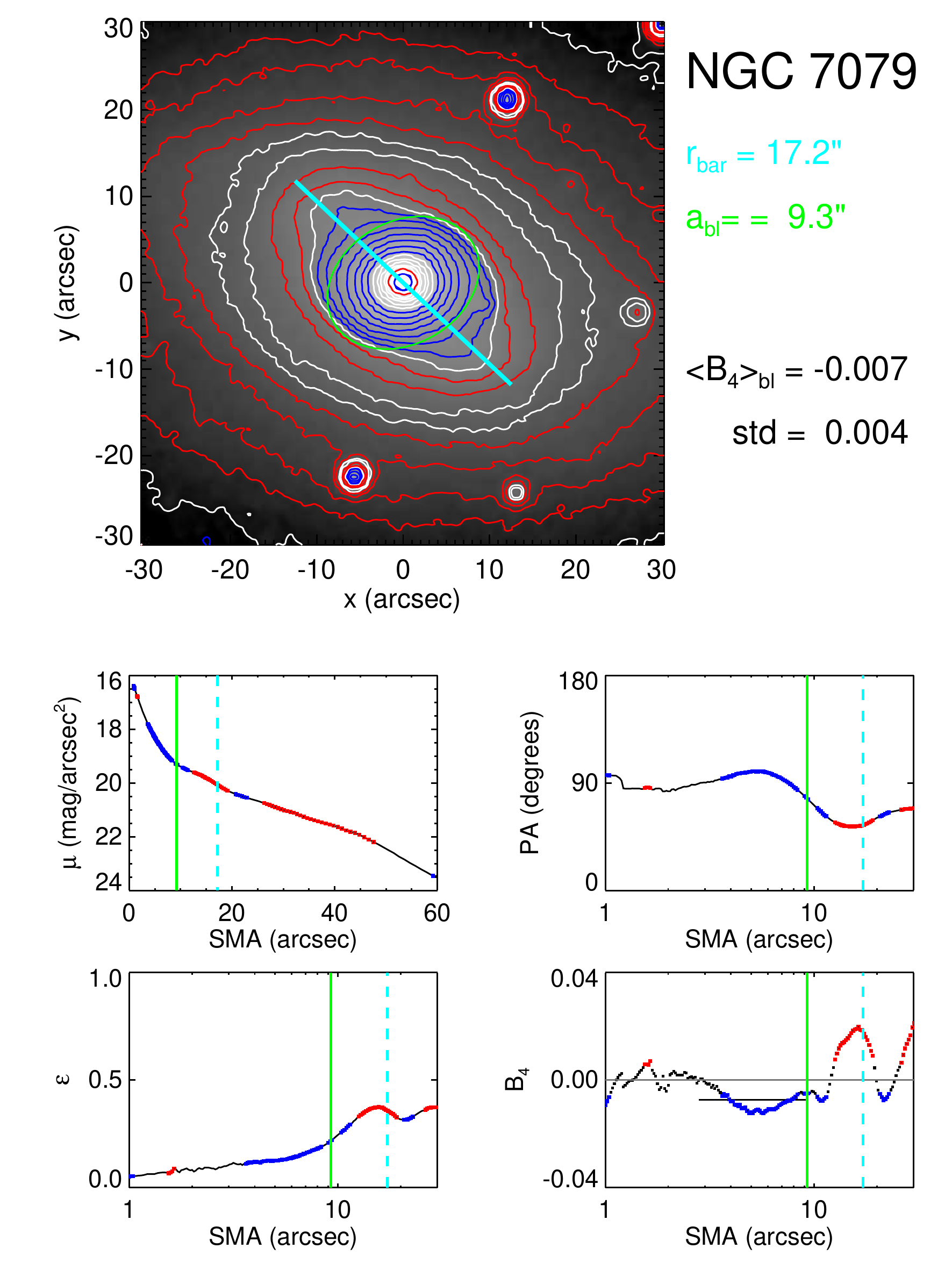} 
      \caption{Isophotal analysis performed for a barlens galaxy NGC
        7079. In the upper panel the 3.6 $\mu$m image is shown in
          the sky plane with North up and East left: overlaid are the
        isophotal contours. The blue line indicates the bar length and
orientation,
        and the green ellipse denotes the ellipse fit to the barlens.
        The four lower panels show the radial profiles from IRAF
        ellipse: the surface brightness  $\mu$ (in mag
          arcsec$^{-2}$) {\it (upper left panel)}, the position angle
        (PA$^{\circ}$) {\it (upper right panel)}, the ellipticity
        ($\epsilon$) {\it (lower left panel)}, and the B$_4$ parameter
        {\it (lower right panel)} as a function of semi-major axis.
        B$_4$ is used as proxy for the boxiness: in the small panels
        the blue and red colors indicate the regions where B$_4 <$
        -0.005 and B$_4 >$ 0.005, respectively. The green vertical
        full line shows the radius of the barlens, and the dashed blue
        line the bar radius (same colours are used on the contour
        plot). The labels in the upper right indicate the bar radius
        (r$_{\rm bar}$), the semi-major axis length of the barlens
        (a$_{\rm bl}$), the mean and standard deviation of B$_4$ in
        the region where the isophotal radius is (0.3 -- 1.0) a$_{\rm
          bl}$. Similar analysis has been carried out for all 84
        barlens galaxies in our sample.}
         \label{fig:b4_measure}
   \end{figure}


\clearpage
\newpage
 \begin{figure}
   \centering
   \includegraphics[width=\hsize]{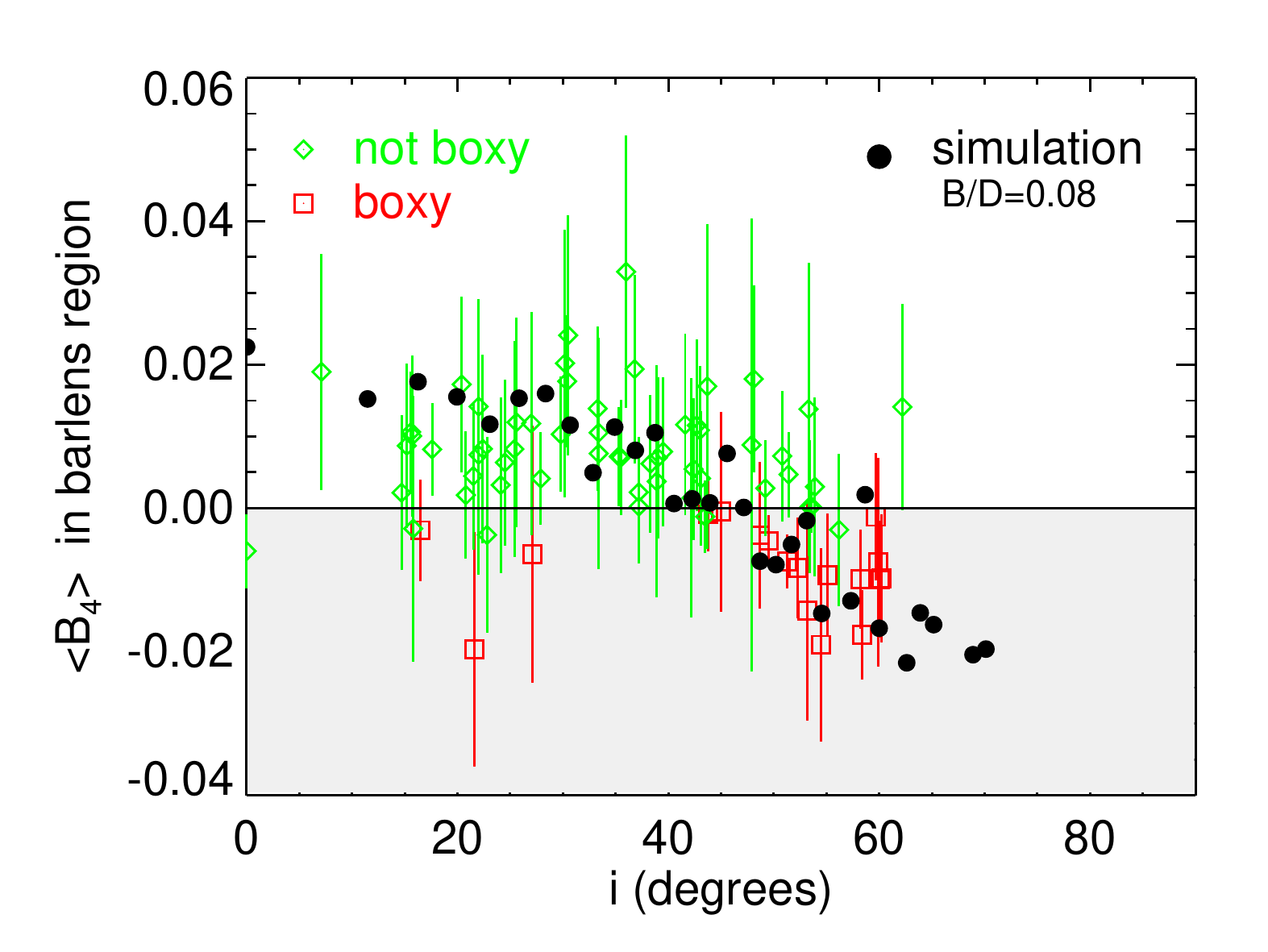}
      \caption{The mean value of B$_4$ parameter in the region of the
        barlens (isophotal radius in range (0.3 -- 1.0) a$_{\rm
          bl}$) is displayed as a function of galaxy inclination. The
        red and green symbols indicate galaxies which have been judged
        boxy and non-boxy, based on visual inspection of the isophotes
        and the B$_4$ profiles. The filled circles show the same
        parameter measured from the synthetic images, for the
        simulation model with B/D = 0.08. The error bars in the
        observation points correspond to $\pm$ one standard deviation of
        $B_4$ in the measurement region. }
         \label{fig:b4_compare}
   \end{figure}

\clearpage
\newpage

%
   \begin{figure}
   \centering
   \includegraphics[width=\hsize]{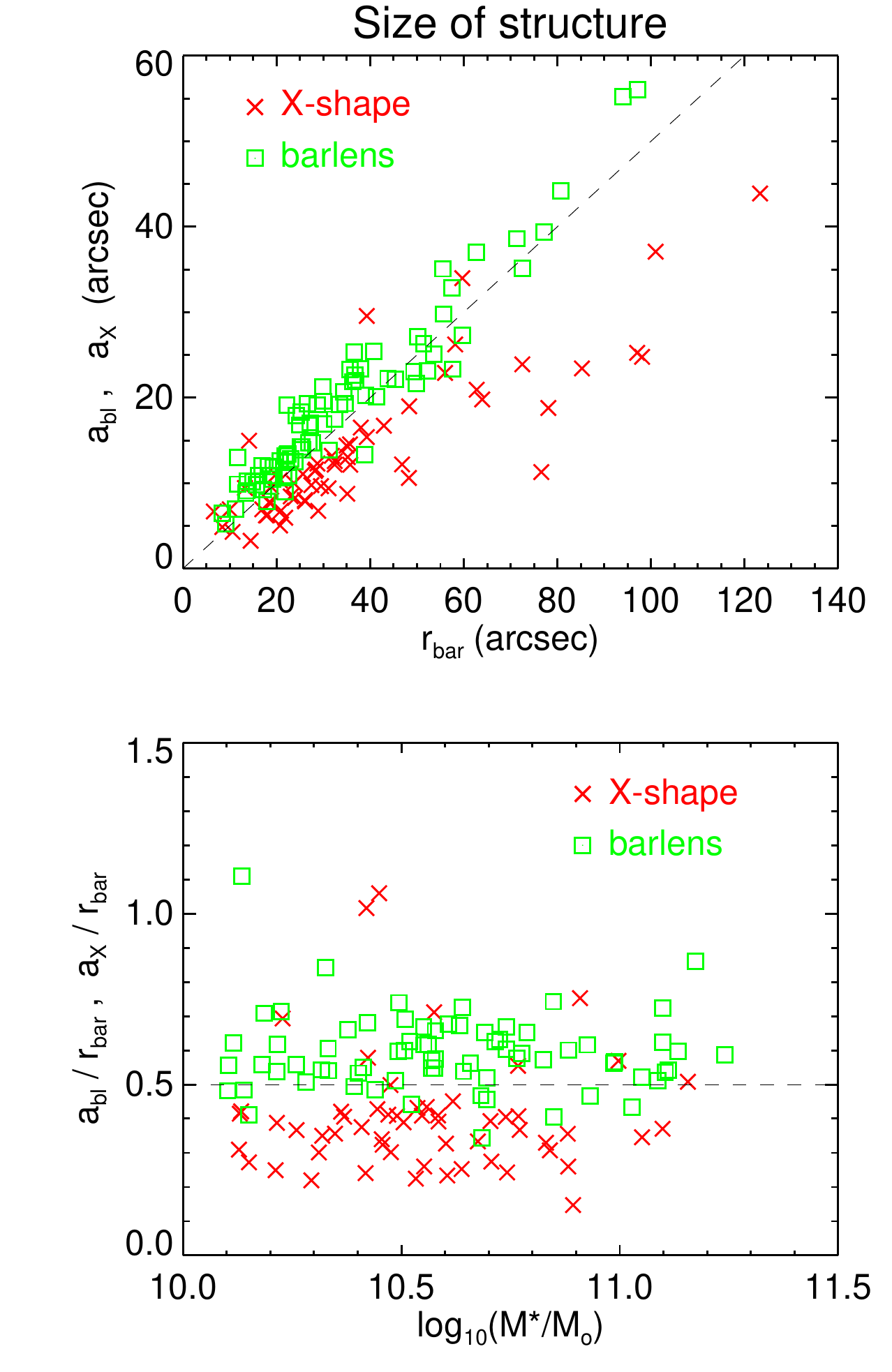} 
      \caption{ {\it Upper panel}: the sizes (a) of barlenses and
        X-shape features are shown as a function of bar radius
        (r$_{bar}$), given in arcseconds. All measurements are in the
        sky plane.  For X-shapes the measurements are from the current study,
        and for barlenses from Laurikainen et al. (2011) and
        HE+2016. {\it Lower panel}: The sizes of barlenses and X-shape
        features are normalized to the barlength, and drawn as a
        function of the parent stellar galaxy mass (M*), taken from S$^4$G
        Pipeline 3 (Mu\'noz-Mateos et al., 2015). The measured uncertainties
are typically less than 0.5 arcsec. }
 \label{fig:size_compare}
   \end{figure}

\clearpage
\newpage
   \begin{figure}
   \centering
   \includegraphics[width=\hsize]{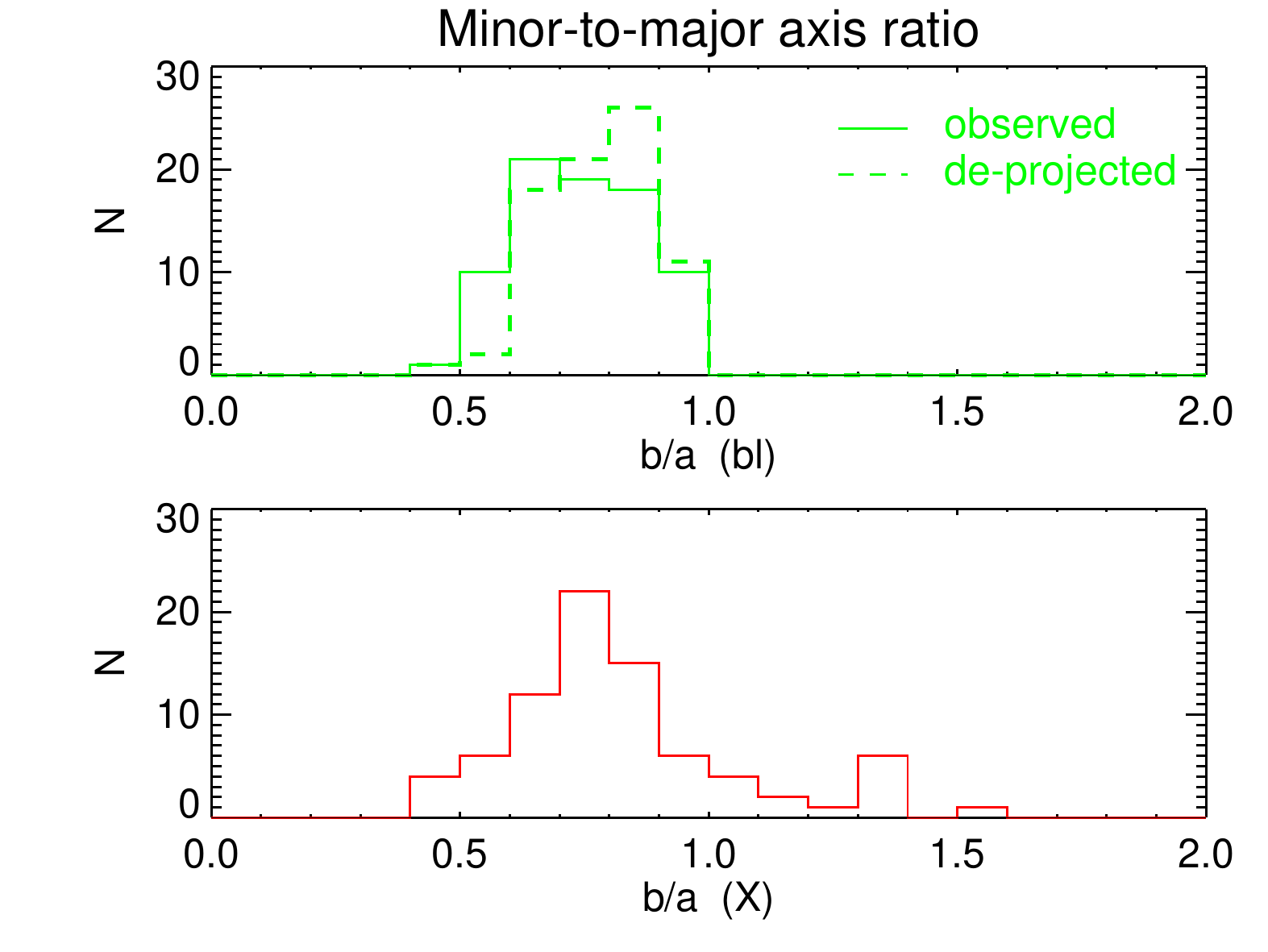} 
      \caption{Distributions of the minor-to-major axis ratios (b/a)
        of barlenses {\it (upper panel)}, both as measured in the
        sky-plane, and after deprojection to disk-plane. The {\it
          lower panel} shows the b/a ratio for the X-shapes (in the
        sky plane).}
 \label{fig:ba_compare}
   \end{figure}

\clearpage
\newpage

%
%


  \begin{figure*}
\includegraphics[angle=0,width=17.0cm]{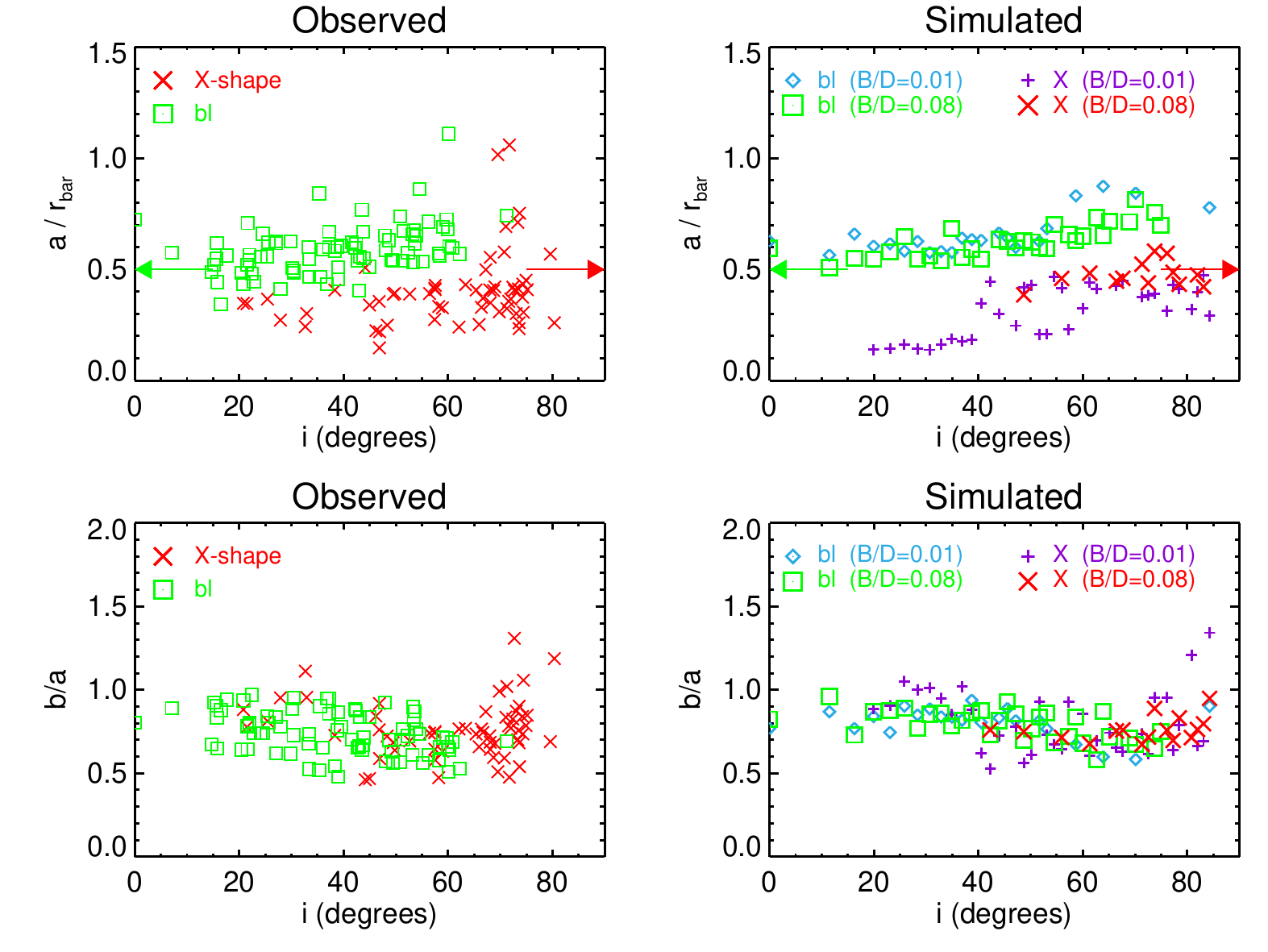} 
      \caption{ Normalized sizes (a/r$_{\rm bar}$) and the
        minor-to-major axis ratios (b/a) of barlenses and X-shape
        features of the galaxies are plotted as a function of parent
        galaxy inclination ({\it left panels}). The same parameters
        for the synthetic images are also shown ({\it right
          panels}). The simulations with small (B/D = 0.01) and large
        bulge (B/D = 0.08) are shown with different symbols. The
        arrows in the two panels use the model with B/D = 0.08,
        indicating the normalized barlens size in face-on view (green
        arrow in the left), and the size of the X-shape when seen the
        same model edge-on (red arrow in the right). The measured uncertainties are
typically less than 0.5 arcsec.  }
 \label{fig:obs_simu_compare}
   \end{figure*}

\clearpage
\newpage
\begin{figure*}
\includegraphics[angle=0,width=15.35cm]{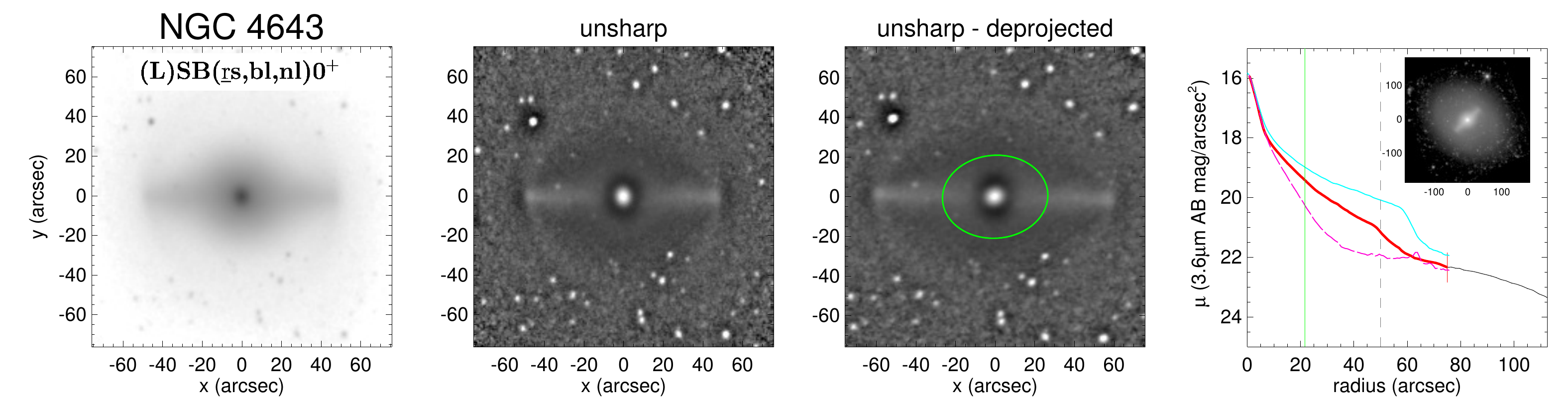}
\includegraphics[angle=0,width=15.35cm]{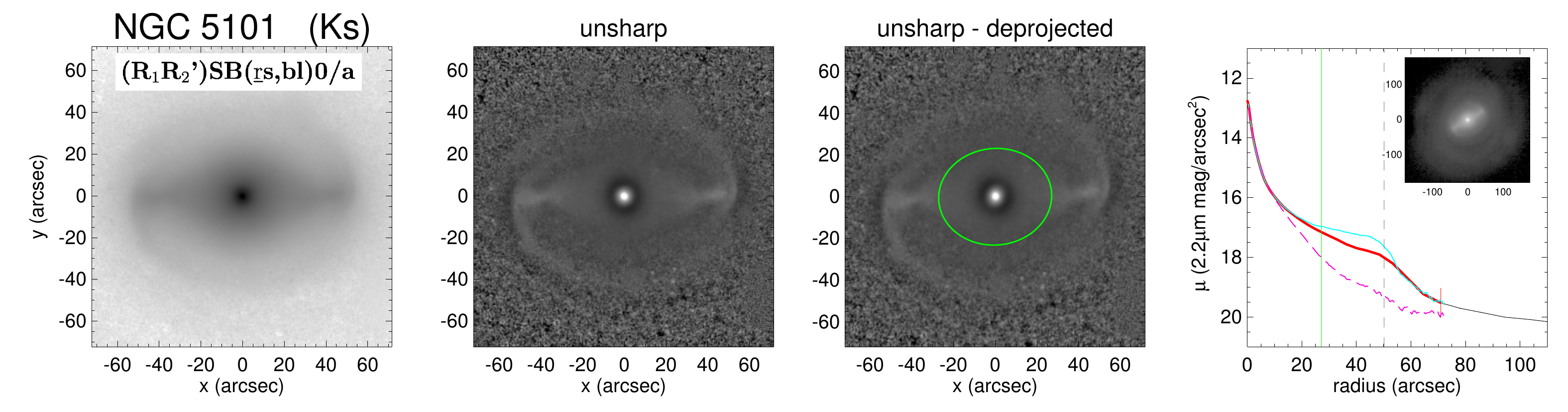}
\includegraphics[angle=0,width=15.35cm]{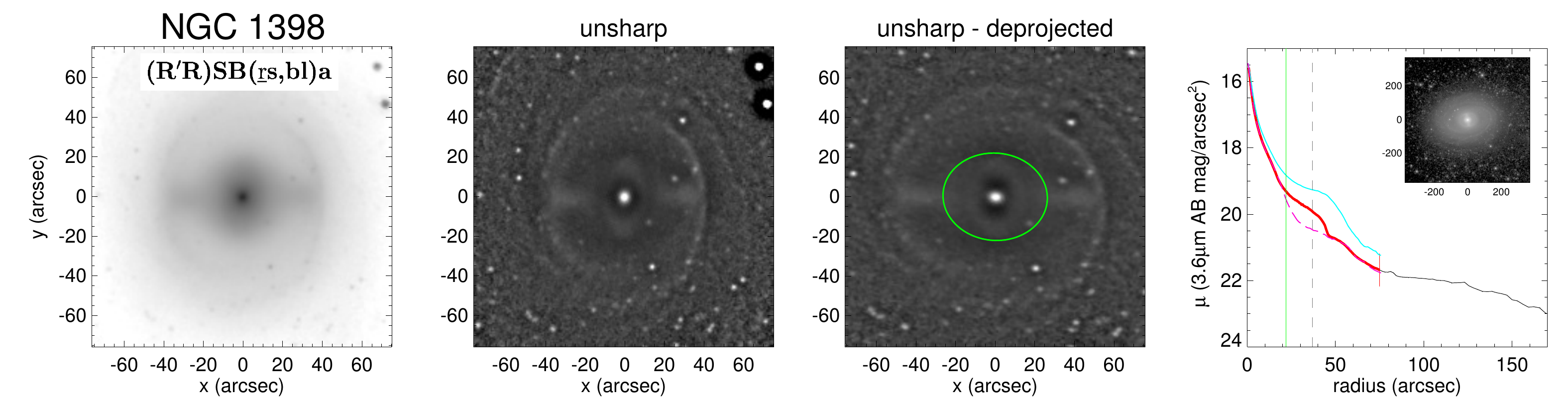}
\includegraphics[angle=0,width=15.35cm]{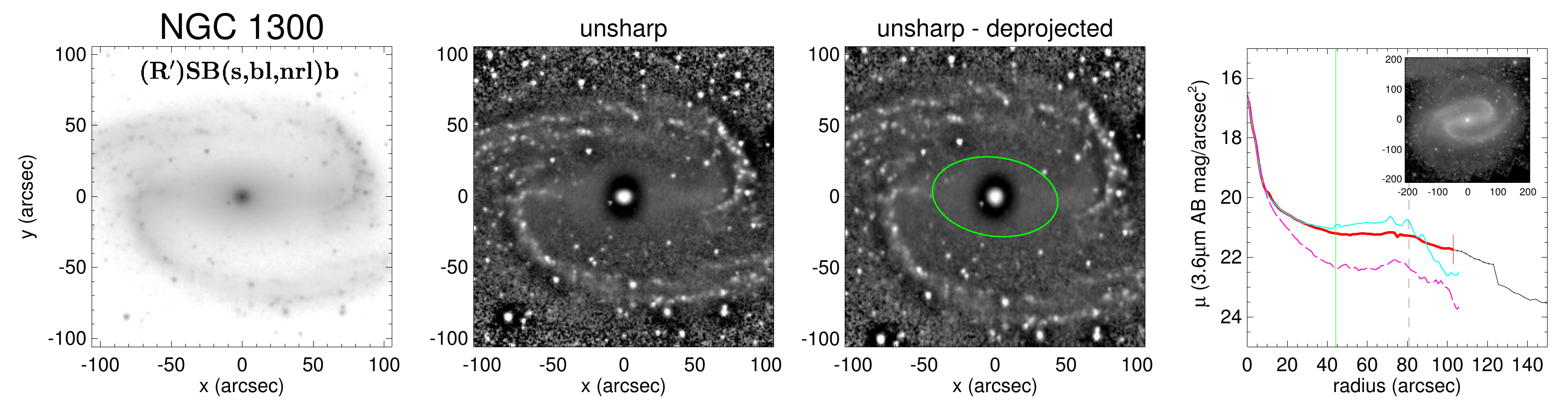}
 \caption{Barlens groups a, b, c, and d, as explained in
   Section 6.1. The {\it left panels} show the 3.6 $\mu$m images and
   the {\it two middle panels} the observed (left) and deprojected
   (right) unsharp mask images.  The images are cut to show only the
   bar region, and the bar is always aligned horizontally. The green
   circles show the barlenses.  In the {\it right panels} the surface
   brightness profiles are shown.  The black lines are the isophotal
   profiles from IRAF ellipse, and the profiles cuts along the bar major
   (blue line) and minor axis (red dashed line) are shown separately
   in the same panel.  However, in case that the galaxy inclination is
   larger than 65$^{\circ}$ only the major axis profile is shown. The
   small panels in the upper corners show the images in full size and
   having original orientations in the sky.  The vertical full and dashed
   lines indicate the sizes of barlenses and bars, respectively. The
   red portion of the surface brightness profile indicates the range
   of the unsharp mask image. In case NIRS0S K$_s$ is shown instead 
of 3.6 $\mu$m image, this is indicated in the label of the
leftmost frame. }
\vspace{-18.5 cm} \hspace{15.5 cm} {\Large  Group a}

\vspace{3.5 cm} \hspace{15.5 cm} {\Large Group b}

\vspace{3.5 cm} \hspace{15.5 cm} {\Large Group c}

\vspace{3.5 cm} \hspace{15.5 cm} {\Large Group d}

         \label{fig:obs_groups1}
\end{figure*}

\clearpage
\begin{figure*}
\includegraphics[angle=0,width=15.35cm]{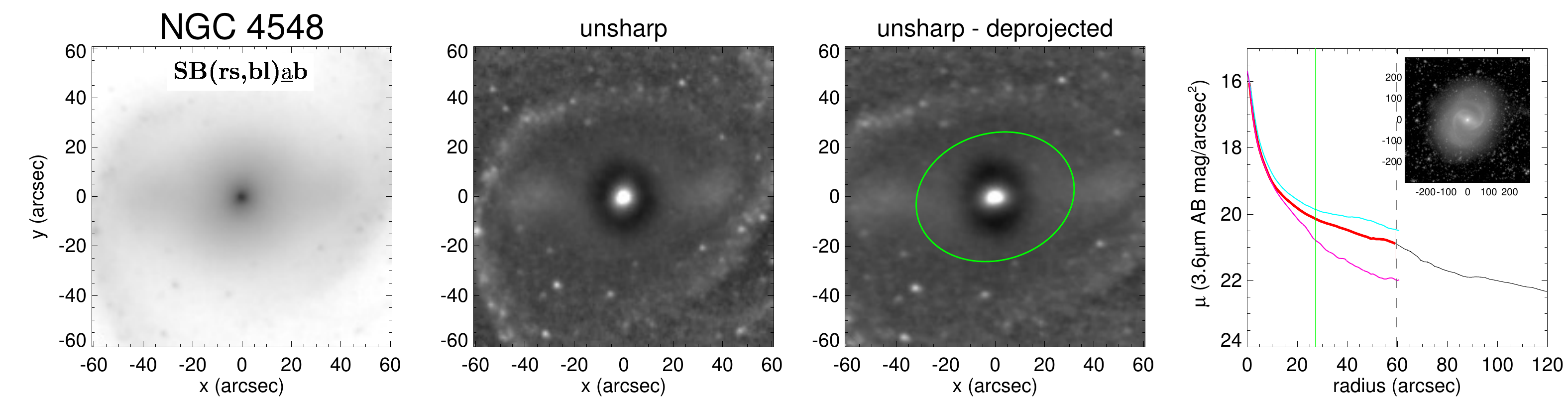}
\includegraphics[angle=0,width=15.35cm]{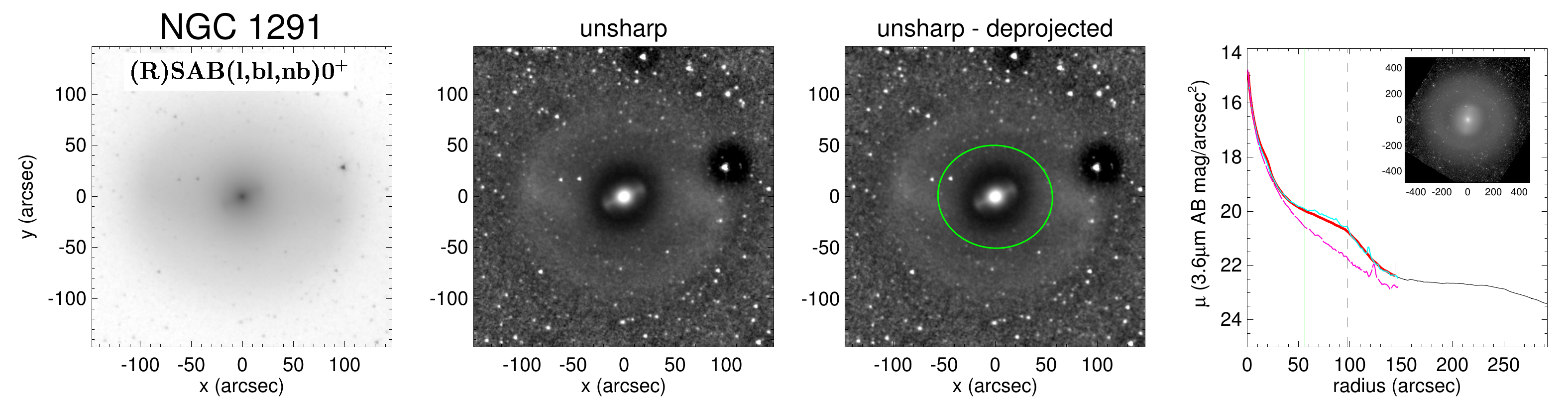}
\includegraphics[angle=0,width=15.35cm]{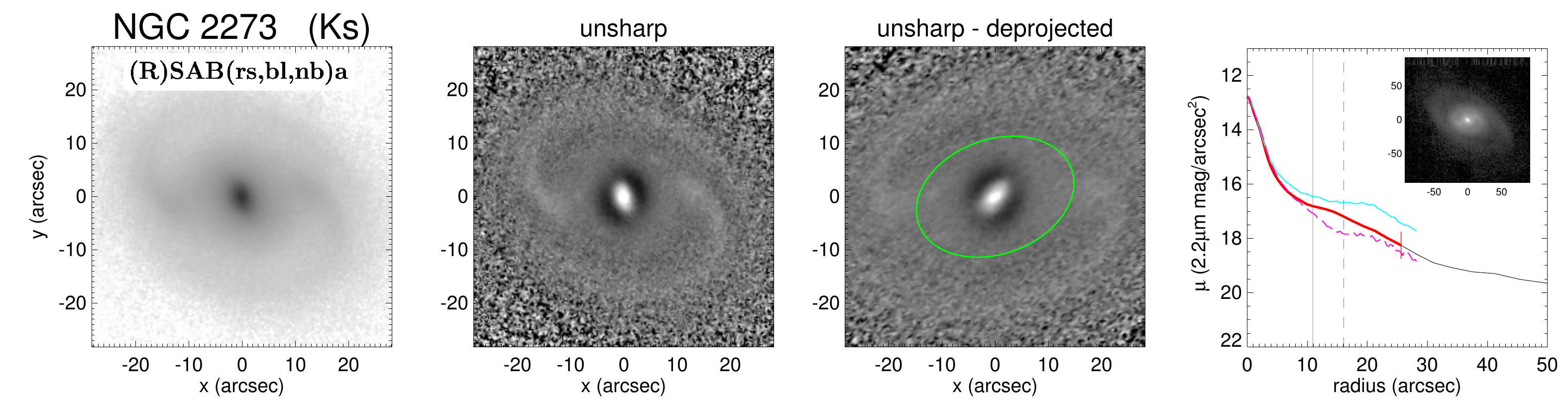}
\includegraphics[angle=0,width=15.35cm]{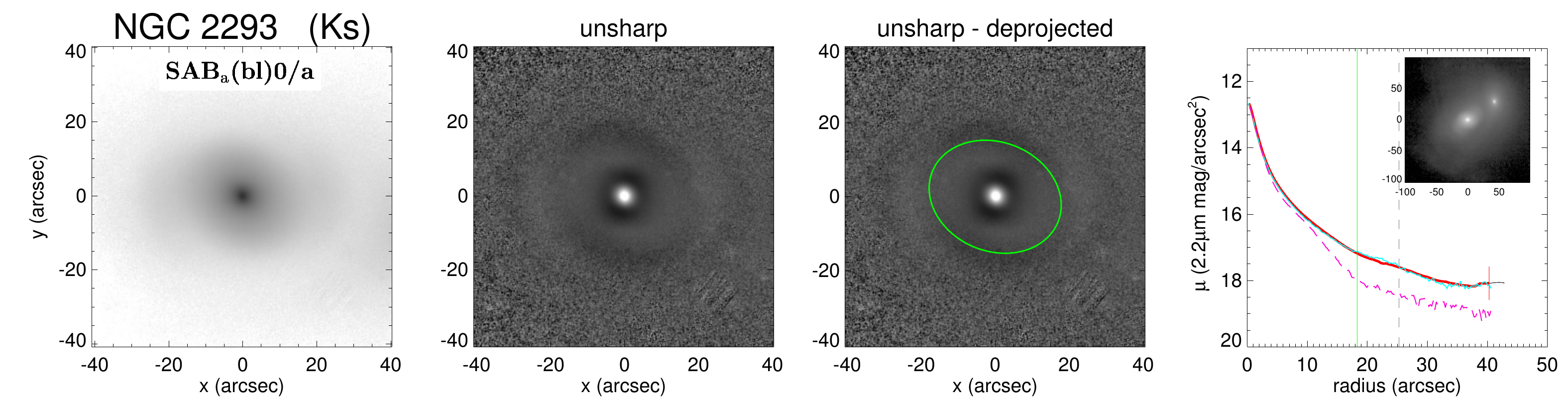}
\caption{Barlens groups e, f, and g, as explained in Section 6.1. The format of the figure is the same as in Fig. 11.}
\vspace{-15 cm} \hspace{15.5 cm} {\Large Group e (B)}

\vspace{3.5 cm} \hspace{15.5 cm} {\Large Group e (AB)}

\vspace{3.5 cm} \hspace{15.5 cm} {\Large Group f}

\vspace{3.5 cm} \hspace{15.5 cm} {\Large Group g}
         \label{fig:obs_groups2}
\end{figure*}
\clearpage





\clearpage
\newpage

  \begin{figure}
   \centering
\subfloat[Parent galaxy group 1a]{
\includegraphics[width=\hsize]{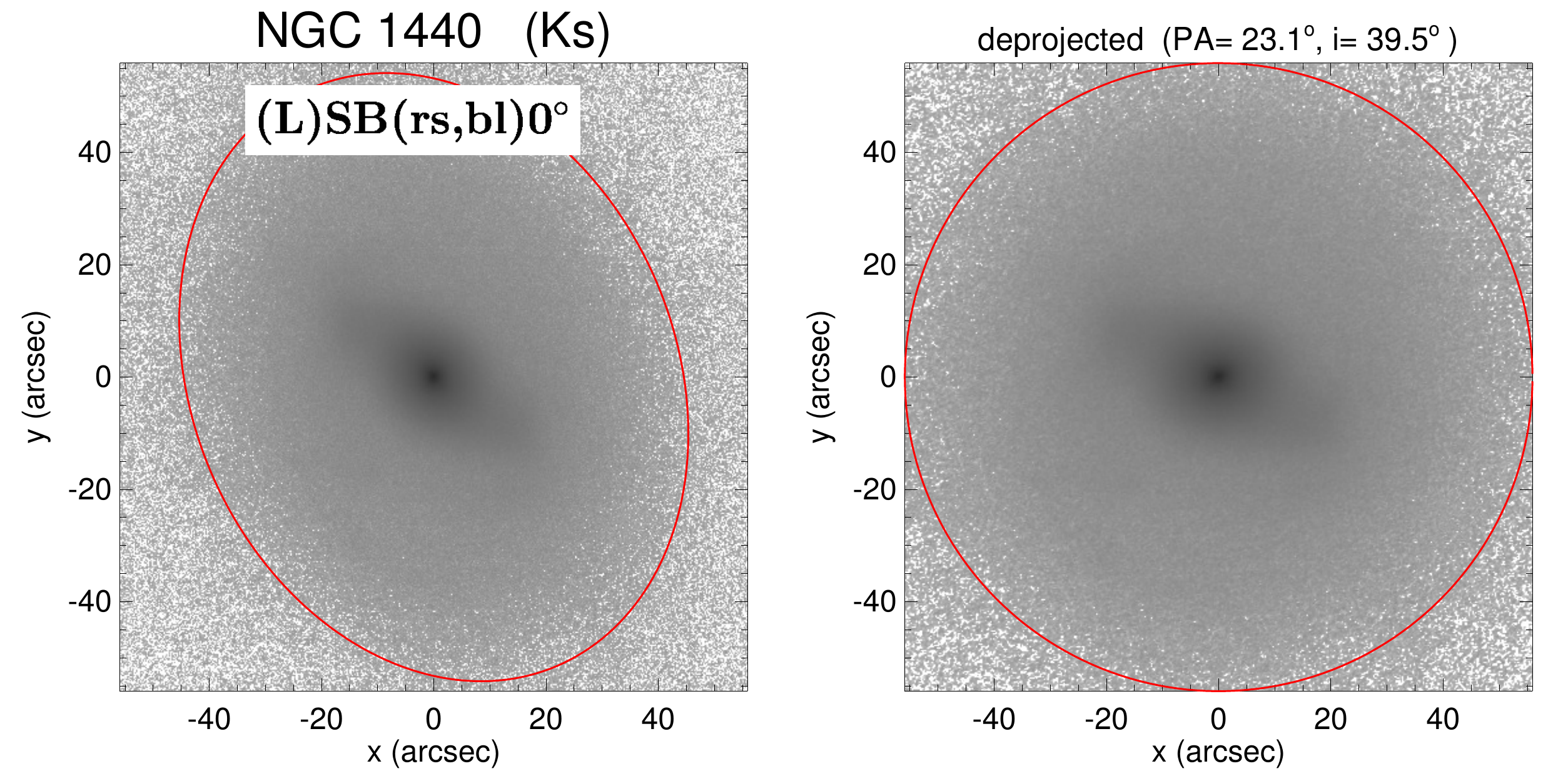}}
\subfloat[Parent galaxy group 1b]{
\includegraphics[width=\hsize]{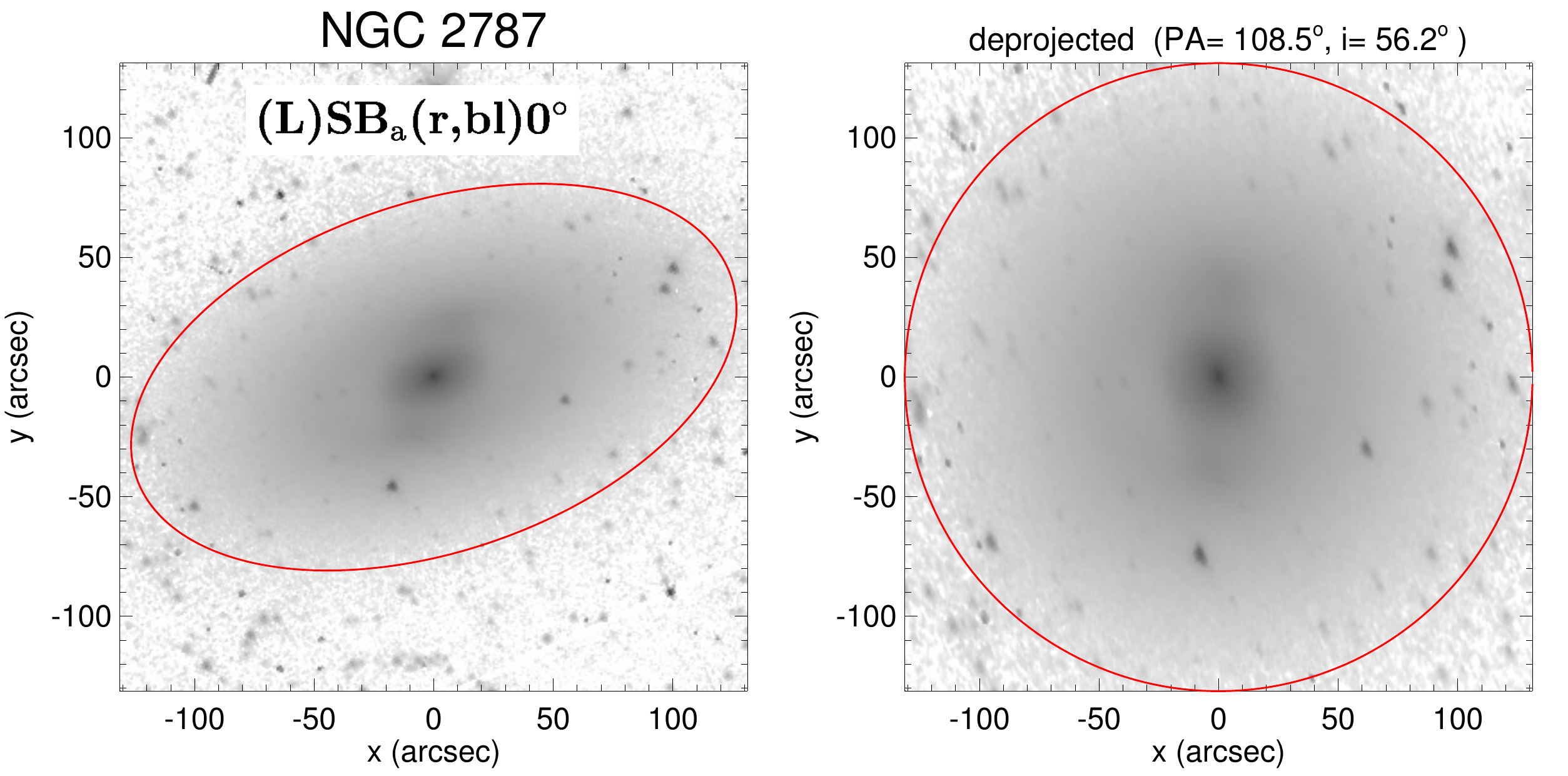}}

\subfloat[Parent galaxy group 2]{
\includegraphics[width=\hsize]{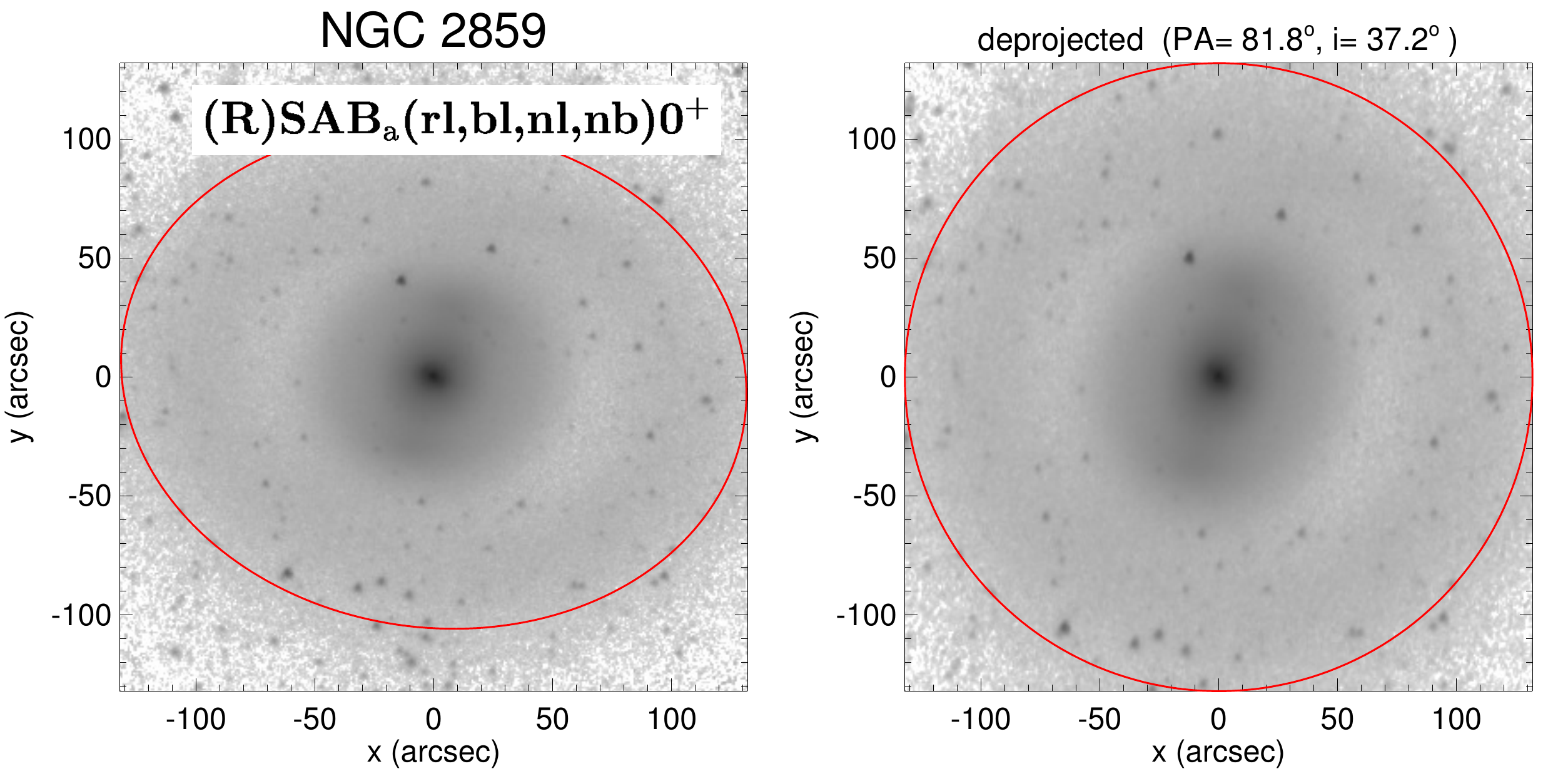}}
\subfloat[Parent galaxy group 3]{
\includegraphics[width=\hsize]{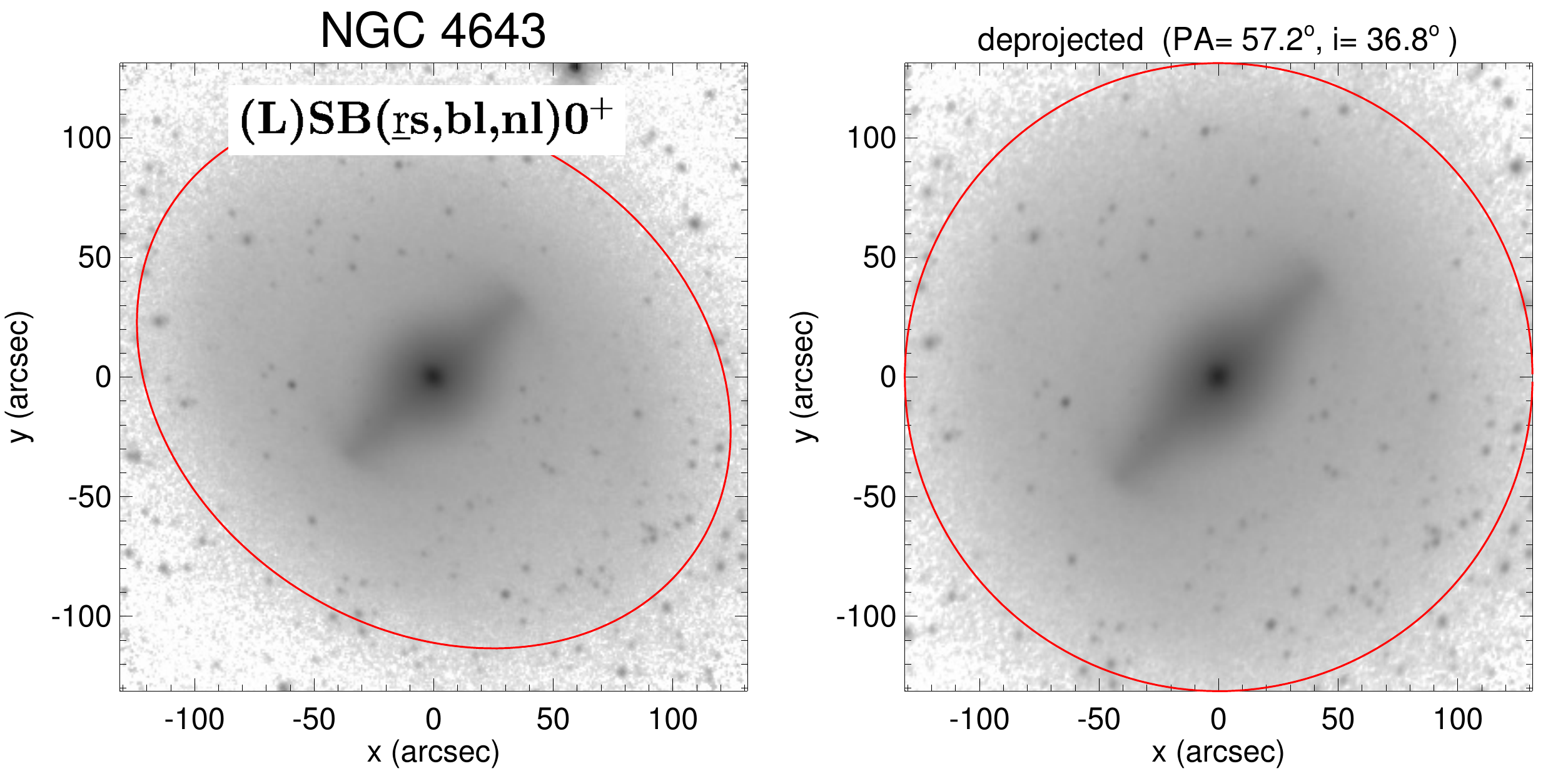}}


\subfloat[Parent galaxy group 4]{
\includegraphics[width=\hsize]{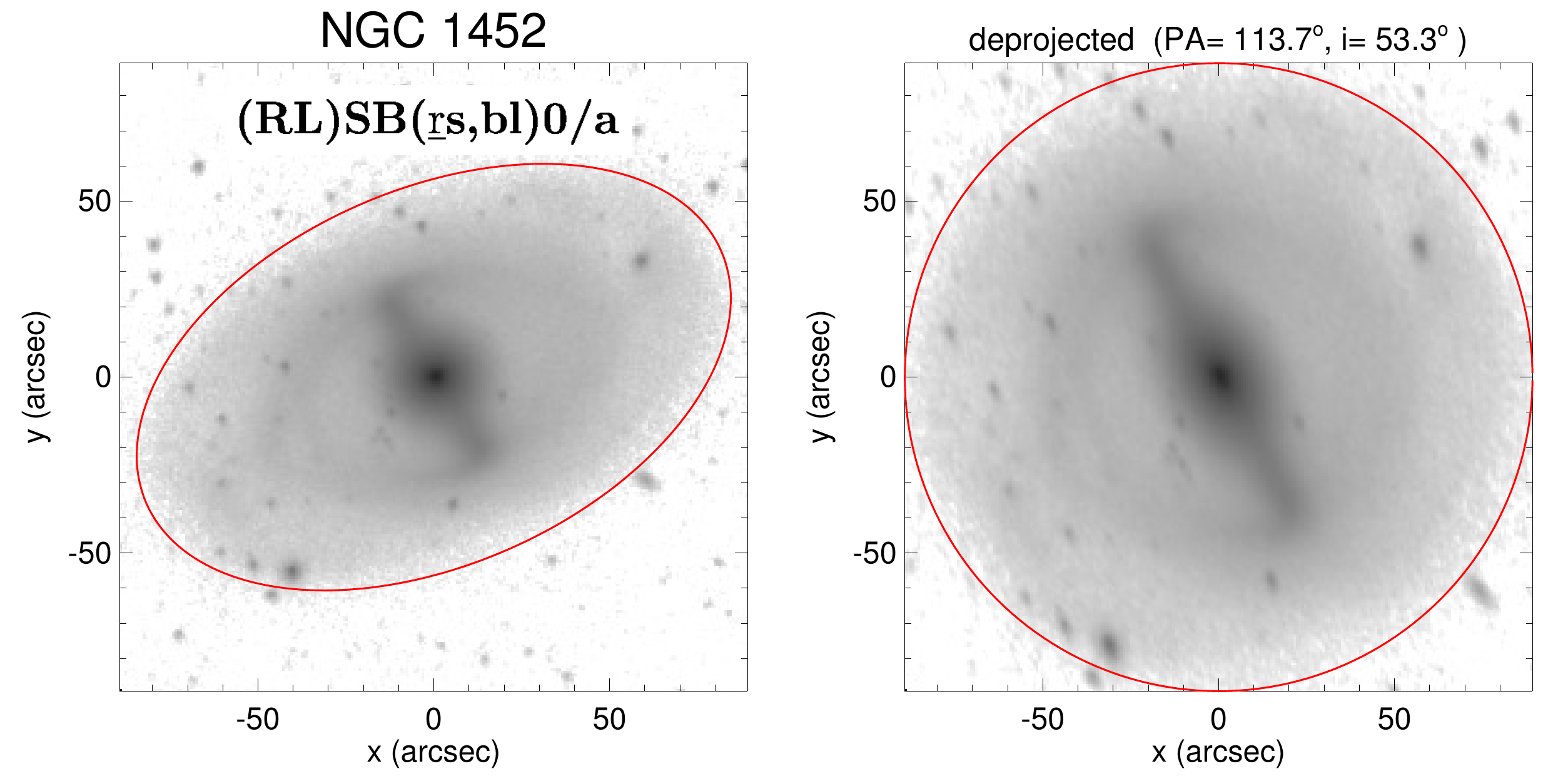}}
\subfloat[Parent galaxy group 5]{
\includegraphics[width=\hsize]{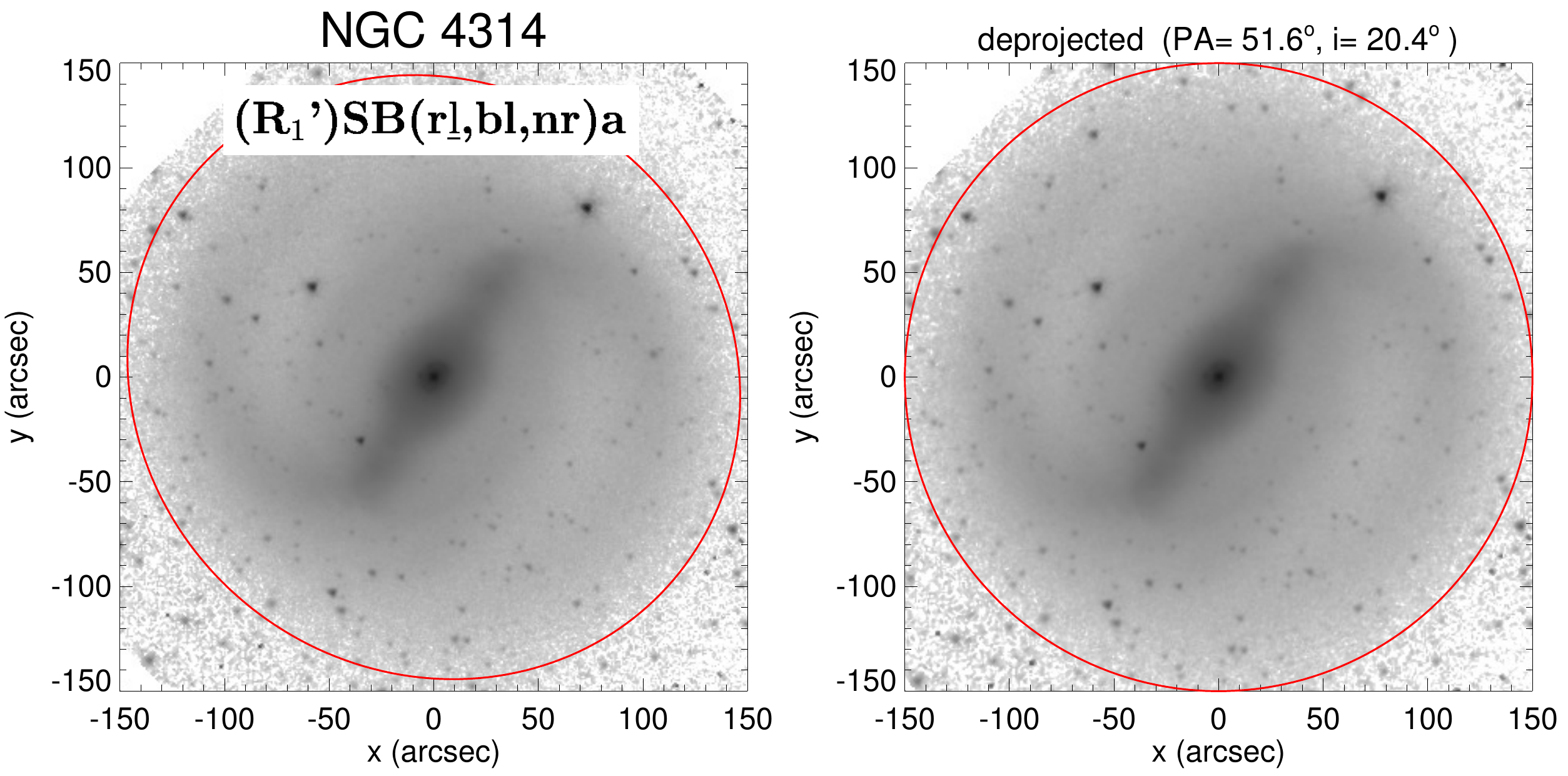}}

\subfloat[Parent galaxy group 6]{
\includegraphics[width=\hsize]{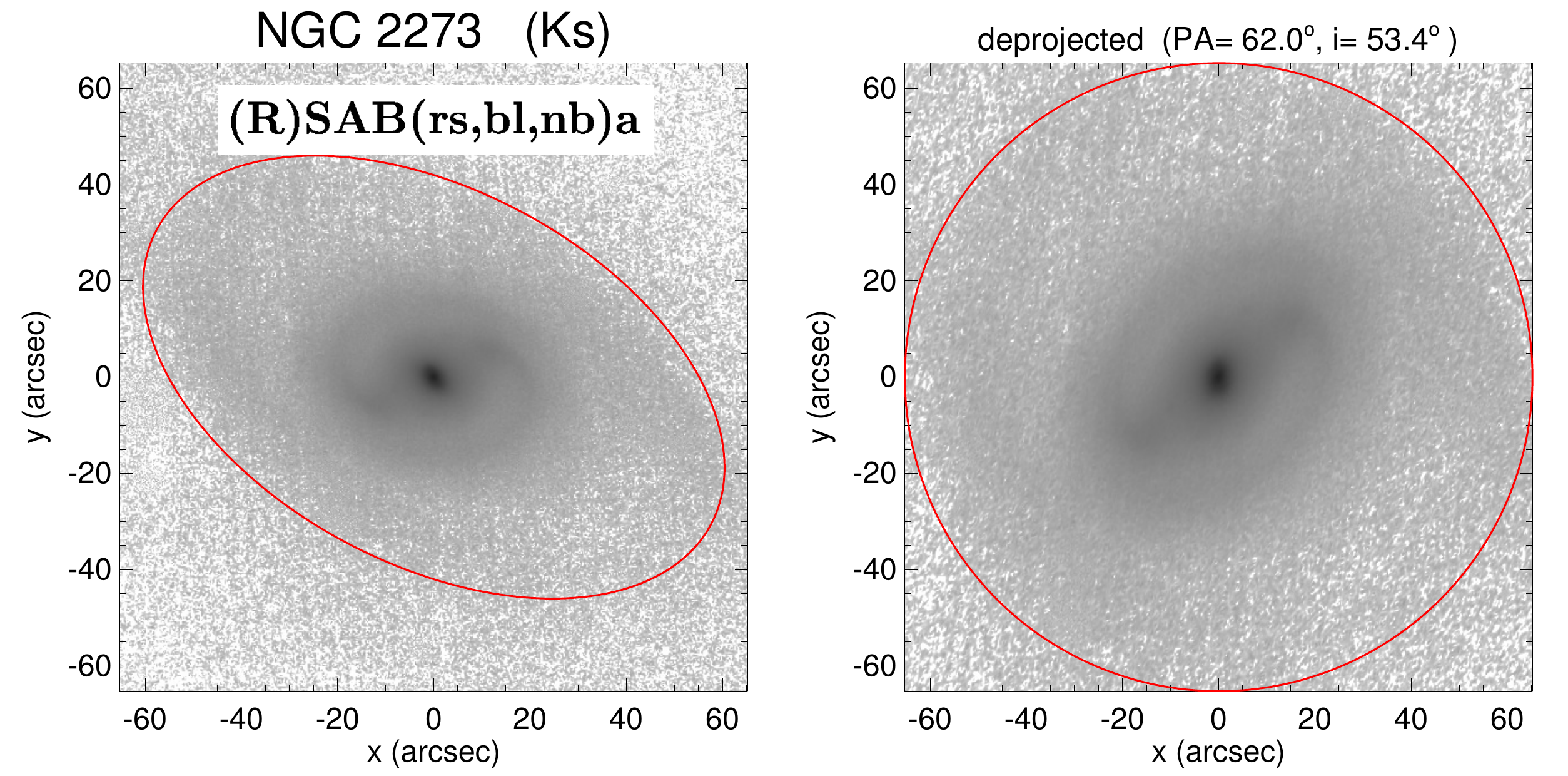}}
\subfloat[Parent galaxy group 7]{
\includegraphics[width=\hsize]{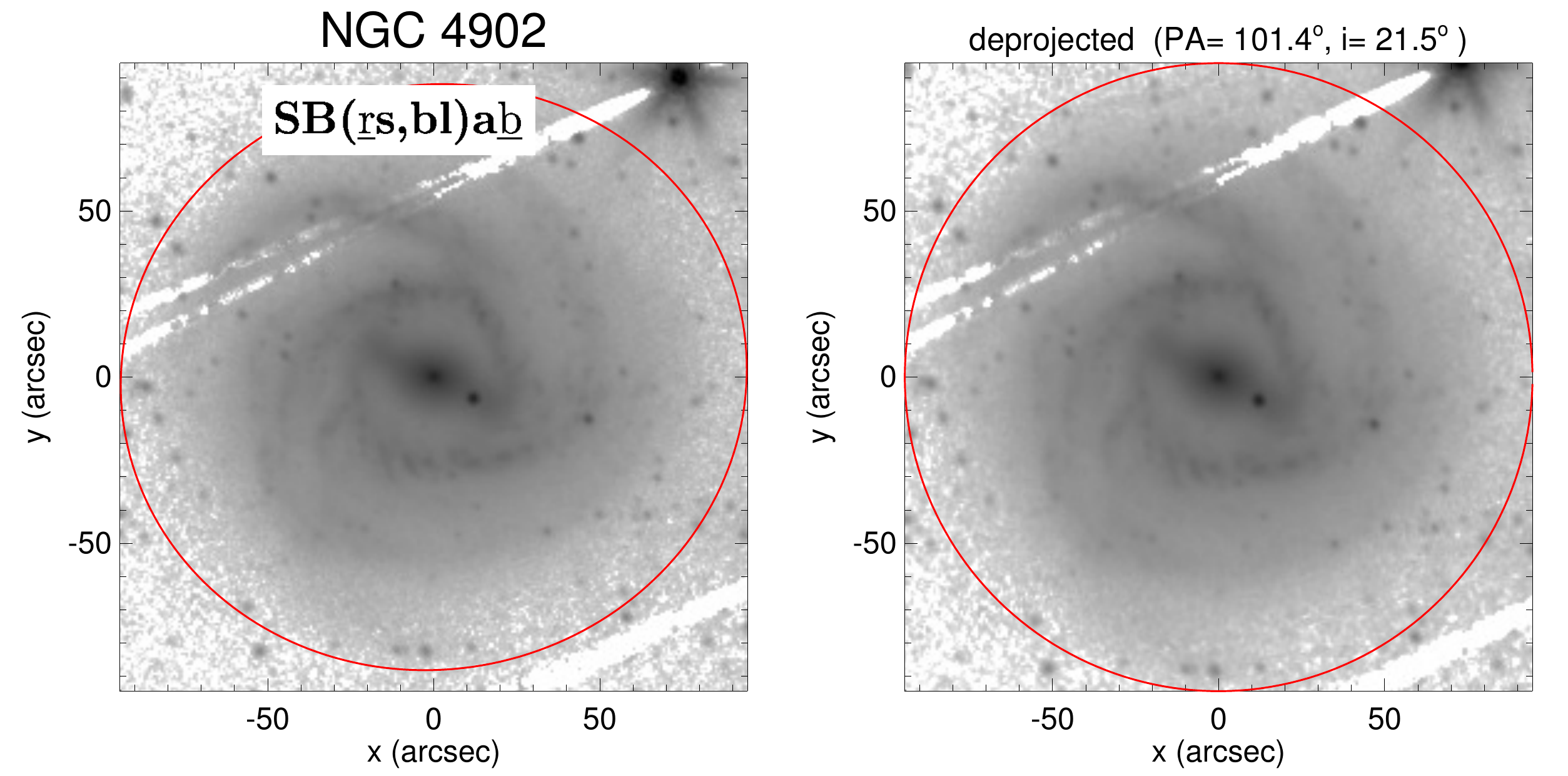}}

\caption{Parent galaxy groups as defined in Section 6.2. The galaxies are shown 
both in the sky {\it (left panels)} and when deprojected to disk plane {\it (right panels)}. 
The red ellipse indicates the orientation of the outer disk; 
this corresponds to the circle in the deprojected image. The images are 3.6 $\mu$m images,
unless indicated by K$_s$ after the galaxy name.}
         \label{fig:obs_parents}
\end{figure}


\clearpage
\newpage
\begin{figure*}
\includegraphics[angle=0,width=17.0cm]{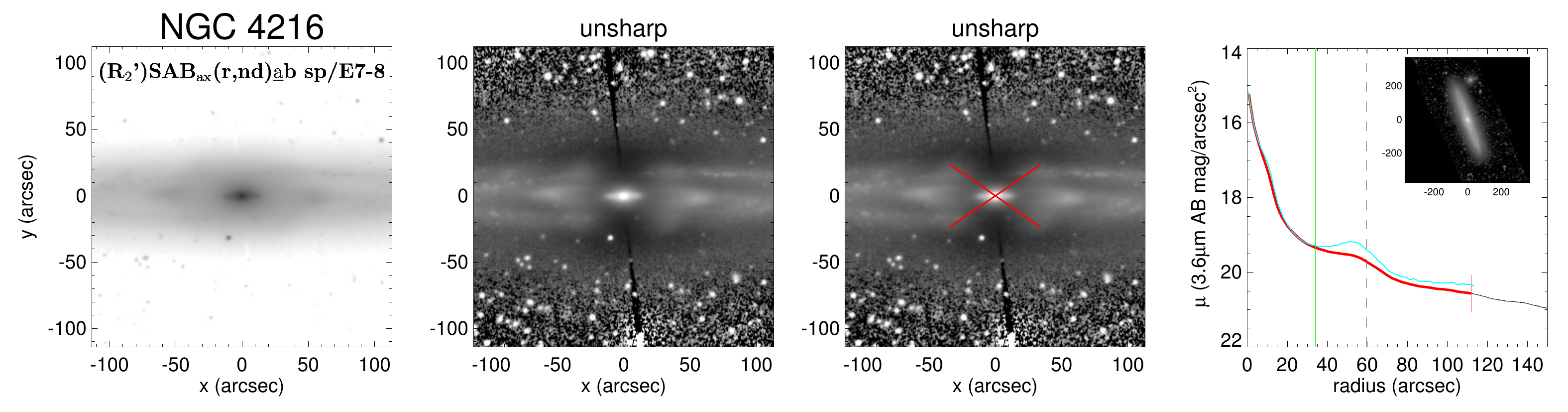}
\includegraphics[angle=0,width=17.0cm]{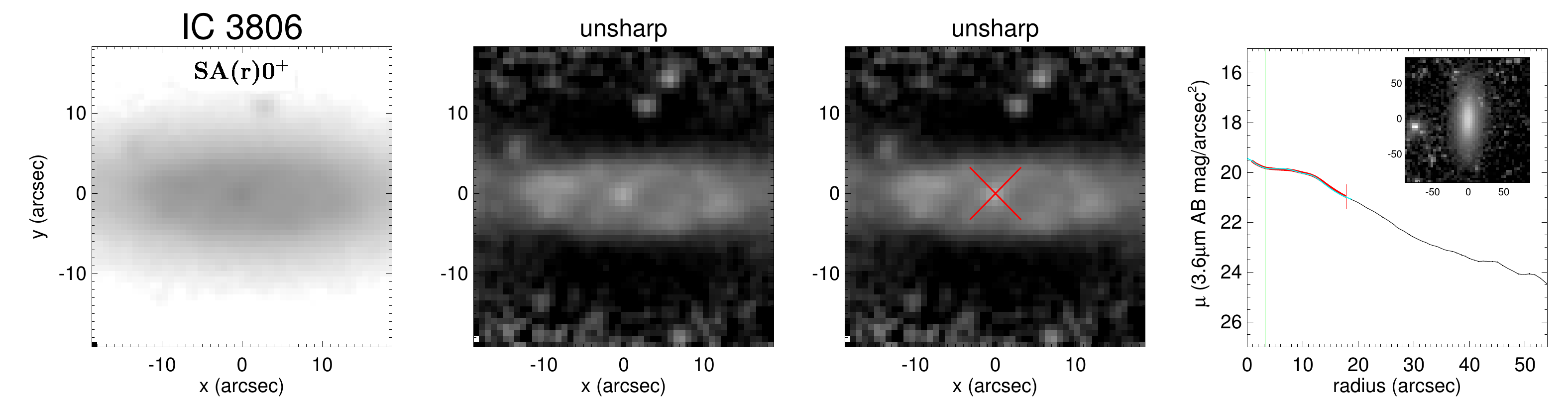}
\includegraphics[angle=0,width=17.0cm]{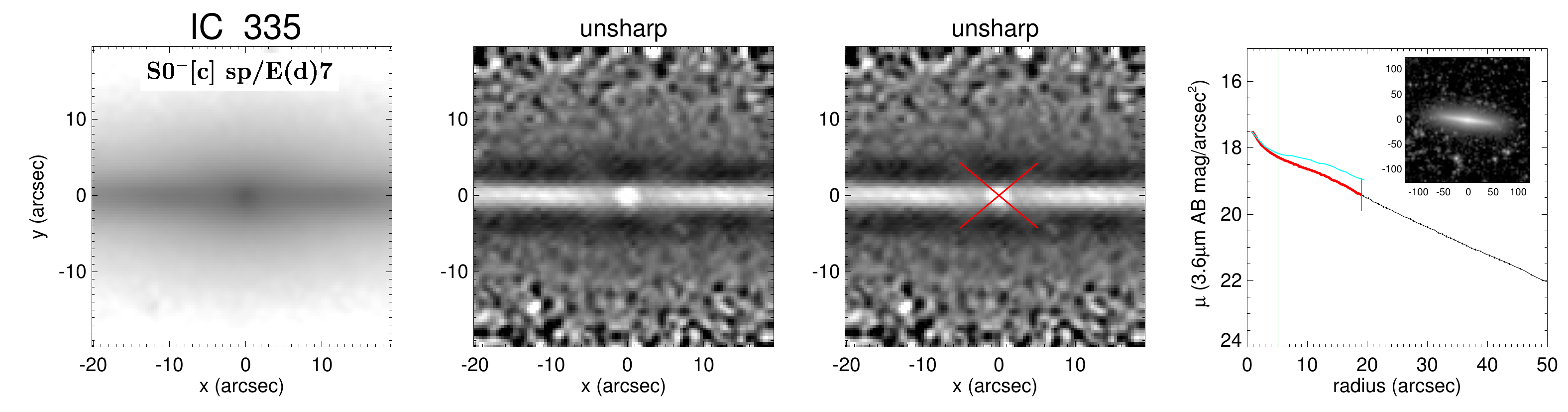}
\includegraphics[angle=0,width=17.0cm]{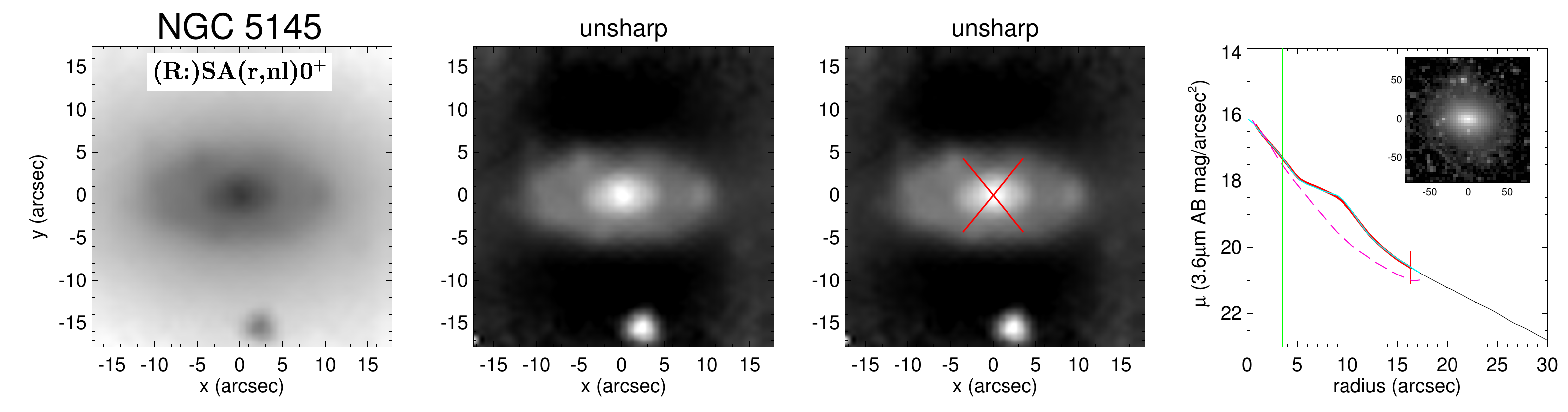}
\caption{Four X-shaped galaxies are shown: NGC 4216 has a prominent inner disk. 
The three other galaxies have the weakest X-shaped features in our sample.  They are shown in
the same format as Figure 11, except that the deprojected images are not
shown. The red cross in the right middle panel shows our measurement of the X-feature.}

        \label{fig:obs_weakests}
\end{figure*}


\clearpage
\newpage
\begin{figure*}
\includegraphics[width=15.35cm]{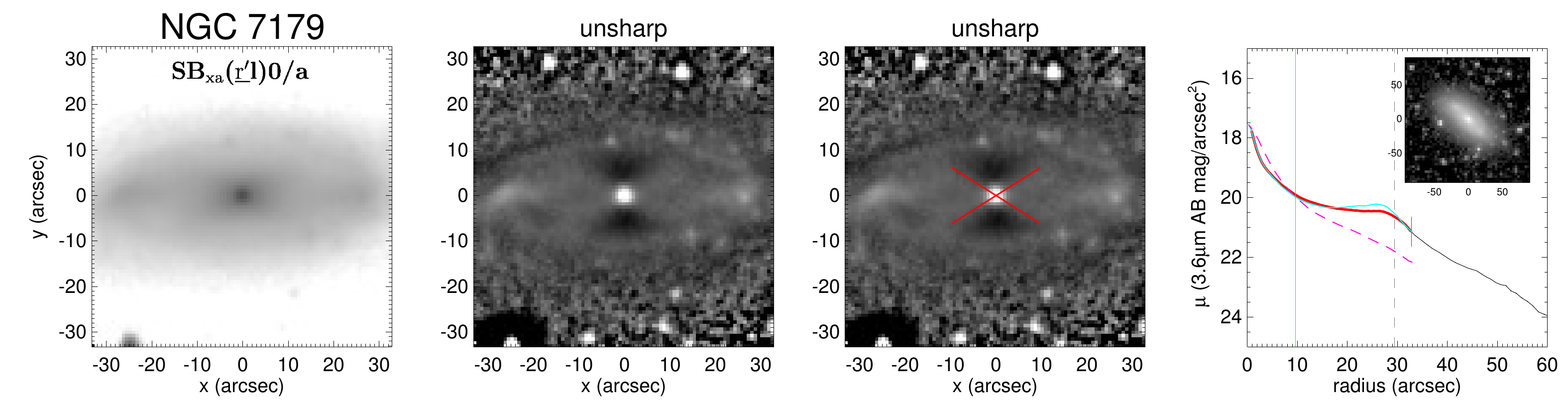}
\vspace{-0.3 cm}
\includegraphics[width=15.35cm]{ngc5101K_bl_B.pdf}
\vspace{-0.3 cm}
\includegraphics[width=15.35cm]{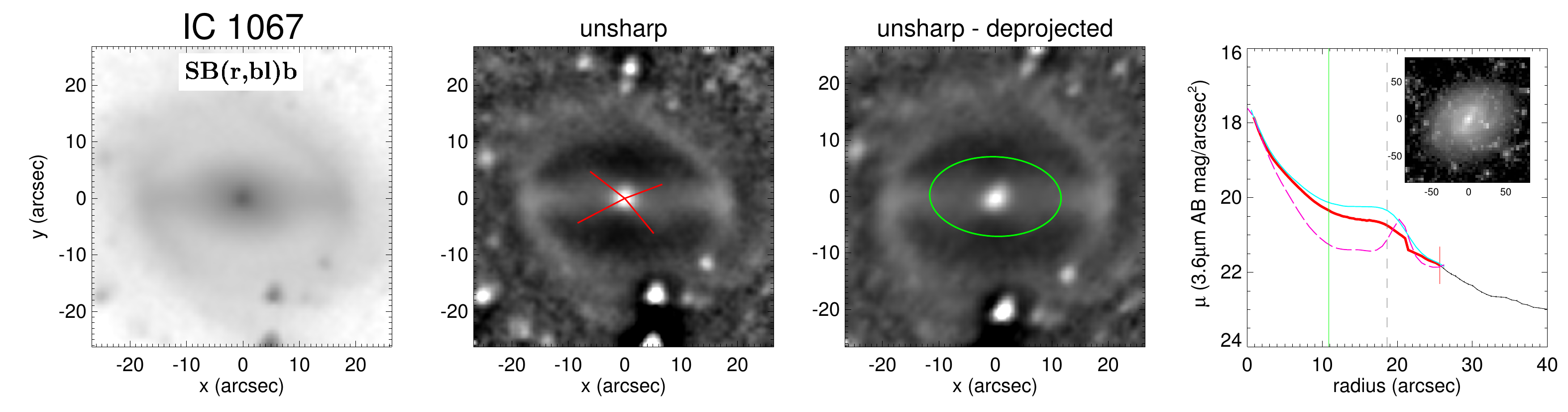}
\vspace{-0.3 cm}
\includegraphics[width=15.35cm]{NGC4643_bl_B.pdf}
\vspace{-0.3 cm}
\includegraphics[width=15.35cm]{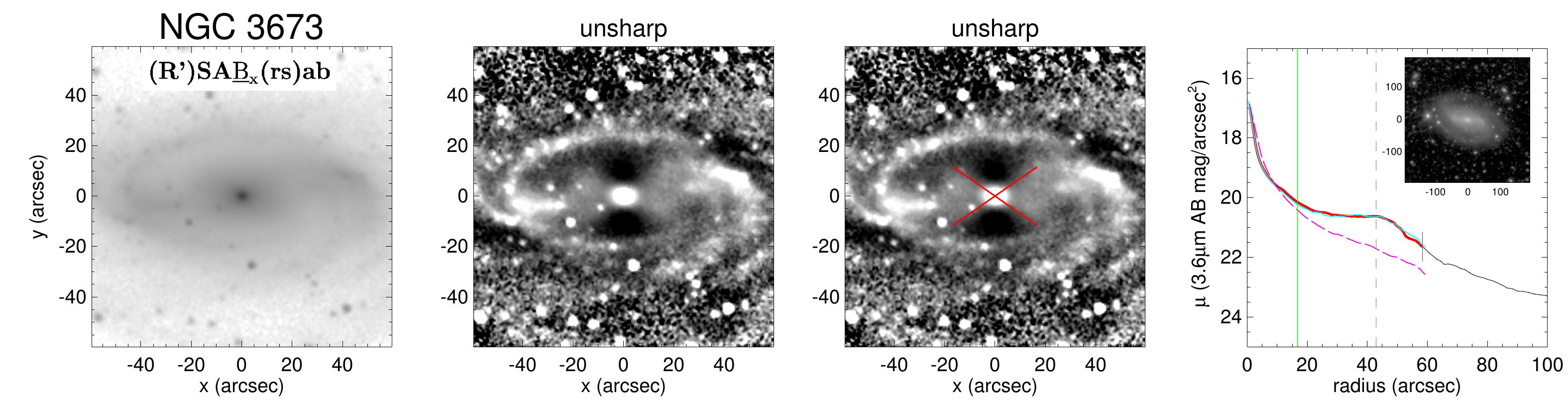}
\vspace{-0.3 cm}
\includegraphics[width=15.35cm]{ngc2273K_bl_AB.pdf}

\caption{ Three galaxy pairs are shown, in which the parent galaxies of a barlens 
and an X-shape feature have similar morphologies: NGC 7179 (X) /NGC 5101 (bl), IC
  1067 (X) / NGC 4643 (bl), and NGC 3673 (X) / NGC 2273 (bl). Because of
  large galaxy inclination, the deprojected images for NGC 7179 and
  NGC 3673 are not shown. IC 1067 has both a barlens and an X-shape
  feature. The format is the same as in Fig. 11.}
\vspace{-23 cm} \hspace{15.5 cm} {\Large X (i=59$^{\circ}$)}

\vspace{3.5 cm} \hspace{15.5 cm} {\Large bl (i=22$^{\circ}$)}

\vspace{3.5 cm} \hspace{15.5 cm} {\Large X (i=38$^{\circ}$)}

\vspace{3.5 cm} \hspace{15.5 cm} {\Large bl (i=11$^{\circ}$)}

\vspace{3.5 cm} \hspace{15.5 cm} {\Large X (i=52$^{\circ}$)}

\vspace{3.5 cm} \hspace{15.5 cm} {\Large bl (i=47$^{\circ}$)}

        \label{fig:obs_similars}

\end{figure*}

\clearpage
\begin{figure*}
\includegraphics[width=17.0cm]{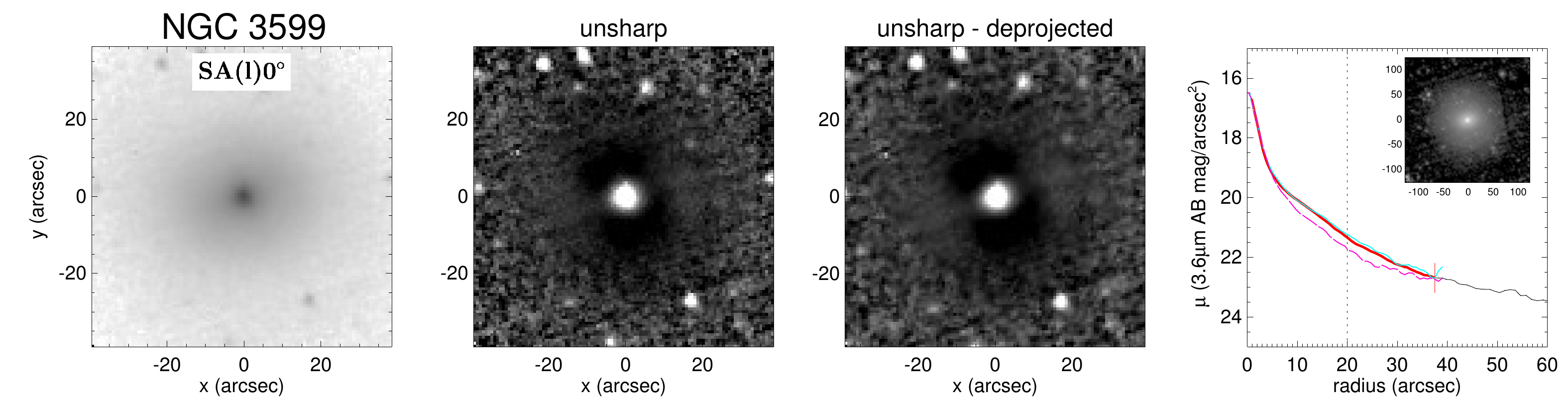}
\includegraphics[width=17.0cm]{NGC4643_bl_B.pdf}
\caption{Our selection criterion for the unbarred galaxies is
demonstrated: original images, unsharp mask images, and the surface
brightness profiles for a unbarred (NGC 3599) and a barlens (NGC
4643) galaxy are compared. Notice the similar central flux
concentrations, and the subsequent, nearly exponential sub-sections in
their surface brightness profiles. The dotted vertical line in the
profile of NGC 3599 marks the extent of the ``barlens-like''
structure. The dashed and full vertical lines in the profile of NGC 4643
show the sizes of the barlens and the bar, respectively. The same format
is adopted as in Fig. 11.}
        \label{fig:obs_nonbars_selection}
\end{figure*}
\clearpage

\begin{figure*}
\includegraphics[angle=0,width=17.0cm]{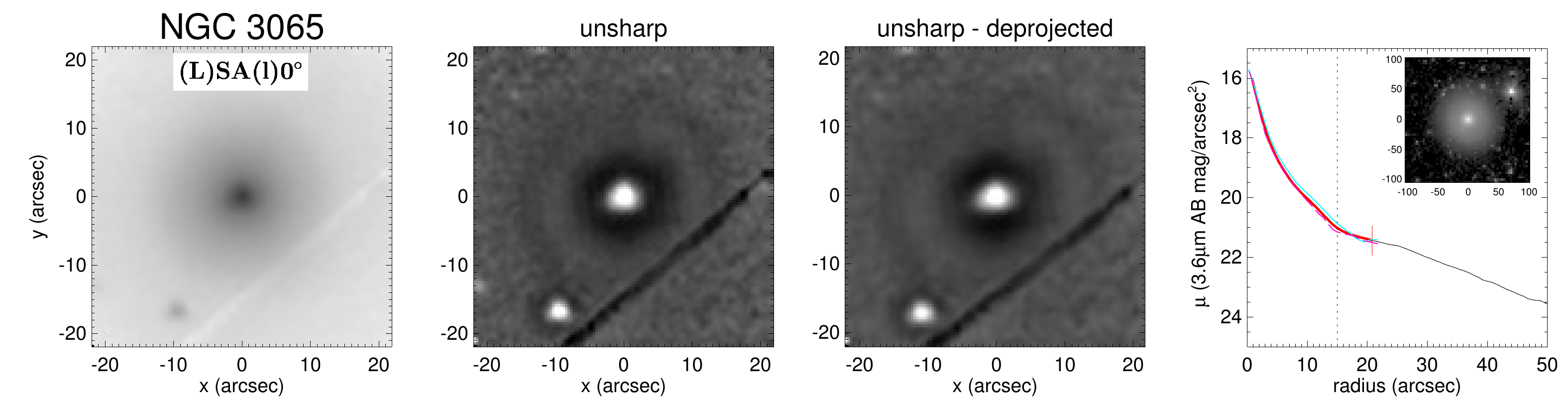}
\includegraphics[angle=0,width=17.0cm]{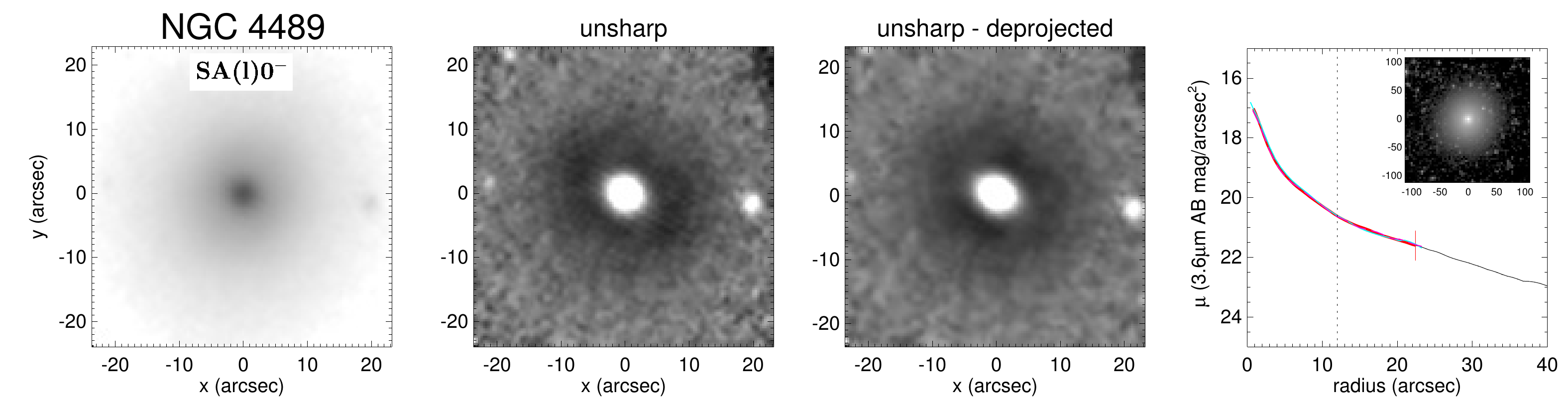}
\includegraphics[angle=0,width=17.0cm]{NGC3599_A_l.pdf}
\includegraphics[angle=0,width=17.0cm]{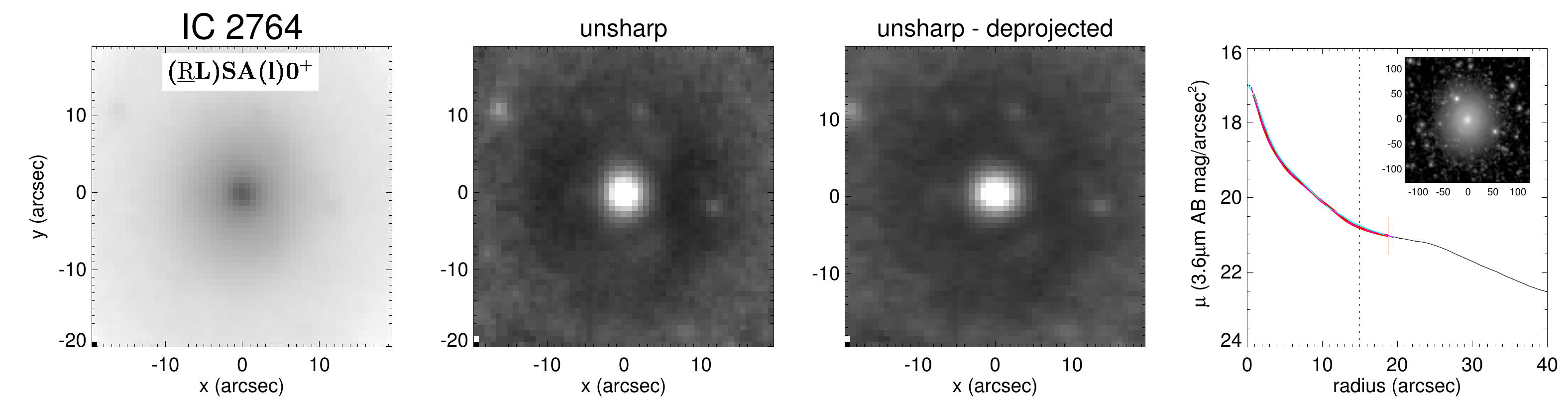}
\includegraphics[angle=0,width=17.0cm]{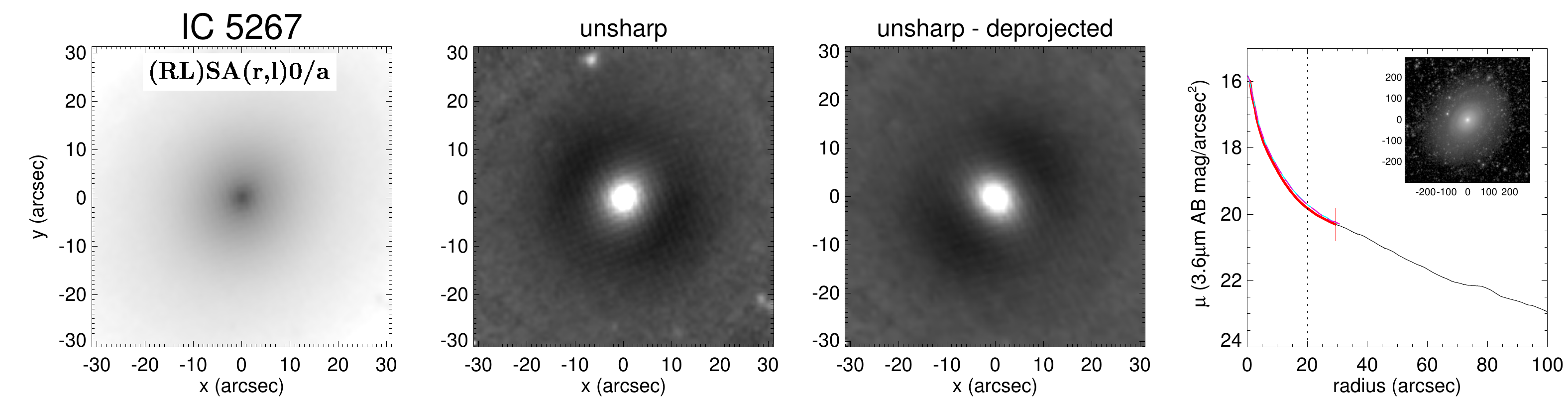}
\caption{A selected sub-sample of the unbarred galaxies,
representative of the surface brightness profiles and innermost
morphologies in the unsharp mask images. 
The same format is adopted as in Fig. 11. Meaning of the dotted
vertical line is the same as in Figure 16.}
\label{fig:obs_nonbars}
\end{figure*}

\newpage
\clearpage

  \begin{figure}
   \centering
   \includegraphics[width=\hsize]{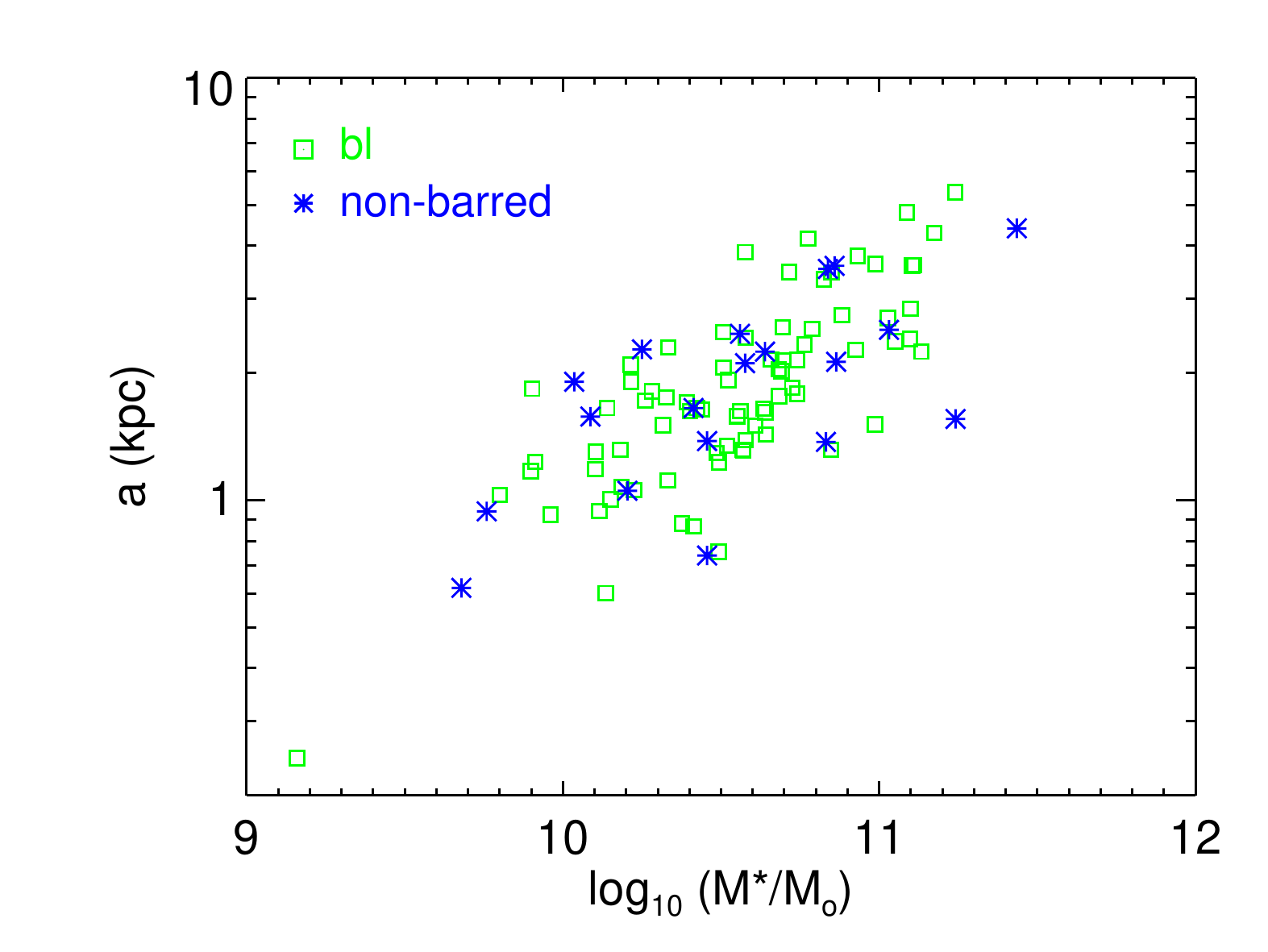}
      \caption{
Sizes of barlenses in barred galaxies (green boxes),
        and in barlens-like structure components in unbarred
        galaxies (blue stars), are plotted as a function of galaxy
        stellar mass (M$^*$). The semimajor-axis lengths are given in kpc. 
              }
        \label{fig:obs_sizes_nonbars}
   \end{figure}


  \begin{figure*}
   \centering
   \includegraphics[width=17.0cm]{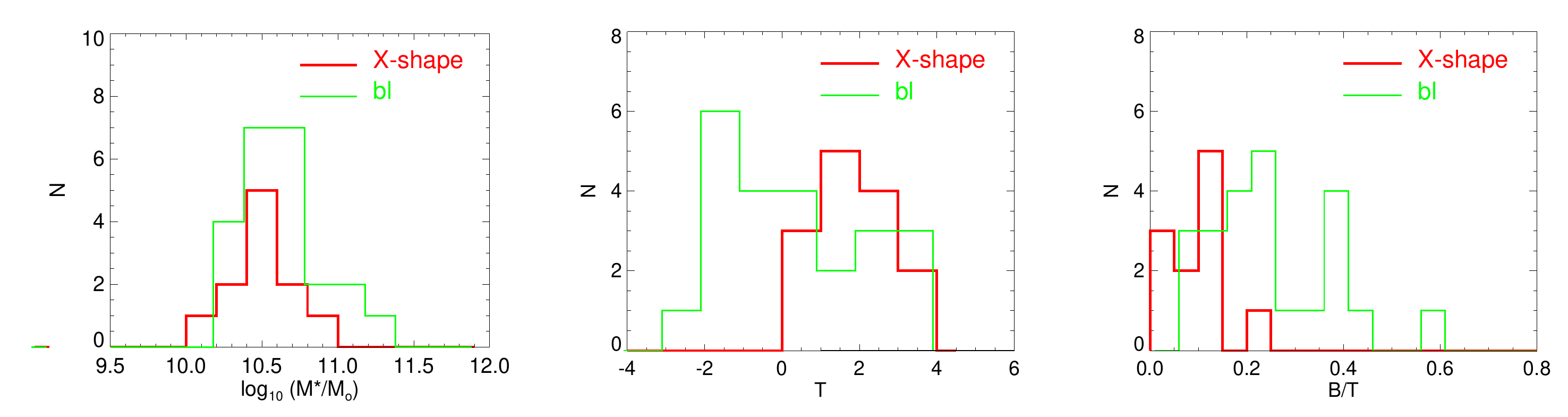} 
      \caption{Number histograms of the parent galaxy stellar masses
        (M$^*$), Hubble stages (T), and bulge-to-total flux
        ratios (B/T, taken from S$^4$G Pipeline 4, Salo et al. (2015), are shown
        separately for barlenses and X-shape features in our
        samples. In this plot shown are only galaxies at
        $i$ = 45$^{\circ}$ -- 60$^{\circ}$.  }
        \label{fig:obs_histo}
   \end{figure*}

\clearpage

\begin{figure*}
\includegraphics[angle=0,width=19.0cm]{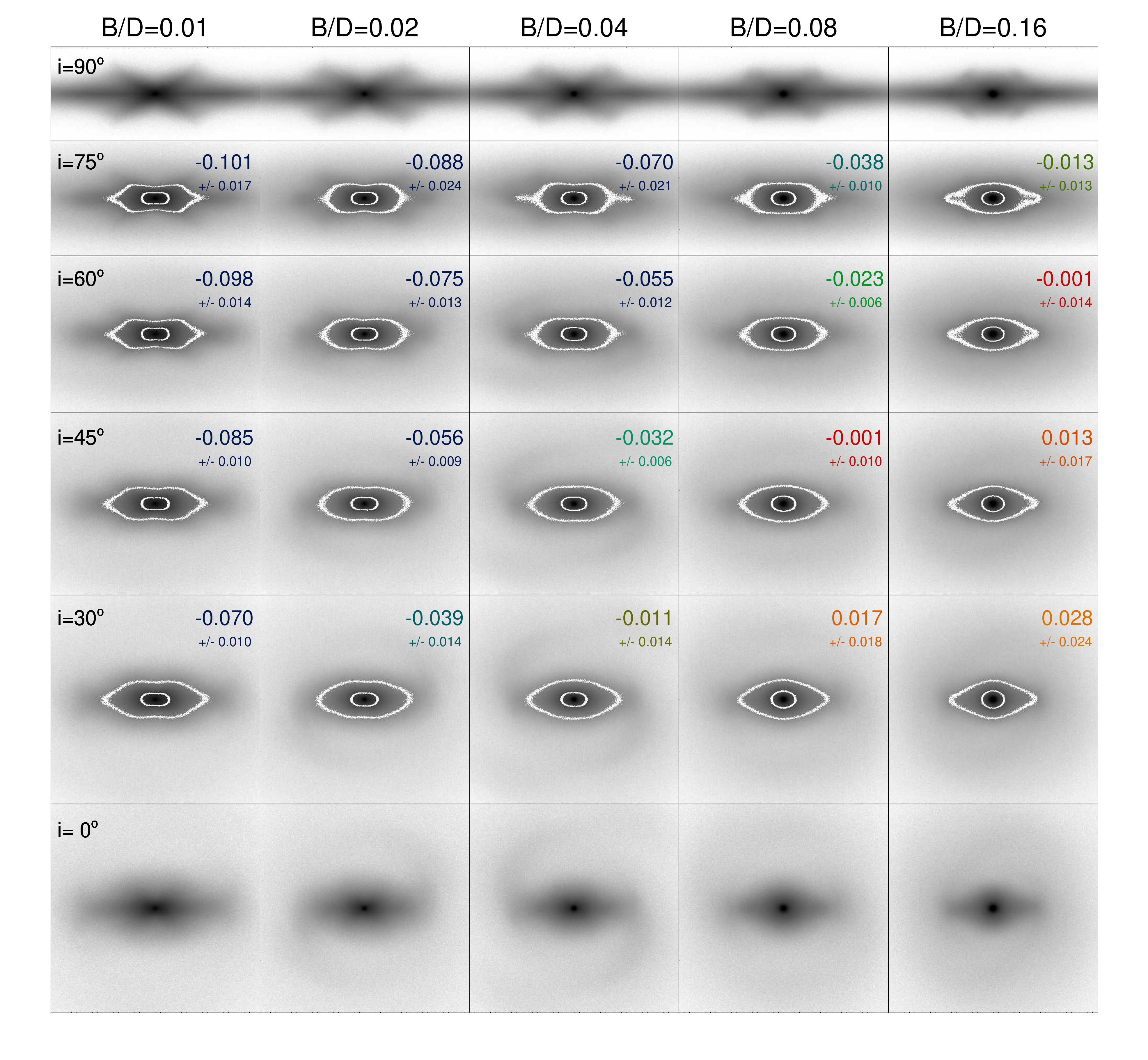}
\caption{Five simulation models from Salo $\&$ Laurikainen (2016) are
  shown at five different galaxy inclinations. The models differ in
  the bulge-to-disk mass ratio which varies from B/D = 0.01 to B/D =
  0.16.  The labels in the frames for $i$ =30$^{\circ}$ -- 75$^{\circ}$ indicate the mean and
  standard deviation of $B_4$ for the B/P/X/bl feature, measured from
  the region between the two marked isophotes. }
\label{fig:simu_all_bd}
\end{figure*}
\clearpage
\begin{figure*}
\includegraphics[angle=0,width=17.0cm]{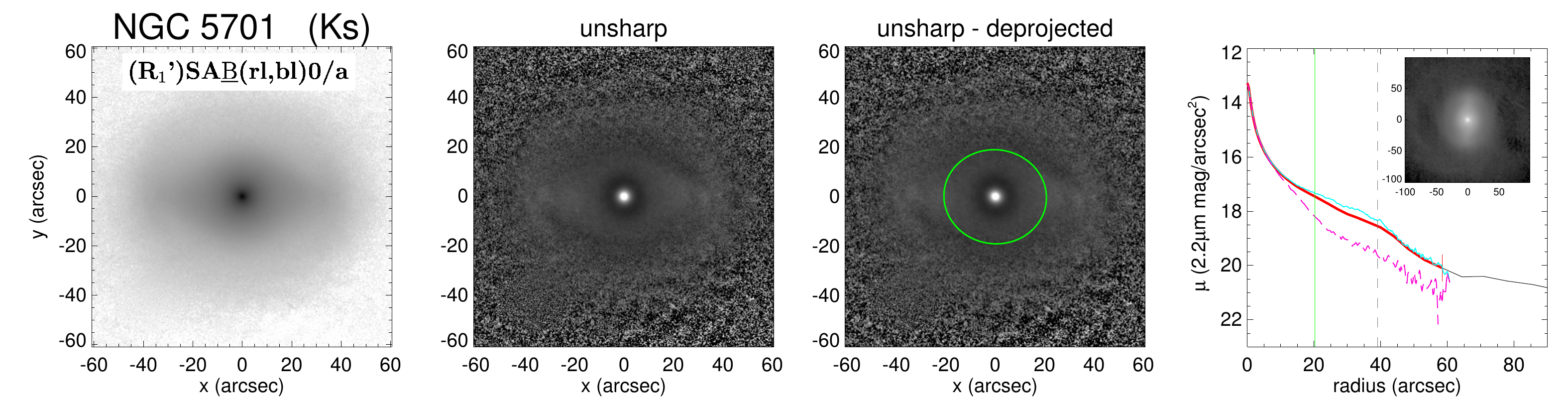}
\includegraphics[angle=0,width=17.0cm]{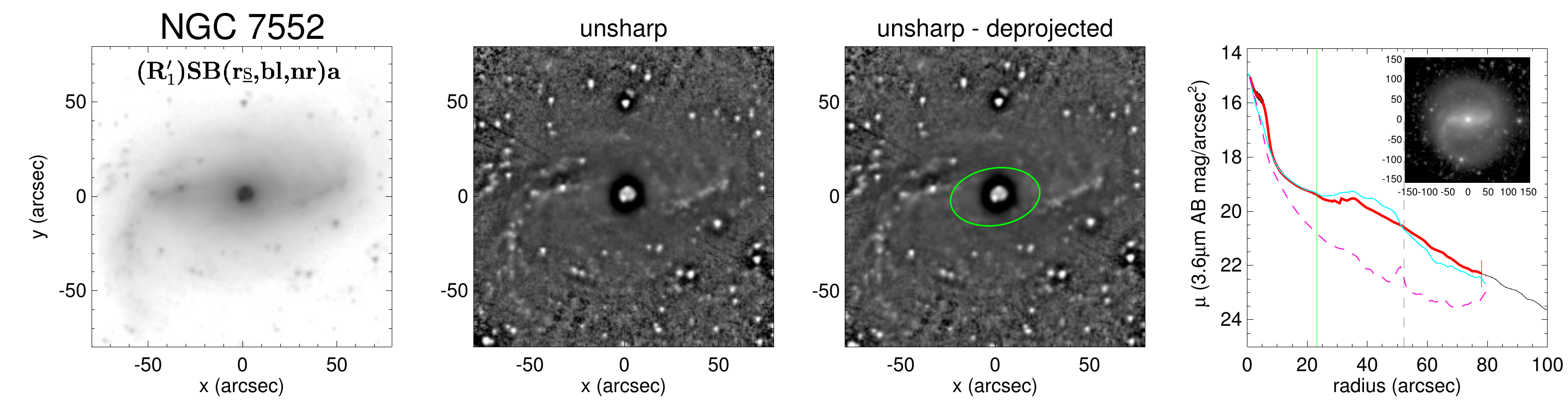}
\caption{Two galaxies in our sample, NGC 5701 and NGC 7552, which
have detailed stellar population analysis in the literature. The
two galaxies are in almost face-on view for which reason
the deprojected image is almost the same as the non-deprojected image.
The format is the same as in Figure 11.}
\label{fig:obs_popu}
\end{figure*}

\clearpage


\newpage
\longtab{
\begin{longtable}{llcc}
\caption{\label{Table:1} Barlens galaxy sample. Indicated are the
  galaxy classifications from Buta et al. (2015), and whether the
  barlens has evidence of boxy isophotes in our isophotal analysis
  (see Section 3.2): shown are our visual evaluation, and the mean and
  standard deviation of B$_4$ within (0.3-1.0) $\cdot$ a$_{\rm
    bl}$. }\\ \hline\hline \noalign{\smallskip} Galaxy & Hubble type &
visual & $<$B$_4>$ \\ & (Buta et al. 2015) & & \\ \noalign{\smallskip}
\hline \endfirsthead
\caption{continued.}\\
\hline\hline
\noalign{\smallskip}
Galaxy & Hubble type         & visual &  $<$B$_4>$ \\
       & (Buta et al. 2015)  &      & \\
\hline
\endhead
\hline
\endfoot
            & & & \\           
Strong bars:& & & \\
\noalign{\smallskip}
NGC 613  & SB($\underline{\rm r}$s,bl,nr)b                 & no   &0.0037$\pm$0.0162 \\
NGC 936  &  (L)SB$_{\rm a}$($\underline{\rm r}$s,bl)0$^+$   & no   &  0.0054$\pm$0.0098 \\
NGC 1015 &  (R$^{\prime}$)SB(r,bl)0/a                      & no   &  0.0241$\pm$0.0167 \\
NGC 1097 &  (R$^{\prime}$)SB(rs,bl,nr)ab pec                & no   & 0.0180$\pm$0.0130  \\
NGC 1300  & (R$^{\prime}$)SB(s,bl,nrl)b                     &no    & 0.0076$\pm$0.0161 \\
NGC 1398 &  (R$^{\prime}$R)SB($\underline{\rm r}$s,bl)a     &no    & 0.0014$\pm$0.0167 \\
NGC 1433 &  (R$_1$')SB(r,p,nrl,nb)a                        & no   & - \\
NGC 1440 &  (L)SB(rs,bl)0$^{\circ}$                         & no   & 0.0032$\pm$0.0070 \\
NGC 1452  & (RL)SB($\underline{\rm r}$s,bl)0/a            & no   & 0.0138$\pm$0.0204 \\
NGC 1512 &  (R$\underline{\rm L}$)SB(r,bl,nr)a            & no   & 0.0116$\pm$0.0120 \\
NGC 1533 &  (RL)SB(bl)0$^{\circ}$                           & no   & 0.0106$\pm$0.0085 \\
NGC 1640  & (R$^{\prime}$)SB$_{\rm a}$(r,bl)ab               & no    & 0.0032$\pm$0.0123 \\
NGC 2787 &  (L)SB$_{\rm a}$(r,bl)0$^{\circ}$                 &  no   & -0.0030$\pm$0.0106 \\
NGC 2968  & (L)SB(s,bl)0$^+$                              &no     & 0.0042$\pm$0.0094 \\
NGC 2983  & (L)SB$_{\rm a}$(s,bl)0$^+$                      & no   & -0.0009$\pm$0.0043 \\
NGC 3266 &  (R$\underline{\rm L}$)SB(bl)0$^{\circ}$        & no     & 0.0117$\pm$0.0147 \\
NGC 3351 &  (R$^{\prime}$)SB(r,bl,nr)a                     & marginal& -0.0005$\pm$0.0139 \\
NGC 3384 &  (L)SA$\underline{\rm B}$(bl)0$^-$             & marginal & -0.0013$\pm$0.0080 \\
NGC 3489 &  (R)SA$\underline{\rm B}$(r,bl)0$^{\circ}$:     &yes       &-0.0099 $\pm$0.0081  \\
NGC 3637 &  (RL)SB$_{\rm a}$(rl,bl)0$^+$                   &no        & 0.0118 $\pm$0.0155 \\
NGC 3941 &  (R)SB$_{\rm a}$(bl)0$^{\circ}$                   & no      & 0.0073 $\pm$0.0091 \\   
NGC 3945  & (R)SB$_{\rm a}$(rl,nl,bl)0$^+$                 &no        & 0.0179$\pm$0.0448 \\
NGC 3953  & SB(r,bl)b                                    & yes      & -0.0176$\pm$0.0063 \\   
NGC 3992  & SB(rs,bl,nb)ab                               & yes      & -0.0094$\pm$0.0086 \\
NGC 4245 &  (RL)SB(r,bl,n$\underline{\rm r}$l)0$^+$      & no       & 0.0139$\pm$0.0115 \\
NGC 4262  & (L)SB$_{\rm a}$(l,bl)0$^-$                     & no       & 0.0064$\pm$0.0116 \\
NGC 4314  & (R$_1$')SB(r$\underline{\rm l}$,bl,nr)a      & no       & 0.0173 $\pm$0.0123 \\
NGC 4340  &  SB$_{\rm a}$(r,nr,nb,bl)0$^{\circ}$            & marginal &-0.0124$\pm$0.0234 \\  
NGC 4371 &  (L)SB$_{\rm a}$(r,bl,nr)0$^+$                  & marginal & -0.0022$\pm$0.0191 \\
NGC 4394  &  ($\underline{\rm R}$L)SB(rs,bl,nl)0/a        &no       & 0.0177 $\pm$0.0092 \\
NGC 4448  & (R)SB(r,bl)0/$\underline{\rm a}$              &-        &  -0.0874$\pm$0.0134 \\
NGC 4548 &  SB(rs,bl)$\underline{\rm a}$b                 & no      & 0.0070$\pm$0.0112 \\
NGC 4579 &  ($\underline{\rm R}$LR')SB(rs,bl)a            & no      & 0.0117 $\pm$0.0126 \\ 
NGC 4593 &  (R$^{\prime}$)SB(rs,bl,AGN)a                    & no      & 0.0105$\pm$0.0093 \\
NGC 4596  &  (L)SB(rs,bl)$\underline{\rm 0}$/a            &no       & 0.0071$\pm$0.0080 \\
NGC 4608 &   SB(r,bl)0$^+$                                & no      & 0.0288$\pm$0.0252\\
NGC 4639  & (R$^{\prime}$)SA$\underline{\rm B}$(rs,bl)ab   & no       &0.0029 $\pm$0.0067 \\
NGC 4643  & (L)SB($\underline{\rm r}$s,bl,nl)0$^+$       & no        & 0.0194$\pm$0.0131 \\
NGC 4659 &  (RL)SAB(l,bl)0$^{\circ}$                      &no         & -0.0012$\pm$0.0050 \\
NGC 4754  & (L)SB$_{\rm a}$(bl)0$^{\circ}$                  & marginal &-0.0012 $\pm$0.0089 \\
NGC 4795  & (R$^{\prime}$)SA$\underline{\rm B}_{\rm a}$(l,bl)a pec &no & 0.0170$\pm$0.0227 \\
NGC 5026  &  (L)SB(rs,nl,bl)a                                & yes  & -0.0189$\pm$0.0137 \\
NGC 5101  &(R$_1$R$_2$')SB($\underline{\rm r}$s,bl)0/a        & no   & 0.0142$\pm$0.0150 \\
NGC 5337 &  SB(rs,bl)0/a                                      &no   & 0.0141$\pm$0.0144\\
NGC 5339  &  SA$\underline{\rm B}$(rs,bl)ab                   &yes  & -0.1745$\pm$0.0261 \\ 
NGC 5347  & SB(rs,bl)a                                        & no  & -0.0037$\pm$0.0137 \\
NGC 5375 &  (R$^{\prime}$)SB$_{\rm a}$(rs,bl)$\underline{\rm a}$b & no & 0.0103$\pm$0.0080 \\ 
NGC 5701  & (R$_1$')SA$\underline{\rm B}$(rl,bl)0/a            & no & 0.0087$\pm$0.011 \\
NGC 5728  & (R$_1$)SB($\underline{\rm r}^{\prime}$l,bl,nr,nb)0/a &no & 0.0109$\pm$0.008 \\
NGC 5850 &  (R$^{\prime}$)SB(r,bl,nr,nb)$\underline{\rm a}$b     &no & 0.0330$\pm$0.0190 \\
NGC 5957 &  (R$^{\prime}$)SA$\underline{\rm B}$(rs,bl)$\underline{\rm a}$b  &no & 0.0082$\pm$0.0150 \\
NGC 6654  & (R$^{\prime}$)SB$_{\rm a}$(s,bl)a                      & no & - \\
NGC 7079  & (L)SA$\underline{\rm B}_{\rm a}$(s,bl)0$^{\circ}$:     & yes & 9.27 -0.0074$\pm$0.0037\\   
NGC 7552  & (R$_1^{\prime}$)SB(r$\underline{\rm s}$,bl,nr)a        & no & 23.13 -0.0028$\pm$0.0185 \\
IC 2051  &  SB($\underline{\rm r}$s,bl)b                         & yes & -0.0143$\pm$0.0153 \\
 & &  & 
\\
Weak bars: & & & \\
\noalign{\smallskip}

NGC 1022 & (RL)SAB($\underline{\rm r}$s,bl,ns)$\underline{\rm 0}$/a  & yes & -0.0196$\pm$0.0163  \\
NGC 1079  & (R$\underline{\rm L}$)S($\underline{\rm A}$B$_{\rm a}$($\underline{\rm r}$s,bl)0$^{\circ}$   & yes & -0.0064$\pm$0.0069 \\
NGC 1201 &  SAB$_{\rm a}$(r$^{\prime}$l,bl,nb)0$^{\circ}$              & no & 0.0047$\pm$0.0060 \\
NGC 1291 &  (R)SAB(l,bl,nb)0$^+$                                  & no & 0.0190$\pm$0.0164 \\
NGC 1302 &  (RLRL)SAB($\underline{\rm r}$l,bl)0$^+$               & no & 0.0101$\pm$0.0113 \\
NGC 1326  & (R$_1$)SAB$_{\rm a}$(r,bl,nr)0$^+$                      & no & 0.0002$\pm$0.0080 \\ 
NGC 1350 &  (R)SAB$_{\rm a}$(r,bl)0/a                              & yes & -0.0099$\pm$0.0069 \\
NGC 2273 &   (R)SAB(rs,bl,nb)a                                   &no   &  0.0057$\pm$0.0225\\
NGC 2293 &  SAB$_{\rm a}$(bl)0/a                                   &no   & -0.0065$\pm$0.004 \\
NGC 2543 &  SAB(s,bl)b                                           & yes  &  -0.0075$\pm$0.0145\\
NGC 2859  & (R)SAB$_{\rm a}$(rl,bl,nl,nb)0$^+$                     & no   & 0.0022$\pm$0.0077 \\
NGC 3368  & ($\underline{\rm R}$L)SAB($\underline{\rm r}$s,bl,nl)0$^{\circ}$   &yes & -0.0038$\pm$0.010 \\
NGC 3380  & (R$\underline{\rm L}$)SAB($\underline{\rm r}$s,bl)0/a &no    & 0.0018$\pm$0.0089 \\
NGC 3892 &  (RL)SAB(rl,bl)0$^+$                                   & yes  & -0.0031$\pm$0.0071  \\
NGC 4143 &  (LR$^{\prime}$L)SAB$_{\rm a}$(s,nb,bl)0$^{\circ}$          & yes  & -0.0027$\pm$0.0024 \\
NGC 4454 &  (RL)SAB(r,bl)0/a                                      & no   & 0.0082$\pm$0.0064 \\
NGC 4503 &  SAB$_{\rm a}$(s,bl)0$^{\circ}$                           &marginal & - \\
NGC 4984  & (R$^{\prime}$R)SAB$_{\rm a}$(l,bl,nl)0/a                 & no   & 0.0030 $\pm$0.0124 \\
NGC 5134  & (R)SAB(rs,bl)a                                         & no   & 0.0022$\pm$0.0108 \\
NGC 5750  & (RL)SAB(r$^{\prime}$l$\underline{\rm r}$,s,bl)0/a       &yes   & -0.0098$\pm$0.0090 \\
NGC 5770  & SAB($\underline{\rm r}$l,bl)0$^+$                         & no   & 0.0083$\pm$0.0132 \\  
NGC 5838  & (L)SAB(nl,bl)0$^{\circ}$                               & yes   & - \\
NGC 6014  & SAB(rs,bl)$\underline{\rm 0}$/a                      &no     & 0.0072$\pm$0.0069 \\
NGC 6684  & (R$^{\prime}$L)SAB(rs,nb,bl)0/a                        & no    & - \\
NGC 6782  & (R)SAB(rl,nr$^{\prime}$,nb,bl)0$^+$                    & no    & 0.0040$\pm$0.0115 \\
 & &  \\
 Barlens+X: & & \\
\noalign{\smallskip}
NGC5957 & (R$^{\prime}$)SB($\underline{\rm r}$s,bl)a              & no   &  0.0082$\pm$0.0150 \\
NGC3380 & (RL)SA$\underline{\rm B}$(rs,bl)0/a                   & no   & 0.0018$\pm$0.0089 \\
NGC3185 & (RL)SAB$_{\rm ax}$ (rs,bl)a                             & yes  & -0.0045$\pm$0.0036 \\ 
NGC4902 & SB($\underline{\rm r}$s,bl)a$\underline{\rm b}$       & no   & 0.0045$\pm$0.0103 \\
NGC7421 & (R$^{\prime}$)SB(rs,bl)ab                               & no   & 0.0041$\pm$0.0064 \\
IC1067 & SB(r,bl)b                                              &  no  & 0.0062$\pm$0.0096 \\
\noalign{\smallskip}
\end{longtable}
}

\clearpage
\newpage

 
\longtab{
\begin{longtable}{llllr}
\caption{\label{Table:2} Sample of galaxies with X-shape
  features. Shown are the measured semimajor (a$_{ \rm X}$) and
  semiminor (b$_{ \rm X}$) axis dimensions of the X-shape
  features. The values are mean values of three measurements and their
  errors are calculated from the standard deviation of the measurents,
  divided by $\sqrt{3}$. PA$_{\rm X}$ indicates the position angle of
  the major axis.}\\ \hline\hline \noalign{\smallskip} Galaxy & Hubble
type & a$_{\rm X}$ & b$_{\rm X}$ & PA$_{\rm X}$ \\ & (Buta et
al. 2015) & (arcsec) & (arcsec) & (degrees) \\ \noalign{\smallskip}
\hline \endfirsthead
\caption{continued.}\\
\hline\hline \noalign{\smallskip}
Galaxy & Hubble type         & a$_{\rm X}$ & b$_{\rm X}$  & PA$_{\rm X}$  \\
       & (Buta et al. 2015)  & (arcsec) & (arcsec)  & (degrees) \\
\noalign{\smallskip}
\hline
\endhead

\hline
\endfoot
\noalign{\smallskip}
  ESO 079-003  &  SAB$_{\rm x}$0/$_{\rm a}$: spw/E(d)5                 &   8.6$\pm$0.1   &  9.1$\pm$0.1    &      311\\  
  ESO 404-027  &  SAB(s)$\underline{\rm a}$b:                        &   7.8 $\pm$0.4 &  5.3$\pm$0.1    &    308\\  
  ESO 443-042  &  S$_{\rm x}$0/$\underline{\rm a}$ spw/E(d)8          &   15.7$\pm$0.0  &   12.0$\pm$0.1  &   308\\  
    IC 0335    &  S0$^-$[c] sp/E(d)7                                 &   5.1$\pm$0.1   & 4.3$\pm$0.1    &      263\\  
    IC 1067    &    SB(r,bl)b                                        &   7.6$\pm$0.1   & 5.5$\pm$0.1    &      327\\  
    IC 1711    &  SA$\underline{\rm B}_{\rm x}$0$^+$: sp              &   5.9$\pm$0.2    &  9.3$\pm$0.1   &      222 \\  
    IC 3806    &  SA(r)0$^+$                                         &    3.2$\pm$0.0  &  3.3$\pm$0.1   &    0 \\  
    IC 4237    &  SB(r)b                                             &    3.3$\pm$0.1  &  2.7$\pm$0.0   &   299\\  
    IC 5240    &  SB$_{\rm x}$(r)0/a                                  &   12.3$\pm$0.3   &  5.7$\pm$0.2   &  272 \\  
    NGC 532    &  SAB$_{\rm xa}$(r)0/a                                &   12.3$\pm$0.1   &  8.3$\pm$0.2   &  208 \\  
    NGC 615    & (R')SA$_{\rm x}$(r)$\underline{\rm a}$b              &    8.9$\pm$0.2   &  5.9$\pm$0.1   &  339 \\  
    NGC 660    &  SAB$_{\rm xa}$ab/PRG                                &   14.4$\pm$0.4   & 10.9$\pm$0.4   &  225 \\  
    NGC 779    & (L)SA$_{\rm x}$(rs)a                                 &    8.5$\pm$0.2   & 5.5$\pm$0.1    &   342\\  
    NGC 955    &  SA$\underline{\rm B}_{\rm x}$0$^+$                  &    7.9$\pm$0.2   & 6.9$\pm$0.2    &   198\\  
    NGC 1461   & S$\underline{\rm A}$B(r)0$^{\circ}$                  &   27.4$\pm$0.9   &17.3$\pm$0.2    &  337 \\  
    NGC 1476   & Im sp                                               &    5.7$\pm$0.3   &  4.1$\pm$0.2   &  266 \\  
    NGC 2549   & SB$_{\rm x}\underline{\rm 0}^{\circ}$ sp              &    4.9$\pm$0.1    & 2.9$\pm$0.1    & 359  \\  
    NGC 2654   & SB$_{\rm x}$(r,nd)0/a sp                             &   16.5$\pm$0.1   & 12.1$\pm$0.2   & 244  \\  
    NGC 2683   & (R'L)SB$_{xa}$(rs)0/a sp                             &  26.2$\pm$0.4    & 20.6$\pm$0.2   &   222\\  
    NGC 3098   & S0$^-$ sp/E8                                        &    5.2$\pm$0.0   &  4.4$\pm$0.1   &  269 \\  
    NGC 3185   & (RL)SAB$_{\rm ax}$(rs,bl)a                          &  12.6$\pm$0.2     &    8.6$\pm$0.1  & 300  \\  
    NGC 3190   & SAB$_{\rm x}$(l,nd)0/$\underline{\rm a}$ sp pec      &  11.3$\pm$0.51   & 10.4$\pm$0.2   &   294\\
    NGC 3227   & SAB$_{\rm x}$(s)a                                    &   20.9$\pm$0.4   &   9.9$\pm$0.2  &  331 \\
    NGC 3254   & SA$\underline{\rm B}_{xa}$b                          &    9.4$\pm$0.1  &   6.6$\pm$0.2  &  230 \\
    NGC 3301   & (R'L)SAB$_{\rm x}$(r)0$^+$ sp                        &   15.0$\pm$0.5   &  7.1$\pm$0.1   &  235 \\
    NGC 3380   & (RL)SA$\underline{\rm B}$(rs,bl)0/a                 &    6.3$\pm$0.2  &  5.5 $\pm$0.1  & 200  \\   
    NGC 3424   & S$_{\rm x}\underline{\rm a}$b: sp pec                &   10.6$\pm$0.3   &  7.3$\pm$0.2   &   291\\
    NGC 3623   & (R')SAB$_{\rm x}$(rs)a                               &   29.6$\pm$0.8   &  15.9$\pm$0.5  &  352 \\  
    NGC 3628   & SB$_{\rm x}$(nd)bc sp/E(b)8 pec                      &   53.0$\pm$2.0   &  33.3$\pm$0.6  &  284 \\
    NGC 3673   & (R')SA$\underline{\rm B}_{\rm x}$(rs)ab              &   16.7$\pm$0.1   &  11.6$\pm$0.2  &   257    \\
    NGC 3692   & (R'L)SA(r)0/a sp                                    &   6.8$\pm$0.1   &  4.5$\pm$0.0   &     274  \\
    NGC 3887   & (RL)SA$\underline{\rm B}_{\rm x}$(rs)bc              &    9.4$\pm$0.4   &  9.0$\pm$0.3   &    359   \\
    NGC 4013   & SA$\underline{\rm B}_{\rm x}$a spw/E(d)7             &   16.1$\pm$0.3   &  13.6$\pm$0.1  &   246   \\
    NGC 4123   & SB$_{\rm x}$(rs)a$\underline{\rm b}$                 &   10.6$\pm$0.2   &  6.2$\pm$0.2   &   285   \\
    NGC 4192   & (R$_{1}$')SAB$_{\rm x}$(rs,nd)$\underline{\rm a}$b    &   37.1$\pm$0.9   & 28.6$\pm$0.7   &  332    \\
    NGC 4216   & (R$_{2}$')SAB$_{\rm ax}$(r,nd)$\underline{\rm a}$b sp/E7-8 & 34.0$\pm$0.6&  23.4$\pm$0.56 &   200    \\
    NGC 4220   & (L)SAB(r)0$^+$                                      &   10.0$\pm$0.1   &  8.2$\pm$0.1   &    316  \\
    NGC 4235   & S$_{\rm x}$0$^+$ sp                                   &   14.9$\pm$0.7   &   12.9$\pm$0.1 &   229    \\
    NGC 4268   & S$\underline{\rm A}$B($\underline{\rm r}$s)0$^+$: sp &  11.6$\pm$0.4   &   8.7$\pm$0.8   &    228   \\
    NGC 4293   & $\underline{\rm R}$(L)SB$_{\rm x}$(r$\underline{\rm s}$)0/a & 18.8$\pm$0.9 &  14.4$\pm$0.5   &   256    \\
    NGC 4302   & SB$_{\rm xa}$?[0/a]bc sp/E7                           &   14.4$\pm$0.3     &  11.0$\pm$0.1   &   359   \\
    NGC 4343   & (R')SAB(r)0/$\underline{\rm a}$ sp/E2                &   6.7 $\pm$0.1    &   3.4$\pm$0.1   &    314   \\
    NGC 4419   & SA$\underline{\rm B}_x$0/$\underline{\rm a}$ sp/E6   &    6.8$\pm$0.2    &   6.9$\pm$0.4   &   312   \\
    NGC 4429   & SAB$_{\rm x}$(r,nl)0$^+$                                 &   23.9$\pm$0.5  &   18.3$\pm$0.4  &    276   \\
    NGC 4435   & S0$^{\circ}$ sp/SB0$^{\circ}$ sp                        &    7.1$\pm$0.1    &   5.2$\pm$0.1   &   189    \\
    NGC 4462   & SAB$_{\rm x}$(rs)a                                    &   11.1$\pm$0.2    &  8.5$\pm$0.0   &    304  \\
    NGC 4488   & SB$_{\rm x}$(s)a                                      &   20.1$\pm$0.7    &   9.7$\pm$0.9   &  327 \\
    NGC 4565   & SB$_{\rm x}$(r)$\underline{\rm a}$b spw               &   25.2$\pm$0.3    &  30.0$\pm$0.7   &  314    \\
    NGC 4569   & (R'L)SA$\underline{\rm B}_{\rm x}$(rs,x$_1$r)a        &   24.8$\pm$0.4    &  16.3 $\pm$0.6   &  196    \\
    NGC 4586   & SA$\underline{\rm B}_{\rm x}$(s,nd)0/a sp             &   19.8 $\pm$0.4   &  19.6$\pm$0.8   &    293    \\
    NGC 4710   & SB$_{\rm xa}$(nd)0$^+$ sp/E(d)7                       &    21.5$\pm$0.4    & 17.4 $\pm$0.3   &  207    \\ 
    NGC 4725   & (R')SAB$_{\rm x}$(r,nb)a                              &   43.9$\pm$0.6    & 33.4 $\pm$0.9    &   217   \\
    NGC 4818   & (RL)SA$\underline{\rm B}_{\rm xa}$(s)0$^{\circ}$        &   10.3$\pm$0.3    &  9.0$\pm$0.2    &  190    \\
    NGC 4845   & (R'L)SAB$_{\rm x}$(r'l,nd)0/a                         &   22.9$\pm$0.1    &  19.4$\pm$0.9   &  257 \\
    NGC 4856   & (RL)SB0$^-$                                          &   11.4$\pm$0.4    &  6.8$\pm$0.2   &   220\\
    NGC 4902    & SB($\underline{\rm r}$s,bl)a$\underline{\rm b}$      &    8.3$\pm$0.1   &   6.4$\pm$0.2   &  245 \\
\noalign{\smallskip}
    NGC 5005   & (R$_{2}$')SAB$_{\rm xa}$(rs)ab                         &   13.0$\pm$0.2     &  9.7$\pm$0.1   &  251 \\
    NGC 5022   & S$_{\rm ax}$a$\underline{\rm b}$: sp                  &    6.0$\pm$0.1     & 5.0$\pm$0.2    &  203 \\
    NGC 5073   & SA$\underline{\rm B}_{\rm xa}$0/a sp                  &   14.3$\pm$0.1     & 11.1$\pm$0.3   & 330  \\
    NGC 5145   & (R:)SA(r,nl)0$^+$                                   &    3.5$\pm$0.0     & 4.3$\pm$0.1    &   264\\
    NGC 5170   & (R')SAB$_{\rm x}$(r$\underline{\rm l}$)0/a sp        &    13.3$\pm$0.8    &  9.9 $\pm$0.3   &  306 \\
    NGC 5297   & SAB$_{\rm x}$(s)bc sp                                &     6.8$\pm$0.1     & 6.1$\pm$0.1    &   333    \\
    NGC 5353   & SB$_{\rm xa}$0$^+$ sp                                 &     9.6$\pm$0.1    &   4.4$\pm$0.2   &  324    \\
    NGC 5377   & (R$_{1}$')SAB$_{\rm xa}$(r'l,nl)0/$\underline{\rm a}$  &   23.4$\pm$0.9    & 17.4$\pm$0.3   &  219    \\
    NGC 5422   & SA$\underline{\rm B}_{\rm ax}$0$^+$ sp                 &   14.7$\pm$0.4     & 8.8$\pm$0.1   &   333    \\
    NGC 5443   & (R'L)SAB$_{\rm x}$(rs)a sp                            &    11.6$\pm$0.5    &  8.4$\pm$0.3   &  215    \\
    NGC 5448   & (R$_1$L)SAB$_{\rm x}$($\underline{\rm r}$s)a          &    14.5$\pm$0.9    & 10.7$\pm$0.5   &  287    \\
    NGC 5529   & SB$_{\rm x}$ab spw                                    &    13.2$\pm$0.1    & 11.8$\pm$0.2   &  295 \\
    NGC 5689   & (R'L)SAB$_{\rm x}$(r'l,nd)$\underline{\rm 0}$/a       &     8.0$\pm$0.3    & 8.5$\pm$0.2   &   267\\
    NGC 5746   & (R')SB$_{\rm x}$(r,nd)0/a sp                          &    22.5$\pm$0.4    &22.7$\pm$0.6   &   351\\
    NGC 5757   & (R')SB(rs)$\underline{\rm a}$b                      &      5.0$\pm$0.3    &  5.6$\pm$0.1   &  341 \\
    NGC 5777   & (R)S$_{\rm x}$(l,nd)$\underline{\rm 0}$/a sp          &      5.9$\pm$0.1    &  5.6$\pm$0.1   &  322 \\
    NGC 5806   & (R'L)SAB(rs,nrl)ab                                  &    15.4$\pm$0.3     &11.4$\pm$0.3   &  358 \\
    NGC 5854   & (RL)SA$\underline{\rm B}_{\rm x}$(rl)0$^+$ sp         &     6.9$\pm$0.2     & 5.8$\pm$0.3   &  238 \\
    NGC 5864   & (R$\underline{\rm L}$)SB$_{\rm xa}$0$^+$ sp           &     9.8$\pm$0.2     & 7.7$\pm$0.1   &  239 \\
    NGC 5878   & SAB$_{\rm xa}$(rs)ab                                  &    11.0$\pm$0.3     & 7.5$\pm$0.4   &  180 \\
    NGC 5916   & SAB(s)a pec                                          &    6.2 $\pm$0.2    & 6.0$\pm$0.0    &   207 \\
    NGC 5957    & (R$^{\prime}$)SB($\underline{\rm r}$s,bl)a            &      8.4$\pm$0.2   &  6.9$\pm$0.2    &  270 \\
    NGC 7140   & (R')SA$\underline{\rm B}_{\rm x}$(rs,nrl)a$\underline{\rm b}$ & 19.0$\pm$0.5 & 12.1$\pm$0.2   &  195 \\
    NGC 7163   & SAB$_{\rm x}$(s)a                                     &    13.2$\pm$0.5     & 8.5$\pm$0.3   &  274 \\
    NGC 7171   & SAB$_{\rm x}$(s)b                                     &     6.9$\pm$0.3     & 5.2$\pm$0.1   &  293 \\
    NGC 7179   & SB$_{\rm xa}$($\underline{\rm r'}$l)0/a               &     9.6$\pm$0.2     & 6.1$\pm$0.3   &  224 \\
    NGC 7183   & SA$\underline{\rm B}_{\rm xa}$0/a sp pec/E(d)7        &    20.0$\pm$0.4     & 13.9$\pm$0.1   &  259 \\
    NGC 7332   & SB$_{\rm x}$0$^{\circ}$                                &    12.12$\pm$0.2    & 10.2$\pm$0.3   &  339 \\
    NGC 7421    & (R$^{\prime}$)SB(rs,bl)ab                             &     6.0$\pm$0.1    &  5.67 $\pm$0.2 & 270  \\
    NGC 7513   & (R'L)SB(rs)a                                        &     8.8$\pm$0.3     &  6.3$\pm$0.1   &  251 \\
    NGC 7531   & SAB$_{\rm x}$(r)a                                     &   12.3$\pm$0.2      & 7.1$\pm$0.1   &  194 \\
    PGC 45650  & SA$\underline{\rm B}_{\rm a}$(s)ab                    &    4.3$\pm$0.1      & 5.6$\pm$0.2  &   268 \\

\end{longtable}/
}

\clearpage
 \begin{table*}
\caption{Galaxies without barlenses or X-shape features, mainly
  unbarred. They are divided to weakly barred (AB), unbarred with
  inner lenses (A$_{\rm l}$), and unbarred without inner lenses
  (A$_{\rm expo}$).  }
\label{table:3}      
\centering          
\begin{tabular}{l l}      
\hline\hline       
\noalign{\smallskip}
Galaxy & Hubble type                  \\
       & (Buta et al. 2015)    \\
\noalign{\smallskip}
\hline              
\noalign{\smallskip}    
AB:     &                                      \\
 \noalign{\smallskip}
    NGC 474  &   (R')SAB(r'l)0/a pec             \\ 
     NGC 584  &  SA(l)0$^-$/E(d)2                   \\
     NGC 1371 &    (RL)SAB(rs,l)a                  \\ 
     NGC 1389 &   SAB(l,nb)0$^-$                     \\
     NGC 2681 &   ($\underline{\rm R}$L)SAB($\underline{\rm r}$s)$\underline{\rm 0}$/a AB    \\ 
     NGC 4267 &   (L)SAB0$^-$                        \\
     NGC 4457 &   (RR)SAB(l)0$^+$                   \\ 
     NGC 7098 &   (R)SAB$_{\rm a}$(r'l,nb)0/a          \\
     IC 2035  &   SAB(s)0$^{\circ}$                    \\
 \noalign{\smallskip}
 \noalign{\smallskip}
A$_{\rm l}$:         &                                      \\ 
 \noalign{\smallskip}
     NGC 524  &   (L)SA(l,nl)0$^{\circ}$                \\ 
     NGC 1297 &   SA(r$\underline{\rm l}$,l)0$^{\circ}$ \\ 
     NGC 1411 &   (L)SA(l,nl)0$^{\circ}$                \\ 
     NGC 2196 &  SA(l)a                                \\
     NGC 2300 &  (R'L)SA(s,l)0$^{\circ}$                 \\
     NGC 2380 &  SA(l,nl)0$^-$                       \\ 
     NGC 3065 &   (L)SA(l)0$^{\circ}$                   \\
     NGC 3599 &   SA(l)0$^{\circ}$                      \\
     NGC 3904 &   SA(l,nl)0$^-$                    \\ 
     NGC 3928 &   SA(l,nl)0$^+$                     \\
     NGC 4339 &  SA(r,l)0$^{\circ}$                  \\
     NGC 4459 &  E2/SA(l)0$^-$                     \\
     NGC 4489 &   SA(l)0$^-$                        \\ 
     NGC 4503 &   $\underline{\rm A}$B(s,l)$^{\circ}$  \\ 
     NGC 4552 &  SA0-/SA(l)0$^-$                      \\
     NGC 5273 &   SA(l,s)0$^{\circ}$                    \\ 
     NGC 5311 &   (L)SA(l,nl)0$^-$                      \\ 
     NGC 5485 &  E(dust lane)/SA(l)0$^-$                 \\ 
     NGC 5631 &   E0-1(S4G)/[SA(l)0$^-$]                \\
     NGC 5638 &  SA(l)0-                                 \\ 
     NGC 5846 &   E+0/[(L)SA(l,nl)0$^{\circ}$]           \\ 
     NGC 5898 &  (L)SA(l,nl)0$^-$                      \\
     NGC 6703 &  (RL)SA(l)0$^{\circ}$                   \\
     NGC 6958 &  SA(l)0$^-$                            \\ 
     NGC 7192 &  (L)SA(l)0$^-$                        \\ 
     NGC 7217 &  (R')SA(l,nl)0/a                       \\ 
     NGC 7377 &  SA(l)0$^-$                            \\
     IC 2764  &   ($\underline{\rm R}$L)SA(l)0$^+$    \\
     IC 5267  &   (RL)SA(r,l)0/a                      \\

 \noalign{\smallskip}
 \noalign{\smallskip}
A:            &                                      \\
 \noalign{\smallskip}
    NGC 3998  &  SA(r)0$^{\circ}$                     \\ 
    IC 4329  &  SA0$^{\circ}$/shells/ripples                  \\
    IC 4991  &  coreE/[SA0$^-$]                       \\

\hline      
 \noalign{\smallskip}        
    \noalign{\smallskip} 
\end{tabular}
\end{table*}

\clearpage

\begin{table*}
\caption{
The fractions of galaxies with various morphological features in the barlens groups a-g.
The classifications of the features are from Buta et al. (2015) and Laurikainen et al. (2011).
The second column indicates the fractions of B and AB families, while the number in
parenthesis is the total number of galaxies in the group. In other columns the percentage of 
the galaxies with features is shown together with a binomial uncertainty. By nuclear features
we mean nuclear bars and rings. The last
column gives the mean galaxy mass in each of the barlens groups. The masses are from \citet{munozmateos2013}.  }   

\label{table:4}      
\centering          
\begin{tabular}{l l l l l l l l l}   
\hline\hline       
\noalign{\smallskip}
 \noalign{\smallskip}
bl-grp & B/AB               & {r+rl}      &  l          & R+RL       & L           & nuclear     & ansae    & $<$logM*/M$_{\odot}$ $>$ \\
       & $\%$ (N)           & $\%$ (N)    & $\%$ (N)    &$\%$ (N)    & $\%$ (N)    & $\%$ (N)    & $\%$ (N) &           \\
\noalign{\smallskip}
\hline              
\noalign{\smallskip}      
a & 100/... (8)  & 100$\pm$0 (8)  & ...            & 37$\pm$17 (3)  & 12$\pm$12  (1)  &  25$\pm$15 (2)  & 12$\pm$12 (1) & 10.45$\pm$0.11    \\
 b & 75/25 (12)  & 100$\pm$0 (12) & ...            & 92$\pm$8 (11)  & 8$\pm$8 (1)    &  42$\pm$14 (5)  & 17$\pm$11 (2) & 10.45$\pm$0.11  \\
 c & 85/15 (7)    & 71$\pm$17 (5)  & 14$\pm$13 (1) & 42$\pm$19 (3)  & 57$\pm$19 (4)  &  28$\pm$17 (2)  & 57$\pm$19 (4) &10.54$\pm$0.13   \\
 d & 100/... (5)   & 80$\pm$17 (4)  & ...            & 60$\pm$21 (3)  & ...             &  100$\pm$0 (5)  & ...            & 10.86$\pm$0.14  \\  
 e & 69/31 (16)  & 63$\pm$12 (10) & 12$\pm$8 (2)  & 38$\pm$12 (6)  & 19$\pm$10 (3)  &  50$\pm$12 (8)  & 44$\pm$12 (7) & 10.66$\pm$0.08  \\
 f & 17/83 (6)    & 83$\pm$15 (5)  & ...            & 67$\pm$19 (4)  & 17$\pm$15 (1)  &  50$\pm$20 (3)  & 33$\pm$19 (2) & 10.75$\pm$0.12  \\
 g & 0/100 (6)     & 50$\pm$20 (3)  & 17$\pm$15 (1) & 83$\pm$15 (5)  & ...             &  33$\pm$19 (2)  & 33$\pm$19 (2) & 10.45$\pm$0.31  \\
\hline      
 \noalign{\smallskip}        
    \noalign{\smallskip} 
\end{tabular}
\end{table*}

\newpage 
\begin{table*}
\caption{ Similar fractions as given in Table 3, but shown for the parent galaxy groups 1-7.}       
\label{table:5}      
\centering          
\begin{tabular}{l l l l l l l l l}     
\hline\hline 
\noalign{\smallskip}      
gal-grp & B/AB            & {r+rl}           &  l           & R+RL           & L         & nuclear          & ansae         & $<$logM*/M$_{\odot}$ $>$ \\ 
       & $\%$ (N)          &  $\%$ (N)         &  $\%$ (N)     &$\%$ (N)         &$\%$ (N)    &$\%$ (N)           & $\%$ (N)       &    \\
\noalign{\smallskip}
\hline 
\noalign{\smallskip}
 1a & 38/62 (8)  &  50$\pm$17 (4)   & 12$\pm$11 (1) &  75$\pm$15 (6)  &  25$\pm$15 (2) &   25$\pm$15 (2)  &   ...             &  10.30$\pm$0.19   \\
 1b & 71/28 (7)  &  14$\pm$13 (1)   & 14$\pm$13 (1) &  14$\pm$13 (1)  &  86$\pm$13 (6) &   14$\pm$13 (1)  &  100$\pm$0 (7)  &  10.48$\pm$0.06  \\
 2  & 33/67 (18)&  78$\pm$10 (14)  & 17$\pm$9 (3)  &  89$\pm$7 (16)  &  5$\pm$5 (1)   &   56$\pm$12 (10) &  50$\pm$12 (9)  &  10.58$\pm$0.07  \\
 3  & 100/.. (5)  &  100$\pm$0 (5)   & ...          &  40$\pm$22 (2)  &  60$\pm$22 (3) &   20$\pm$18 (1)  &  20$\pm$18 (1)& 10.83$\pm$0.13  \\
 4  & 83/16 (6)  &  100$\pm$0 (6)   & ...           &  33$\pm$19 (2)  &  17$\pm$15 (1) &   50$\pm$20 (3)  &  33$\pm$19 (2)& 10.22$\pm$0.10  \\
 5  & 85/12 (7)  &  100$\pm$0 (7)   & ...           &  86$\pm$13 (6)  &  ...          &   86$\pm$13 (6)  &  ...       & 10.71$\pm$0.15  \\
 6  & 33/67 (6)  &  50$\pm$20 (3)   & 16$\pm$15 (1) &  50$\pm$20 (3)  &  17$\pm$15 (1) &   50$\pm$20 (3)  &  17$\pm$15 (1)   & 10.70$\pm$0.10  \\
 7  & 67/33 (12) & 100$\pm$0 (12)   & ...          &  50$\pm$14 (6)  &  ...          &   33$\pm$13 (4)  &  ...      & 10.55$\pm$0.11  \\

\hline  
       \noalign{\smallskip} 
           \noalign{\smallskip} 
 \noalign{\smallskip} 
 \noalign{\smallskip} 
 \noalign{\smallskip} 

\end{tabular}
\end{table*}

\begin{table*}
\caption{ Cross checking the parent galaxy (1 -- 7) and barlens (a -- g) groups for the galaxies in which both definitions exist.
Notice that both group definitions appear only for a small number of galaxies in our sample.}       
\label{table:6}      
\centering          
\begin{tabular}{lllllllll}      
\hline\hline       
\noalign{\smallskip}
 & 1a & 1b &  2 & 3 & 4 & 5 & 6 & 7 \\ 
\hline
\noalign{\smallskip}
a: &  &    & 1& 2& 2&  &  & 2\\
b: &  &    & 3& 3& 2& 1&  & 2\\
c: & 1&  3 & 1&  & 1&  &  &   \\
d: &  &    & 1&  &  & 3& 1&   \\
e: & 2&  2 & 5& 1& 2&  & 1& 1 \\  
f: &  &    & 2&  &  & 1& 3&   \\
g: & 3&  1 & 2&  &  &  &  &   \\
 \hline                  
\end{tabular}
\end{table*}

\end{document}